\documentclass{statsoc}
\usepackage[a4paper]{geometry}
\usepackage{amsmath, amssymb,mathtools}
\usepackage{amsfonts, multirow, epsfig, floatrow}
\usepackage{graphicx, pdflscape, verbatim, enumerate, colortbl, setspace}
\usepackage{setspace, color,bm}
\usepackage[normalem]{ulem}
\usepackage{natbib}
\usepackage{multirow}
\usepackage{booktabs,array}
\usepackage{hyperref}
\usepackage{float}
\usepackage{caption}
\usepackage{subcaption}
\usepackage{tikz}
\usepackage{algorithm, algorithmicx, algpseudocode}
 \usetikzlibrary{shapes.misc}
\usepackage{makecell}
\usepackage{multibib}
\newcites{supp}{Supplementary References}
\usepackage{afterpage}

\newtheorem{Proposition}{Proposition}
\newtheorem{Condition}{Condition}
\newtheorem{Lemma}{Lemma}

\newtheorem{Theorem}{Theorem}
 
\newtheorem{Remark}{Remark}

\DeclareMathOperator*{\argmin}{arg\,min}

\newcommand{\pz}{p_{\rm z}}
\newcommand{\px}{p_{\rm x}}
\newcommand{\R}{\mathbb{R}}
\newcommand{\E}{{\mathbf{E}}}

\newcommand{\cip}{\overset{p}{\to}}
\newcommand{\cid}{\overset{d}{\to}}

\newcommand{\V}{\mathcal{V}}

\newcommand{\PP}{\mathbb{P}}

\newcommand{\vb}{v}
\newcommand{\T}{\mathbf{M}}
\newcommand{\betaRF}{\widetilde{\beta}_{\rm RF}}
\newcommand{\betaHe}{\widehat{\beta}_{\rm RF}}
\newcommand{\conpara}{\mu(V)}
\newcommand{\Err}{{\rm Err}}
\newcommand{\Pm}{P_{BW}}
\newcommand{\VP}{V_{BW}}
\newcommand{\Pb}{P_{\VP, W}^{\perp}}
\newcommand{\aw}{\overrightarrow{\bf w}}

\newcommand\Peter[1]{{\color{red}Peter: ``#1''}}
\newcommand\Zijian[1]{{\color{blue}Zijian: ``#1''}}

\title[Two-stage Curvature Identification]{Robustness Against Weak or Invalid Instruments: Exploring Nonlinear Treatment Models with Machine Learning}

\author{Zijian Guo, Mengchu Zheng}
\address{Department of Statistics, Rutgers University, USA}
\author[Guo, Zheng and B\"{u}hlmann]{Peter B\"{u}hlmann}
\address{Seminar for Statistics, ETH Z\"{u}rich, Switzerland.}

\begin{document}

\maketitle 

\begin{abstract}
We discuss causal inference for observational studies with possibly invalid instrumental variables. We propose a novel methodology called two-stage curvature identification (\texttt{TSCI}) by exploring the nonlinear treatment model with machine learning. {The first-stage machine learning enables improving the instrumental variable's strength and adjusting for different forms of violating the instrumental variable assumptions.} The success of \texttt{TSCI} requires the instrumental variable's effect on treatment to differ from its violation form. A novel bias correction step is implemented to remove bias resulting from the potentially high complexity of machine learning. Our proposed \texttt{TSCI} estimator is shown to be asymptotically unbiased and Gaussian even if the machine learning algorithm does not consistently estimate the treatment model. Furthermore, we design a data-dependent method to choose the best among several candidate violation forms. We apply \texttt{TSCI} to study the effect of education on earnings.
\end{abstract}

%%%%%%%%%%%%%%%%%%%%%%%%%%%%%%%%%%%%%
\section{Introduction}
\label{sec:intro}
%%%%%%%%%%%%%%%%%%%%%%%%%%%%%%%%%%%%%

Observational studies are major sources for inferring causal effects when randomized experiments are not feasible. But such causal 
inference from observational studies requires strong assumptions and may be invalid due to the presence of unmeasured confounders. The instrumental variable (IV) regression is a practical and highly popular causal inference approach in the presence of unmeasured confounders. %{As illustrated in Figure \ref{fig: SEM valid},} 
The IVs are required to satisfy three assumptions: conditioning on the baseline covariates, (A1) the IVs are associated with the treatment;
(A2) the IVs are not associated with the unmeasured confounders; (A3) the IVs do not directly affect the outcome.
Despite the popularity of the IV method, there is a significant concern about whether the used IVs satisfy (A1)-(A3) in practice. Assumption (A1) requires the IV to be strongly associated with the treatment variable, which can be checked with an F-test in a first-stage linear regression model. 
Inference with the assumption (A1) being violated has been actively investigated under the name of weak IV \citep[e.g.]{stock2002survey,staiger1997instrumental}. 
Assumptions (A2) and (A3) ensure that the IV only affects the outcome through the treatment. {If an IV violates either (A2) or (A3), we call it an invalid IV and define its functional form of violating (A2) and (A3) as the violation form, e.g., {a} linear violation.} In the just-identification regime, most empirical {analyses} rely on external knowledge to {argue about the validity of} (A2) and (A3). {However, there is a pressing need to develop a robust causal inference method against the proposed IVs violating the classical assumptions. }  

%\Peter{Sounds good!}
%where machine learning algorithms are employed to provide valid causal inference even if some proposed IV.  
%\Zijian{Peter, please check the above change and I think the AE has a wrong impression that we can use any covariate as invalid IV. So I think we may need to clarify this and we might remove the following part?}
%We propose here a framework that does not rely on such additional assumptions. %\Peter{Are you suggesting to delete the previou sentence: I think it would be good.}

%%%%%%%%%%%%%%%%%%%%%%%%%%%%%%%%%%%%%%%%%%%%%%%%%%%%%%%%%%%%%%%%%%%%%%%%%%%%%%%%%%%%%%%%%%%%%%%%%%%%%%%%%%%%%%%%%%
\subsection{Our results and contribution}%\Zijian{Peter, may u double check this subsection?}}
%%%%%%%%%%%%%%%%%%%%%%%%%%%%%%%%%%%%%%%%%%%%%%%%%%%%%%%%%%%%%%%%%%%%%%%%%%%%%%%%%%%%%%%%%%%%%%%%%%%%%%%%%%%%%%%
We aim to devise a robust IV framework that leads to causal identification even if the IVs {proposed by domain experts} may violate assumptions (A1) to (A3). {Our framework provides a robustness guarantee even when all proposed IVs are invalid, including the most common regime with a single IV that is possibly invalid.} It is well known that the treatment effect is not identifiable when there is no constraint on how the IVs violate assumptions (A2) and (A3).  Our key identification assumption is that the violations of (A2) and (A3) arise from simpler forms than the association between the treatment and the IVs; that is, we exclude ``special coincidences" such as {the IVs violate (A2) and (A3) by linear forms and the conditional mean model of the treatment given IVs is also linear. } %the IVs' violations as well as the association between treatment and IVs are all linear. 
{This identification condition can be evaluated with the generalized IV strength measure introduced in Section \ref{sec: IV strength} for a user-specific violation form.}

We propose a novel two-stage curvature identification (\texttt{TSCI}) method to infer the treatment effect. An important operational step is to fit the treatment model with a machine learning (ML) algorithm, e.g., random forests, boosting, or deep neural networks. {This first-stage ML model enhances the IV's strength by learning a general conditional mean model of treatment given the IVs, instead of restricting to a linear association model.} {In the second stage, we leverage the ML prediction and adjust for different IV violation forms.} {A} main novelty of our proposal is to estimate the bias resulting from {high complexity of the first-stage ML} and implement {a corresponding} bias correction step; see more details in Section \ref{sec: second stage}.

% We show that our developed TSCI methodology yields a consistent estimator of the treatment effect assuming the above mentioned ``different functional forms" of instrument-treatment associations and violations. In addition, 
We show that the \texttt{TSCI} estimator is asymptotically Gaussian when the generalized IV strength measure is sufficiently large. We further devise a data-dependent way of choosing the most suitable violation form among a collection of IV violation forms. {By including the valid IV setting into the selection,} our proposed general \texttt{TSCI} methodology does not exclude the possibility of the IVs being valid but provides a robust IV method {against} different invalidity forms.

To sum up, the contribution of the current paper is {two-fold:}
%two-folded:
\begin{enumerate}
\item %Our proposed \texttt{TSCI} method is a robust IV approach that provides valid causal inference even if the IVs violate the classical assumptions . 
\texttt{TSCI} explores a nonlinear treatment model with ML and leads to more reliable causal conclusions than existing methods assuming valid IVs. 
\item {\texttt{TSCI} provides an efficient way of integrating the first-stage ML.} {A}  
methodological novelty of \texttt{TSCI} is its  
bias correction in {its} second stage, which addresses the issue of {high complexity and potential overfitting of} ML. %Interestingly, {the asymptotic normality and validity of the TSCI confidence interval do not require the ML algorithm to consistently estimate the conditional mean of the treatment model.}
%does not need to be consistent for its classical prediction task. 
%algorithm.
\end{enumerate}

\subsection{Comparison to existing literature}
%%%%%%%%%%%%%%%%%%%%%%%%%%%%%%%%%%
There is relevant literature on causal inference when assumptions (A2) and (A3) are violated. When all IVs are invalid, 
\citet{lewbel2012using,tchetgen2017genius} proposed an identification strategy by assuming that the treatment model has heteroscedastic error but the covariance between the treatment and outcome model errors is 
homoscedastic. Our proposed \texttt{TSCI} is based on a different idea by exploring the nonlinear structure between the treatment and the IVs, not requiring anything on homo- and heteroscedasticity. 
\citet{bowden2015mendelian} and \citet{Kole15} assumed that the direct effect of the IVs on the outcome is nearly orthogonal to the effect of the IVs on the treatment.  \citet{han2008detecting,Bowden16,kang2016instrumental,windmeijer2019use,guo2018confidence,windmeijer2019confidence} considered the setting that a proportion of IVs are valid and conducted causal inference by selecting valid IVs in a data-dependent way. Along this line, \citet{guo2021causal} proposed a uniform inference procedure that is robust to the IV selection error. {More recently, \citet{sun2021semiparametric} leveraged the number of valid IVs and used the interaction of genetic markers to identify the causal effect.} In contrast to assuming the orthogonality and existence of valid IVs, our proposed \texttt{TSCI} is effective even if \emph{all} IVs are invalid and the effect orthogonality condition does not hold. {Such a setting is especially useful in handling econometric applications with a single IV.}

The nonlinear structure of the treatment model has been explored in the econometric and the more recent ML literature. In the IV literature, \citet{kelejian1971two,amemiya1974nonlinear,newey1990efficient,hausman2012instrumental} considered constructing a non-parametric treatment model and
\citet{belloni2012sparse} proposed constructing the optimal IVs with a Lasso-type first-stage estimator. %\Mengchu{\citet{hausman2012instrumental} also used a non-parametric treatment model and proposed a jackknife version of LIML to correct the inconsistency with many weak IVs and heteroskedasticity.}
 More recently, \citet{chen2020mostly,liu2020deep} proposed constructing the IV as the first-stage prediction given by the ML algorithm. All of the above works are focused on the valid IV setting. In contrast, {our current paper applies to the more robust regime where the IVs are allowed to violate assumptions (A2) or (A3) in a range of forms.} Even in the context of valid IVs, our proposed \texttt{TSCI} may lead to a more accurate estimator than directly using the predicted value for the treatment model as the new IV \citep{chen2020mostly,liu2020deep}; see Section \ref{sec: dml comp} for details.

ML algorithms integrated into the IV analysis to better accommodate the complicated outcome model. 
In Sections \ref{sec: dml comp} and \ref{sec: dml valid IV sim}, we provide detailed comparisons to Double Machine Learning (\texttt{DML}) proposed in \citet{chernozhukov2018double}. \citet{athey2019generalized} proposed the generalized random forests to infer heterogeneous treatment effects. Both works are based on valid IVs, while {our current paper assumes a homogeneous treatment effect} and focuses on the different robust IV framework with {possibly invalid IVs.}

{Finally, our proposed TSCI methodology can be used to check the IV validity for the just-identification case.  This is distinct from the Sargan test, J test, or specification test \citep{hansen1982large,sargan1958estimation,woutersen2019increasing} which are used to test IV validity in the over-identification case.}%\citet{hartford2017deep} proposed the Deep IV method by fitting the outcome and treatment models with deep neural networks; \citet{bennett2019deep} studied the deep generalized {method} of moments with IVs; \citet{xu2020learning} applied the deep neural networks to define the nonlinear features of the treatments and IVs.
 %\Peter{just to understand this. You want to add another point of description here? And do we really have a rigorous test?} \Zijian{We do have a rigorous test. The question is that there are IV validity tests in the econ literature. The reviewers brought this up. So if we have more IVs than the treatments, then we know whether all IVs are valid as a total. } %In any case, we may start this paragraph with:\\
%... \\
%And we could replace "can be used to test" with "can be used to check"?}
%\Peter{OK -- it's fine now.}

\noindent {\bf Notation.} For a matrix $X\in \R^{n\times p}$, a vector $x\in \R^{n}$, and a set $\mathcal{A}\subset \{1,\cdots,n\}$, we use $X_{\mathcal{A}}$ to denote the submatrix of $X$ whose row {indices}
%indexes 
belong to $\mathcal{A}$, and $x_{\mathcal{A}}$ to denote the sub-vector of $x$ with indices in $\mathcal{A}.$ For a set $\mathcal{A}$, $\left|\mathcal{A}\right|$ denotes its cardinality. {For a vector $x\in \R^{p}$, the $\ell_q$ norm of $x$ is defined as $\|x\|_{q}=\left(\sum_{l=1}^{p}|x_l|^q\right)^{\frac{1}{q}}$ for $q \geq 0$ with $\|x\|_0=\left|\{1\leq l\leq p: x_l \neq 0\}\right|$ and $\|x\|_{\infty}=\max_{1\leq l \leq p}|x_l|$.} We use $c$ and $C$ to denote generic positive constants that may vary from place to place. For a sequence of random variables $X_n$ indexed by $n$, we use $X_n \cip X$ and $X_{n} \cid X$ to represent that $X_n$ converges to $X$ in probability and in distribution, respectively. For two positive sequences $a_n$ and $b_n$,  $a_n \lesssim b_n$ means that $\exists C > 0$ such that $a_n \leq C b_n$ for all $n$; $a_n \asymp b_n $ if $a_n \lesssim b_n$ and $b_n \lesssim a_n$, and $a_n \ll b_n$ if $\limsup_{n\rightarrow\infty} {a_n}/{b_n}=0$. For a matrix $M$, we use ${\rm Tr}[M]$ to denote its trace, ${\rm rank}(M)$ to denote its rank, and $\|M\|_{F}$, $\|M\|_2$ and $\|M\|_{\infty}$ to denote its Frobenius norm, spectral norm and element-wise maximum norm, respectively. For a square matrix $M$, $M^2$ denotes the matrix multiplication of $M$ by itself. For a symmetric matrix $M$, we use $\lambda_{\max}(M)$ and $\lambda_{\min}(M)$ to denote its maximum and minimum eigenvalues, respectively. 
%%%%%%%%%%%%%%%%%%%%%%%%%%%%%%%%%%%%%%%%%%%%%%%%%%%%%%%%%%%%%%%%%%%%%%%%%%%%%
\section{Invalid instruments: modeling and identification}
\label{sec: model}
%%%%%%%%%%%%%%%%%%%%%%%%%%%%%%%%%%%%%%%%%%%%%%%%%%%%%%%%%%%%%%%%%%%%%%%%%%%%%
We consider i.i.d. data $\{Y_i,D_i,Z_i,X_i\}_{1\leq i\leq n},$ where for the $i$-th observation, $Y_i\in \R$ and $D_i\in \R$ respectively denote the outcome and the treatment and $Z_i\in \R^{p_z}$ and $X_i\in \R^{p_x}$ respectively denote the IVs and measured covariates. 
%{In Section \ref{sec: example}, we start with an example motivating the identification strategy. We introduce in Section \ref{sec: general} the general invalid IV model and provide the causal interpretation in Section \ref{sec: causal interp}. We provide an overview of our proposed \texttt{TSCI} methodology in Section \ref{sec: overview}.} 

%%%%%%%%%%%%%%%%%%%%%%%%%%%%%%%%%%%%%%%%%%%%%%%%%%%%%%%%%%%%%%%%%%%%%%%%%%%%%
\subsection{Examples for effect identification with nonlinear treatment models}
\label{sec: example}
%%%%%%%%%%%%%%%%%%%%%%%%%%%%%%%%%%%%%%%%%%%%%%%%%%%%%%%%%%%%%%%%%%%%%%%%%%%%%
{To explain the main idea,}  we start with the special case with no covariates, %we consider the following outcome and treatment models with no covariates,
\begin{equation}
Y_i=D_i \beta + Z_i^{\intercal}\pi+\epsilon_i, \quad \text{and}\quad D_i=f(Z_i)+\delta_i, \quad \text{for}\quad 1\leq i\leq n,
\label{eq: examples}
\end{equation}
where the errors $\epsilon_i$ and $\delta_i$ satisfy $\E(\epsilon_i \mid Z_i)=0$ and $\E(\delta_i\mid Z_i)=0.$ {These errors $\epsilon_i$ and $\delta_i$ are typically correlated due to unobserved confounding.} The outcome model in \eqref{eq: examples} is commonly used in the invalid IV literature \citep{small2007sensitivity,kang2016instrumental,guo2018confidence,windmeijer2019use}, with $\pi \neq 0$ standing for the violation of IV assumptions (A2) and (A3). The treatment model in \eqref{eq: examples} is not a causal generating model between $D_i$ and $Z_i$ but only stands for a probability model between $D_i$ and $Z_i$. {Specifically, the treatment assignment might be directly influenced by unmeasured confounders, but the treatment model in \eqref{eq: examples} does not explicitly state this generating process. Instead, for the joint distribution of $D_i$ and $Z_i$, we define the conditional expectation $f(Z_i)=\E(D_i\mid Z_i),$ which leads to the treatment model in \eqref{eq: examples}. We discuss the structural equation model {interpretation} of the model \eqref{eq: examples} in Section \ref{sec: causal interp}, which explicitly {characterizes}
how unmeasured confounders may affect the treatment assignment.}

{We introduce some projection notations to facilitate the discussion.} For a random variable $U_i$, define its best linear approximation by $Z_i$ as $\gamma^*=\argmin_{\gamma\in \R^{p_z+1}}\E(U_i-\widetilde{Z}_i^{\intercal}\gamma)^2$ with $\widetilde{Z}_i=(1,Z_i^{\intercal})^{\intercal}$. For $1\leq i\leq n$, write $\mathcal{P}_{Z_i} U_i=\widetilde{Z}_i^{\intercal}\gamma^*$ and $\mathcal{P}^{\perp}_{Z_i} U_i=U_i-\widetilde{Z}_i^{\intercal}\gamma^*.$ We apply $\mathcal{P}^{\perp}_{Z_i}$ to the model \eqref{eq: examples} and obtain 
$
\mathcal{P}^{\perp}_{Z_i}Y_i=\mathcal{P}^{\perp}_{Z_i}D_i \beta+\mathcal{P}^{\perp}_{Z_i}\epsilon_i$, where {the linear violation form $Z_i^{\intercal}\pi$ in \eqref{eq: examples} is removed after applying the transformation $\mathcal{P}^{\perp}_{Z_i}$.} If $f$ is nonlinear, then ${\rm Var}\left(\mathcal{P}^{\perp}_{Z_i}f(Z_i)\right)>0$ and the adjusted variable $\mathcal{P}^{\perp}_{Z_i}f(Z_i)$ can be used as a valid IV to identify the effect $\beta$ via the following estimating equation
\begin{equation}
\E\left[\mathcal{P}^{\perp}_{Z_i}f(Z_i)(Y_i-D_i\beta)\right]=\E\left[\mathcal{P}^{\perp}_{Z_i}f(Z_i)(Z_i^{\intercal}\pi+\epsilon_i)\right]=0.
\label{eq: adjustment}
\end{equation}
{Note that the last equality follows from the population orthogonality between $\mathcal{P}^{\perp}_{Z_i}f(Z_i)$ and $Z_i$, together with $\E(\epsilon_i \mid Z_i)=0.$}

The estimation equation in \eqref{eq: adjustment} reveals that the treatment effect $\beta$ can be identified by exploring the nonlinear structure of $f$ under the model \eqref{eq: examples}. {{We propose to learn $f$ using a ML prediction model and measure the variability of $\mathcal{P}^{\perp}_{Z_i}f(Z_i)$ by defining a notation of generalized IV strength in Section \ref{sec: IV strength}.} {We shall emphasize that using ML algorithms to learn the possibly nonlinear function $f$ is critical in the current framework, allowing for invalid IVs. The ML algorithms help capture complicated structures in $f$ and {typically} retain enough variability contained in the adjusted variable  $\mathcal{P}^{\perp}_{Z_i}f(Z_i)$.}  {In Section \ref{sec: noncompliance} in the supplement, we present another example for causal identification in the context of the randomized experiments with non-compliance and invalid IVs.}

\begin{Remark} \rm %\Zijian{Peter, may you help carefully check the following remark as both reviewers have some questions related to this? Any suggestion on where to put this remark? It may lead to a relatively long introduction.} \Peter{I think it is appropriate here. W are i section 2, not the introductory section: and it should appear in section 2. Just leave it as it is.} 
Nonlinear functional form identification has been proposed in various existing literature; see \cite{lewbel2019identification}[Sec.3.7] for a review. More concretely,  
{\citet{angrist2009mostly}[Sec.4.6.1] discussed the identification of when the IV has a linear direct effect on the outcome, but the treatment model is a nonlinear parametric model of the IV. The Heckman selection model \cite{heckman1976common} offered a practical and simple solution for estimating effects when the samples are not randomly selected. When no variables satisfy the exclusion restriction, one possible identification relies on the nonlinearity of the inverse Mills ratio. As pointed by \cite{puhani2000heckman}, such an identification is less reliable when the nonlinear model form is relatively weak in applications. Additionally, \cite{shardell2016instrumental,ten2007causal} proposed the causal identification using the interaction of the (binary) IV and baseline covariates as the new IV.  \cite{lewbel2019identification} commented that the proposed identification fails when the nonlinear functional form does not exist or is relatively weak. As a main difference, our proposal does not use hand-picked instruments or assume the data-generating distribution has some pre-specified nonlinear functional forms. Operationally, we apply the first-stage ML algorithm to learn the complex nonlinear treatment model, which is more powerful in detecting the nonlinear dependence based on exploring the data. However, including the first-stage ML requires a more delicate following-up statistical procedure, such as the bias correction step in \eqref{eq: ML corrected hetero}. More importantly, instead of betting on the existence of nonlinearity, our proposed generalized IV strength measure in \eqref{eq: IV strength homo} guides whether there is enough nonlinear structure after adjusting for invalid forms.} 
\end{Remark}
\vspace{-5mm}
%%%%%%%%%%%%%%%%%%%%%%%%%%%%%%%%%%%%%%%%%%%%%%%%%%%%%%%%%%%%%%%%%%%%%%%%%%%%%
\subsection{Model with a general class of invalid IVs}
\label{sec: general}
%%%%%%%%%%%%%%%%%%%%%%%%%%%%%%%%%%%%%%%%%%%%%%%%%%%%%%%%%%%%%%%%%%%%%%%%%%%%%
We now introduce the general outcome model, allowing for more complicated violation forms than the linear violation in \eqref{eq: examples}, 
\begin{equation}
Y_i=D_i\beta+g(Z_i,X_i)+\epsilon_i,\quad \text{with} \quad \E(\epsilon_i\mid X_i, Z_i)=0, \quad \text{for}\quad 1\leq i\leq n
\label{eq: outcome model}
\end{equation} 
where $\beta\in \R$ is the homogeneous treatment effect  and the function  {$g:\R^{p_x+p_z}\rightarrow \R$ encodes how the IVs violate the assumptions (A2) and (A3).} The classical valid IV setting requires that the function $g(z,x)$ does not change with different assignment of $z$. In the invalid IV literature, the commonly used outcome model with a linear violation form can be viewed as a special case of \eqref{eq: outcome model} {when} taking  $g(Z_i,X_i)=Z_i^{\intercal}\pi_z+X_i^{\intercal}\pi_x.$

For the treatment, we define its conditional mean function $f(Z_i,X_i)\coloneqq \E(D_i\mid Z_i,X_i)$ for $1\leq i\leq n,$ leading to the following treatment model:  \begin{equation}
D_{i}=f(Z_{i},X_{i})+\delta_i \quad \text{with}\quad \E(\delta_i\mid Z_i, X_{i})=0.
\label{eq: treatment model}
\end{equation}
The model \eqref{eq: treatment model} is flexible as $f$ might be any unknown function of $Z_i$ and $X_i$, and the treatment variable can be continuous, binary, or {a} count variable. Similarly to \eqref{eq: examples}, the model \eqref{eq: treatment model} is only a probability {relation}
instead of a causal generating model and the treatment assignment is allowed to be {influenced} by unmeasured confounders. %{In Section \ref{sec: causal interp}, we interpret the outcome and treatment models \eqref{eq: outcome model} and \eqref{eq: treatment model} from both {the} structural equation model and the {potential outcome} perspectives.} 

{It is well known that, for the outcome model \eqref{eq: outcome model}, the identification of $\beta$ is generally impossible without additional identifiability conditions on the function $g$. {In this paper, we allow the existence of invalid IVs but constrain the functional form of violating (A2) and (A3) to a pre-specified class. An operational step is to specify a set of basis functions for $g(\cdot)$ in the outcome model \eqref{eq: outcome model}.}

%to a pre-specified class. %\Zijian{Particularly, we assume in the following that the function $g(\cdot)$ can be approximated by {a known} set of basis functions.} 
%Particularly, we allow the function $g(z,x)$ to change with the assignment of $z$ but assume that the function $g(z,x)$ can be approximated by \Zijian{a known} set of basis functions \Peter{We can almost always always do such a basis function approximation... I think; what you might mean is approximated by a known set of basis functions??? I am not sure...}, \Zijian{I fully agree and have added the word.} denoted as $\mathcal{V}$.}

%In the following, we assume that the function $g(\cdot)$ can be approximated by {a known} set of basis functions, denoted as $\mathcal{V}$.  The 
%set $\mathcal{V}$ {is temporarily assumed to be known} and a data-dependent basis set selection procedure is presented in Section \ref{sec: vio sel}. 

%{In the remaining {part} of this subsection, we discuss the specification of $g(\cdot)$ , which encodes the IV violation form.}

Define $\phi(X_i)=\E[g(Z_i,X_i)\mid X_i]$ and decompose $g$  as $g(Z_i,X_i)=h(Z_i,X_i)+\phi(X_i)$ with  $h(Z_i,X_i)=g(Z_i,X_i)-\phi(X_i).$  
 When the IVs $Z_i$ are valid, $g(Z_i,X_i)$ does not directly depend on the assignment of $Z_i$, which implies $\phi(X_i)=g(Z_i,X_i)$ and $h(\cdot)=0.$ %We refer to this $h$ function as the violation function, and a non-zero $h$ function implies a violation of (A2) and (A3). 
%When the IV is valid, the violation function $h(\cdot)\equiv 0,$ and 
In this case, we just need to approximate $g(z,x)=\phi(x)$ by a set of basis functions $\aw(x)\in \R^{p_w}$.  With $\aw(x)$, we approximate $g(Z_i,X_i)=\phi(X_i)$ by a linear combination of $W_i=\aw(X_i) \in \R^{p_w}$ for $1\leq i\leq n$ and $\{\phi(X_i)\}_{1\leq i\leq n}$ by the matrix $W = (W^{\intercal}_1,\ldots ,W^{\intercal}_n)^{\intercal}$. %We now give examples of the set of basis functions $\aw(x).$ 
{For example, if $\phi(x)$ is linear, we set $\aw(x)=\{1,x_1,x_2,\cdots,x_{p_x}\}$. Furthermore, we consider the additive model $\phi(x)=\sum_{j=1}^{p_x} \phi_j(x_j)$ with smooth functions $\{\phi_j(\cdot)\}_{1\leq j\leq p_x}.$ For $1\leq j\leq p_x,$ we construct a set of basis functions  $\overrightarrow{{\mathbf b}_j}=\{b_{j,l}(\cdot)\}_{1\leq l\leq M_j}$ for $\phi_j(x_j)$ with $M_j\geq 1$ denoting the number of basis functions. Examples of the basis functions include the polynomial 
%basis 
or 
%the 
B spline basis. We then approximate $\phi(x)$ with $\aw(x)=\{1,\overrightarrow{{\mathbf b}_1}(x_1),\cdots,\overrightarrow{{\mathbf b}_{p_x}}(x_{p_x})\}.$} %as a long vector stacking the basis functions for all coordinates. 

{When the IV is invalid with $h \not \equiv 0$, in addition to approximating $\phi(x)$ by $\aw(x)$, we further} approximate the function $h(z,x)$ by a set of pre-specified basis functions $v_1(z,x), \cdots, v_L(z,x)$ and then approximate $g(z,x)$ by 
\begin{equation}
\mathcal{V}\coloneqq \{v_1(z,x),\cdots,v_L(z,x), \aw(x)\}\quad \text{with} \quad v_{l}: \R^{p_x+p_z} \rightarrow \R \quad \text{for} \quad 1\leq l\leq L.
\label{eq: basis function}
\end{equation}
We now present different choices of $g(\cdot)$ in \eqref{eq: outcome model}, or equivalently, the basis set $\mathcal{V}$ in \eqref{eq: basis function}. With them, the model \eqref{eq: outcome model} can accommodate a wide range of valid or invalid IV forms.  
\begin{enumerate}
%\item[(1)] \noindent{\bf Valid IV.} We take $g(z,x)=\phi(x)$ and $\mathcal{V}=\{\aw(x)\}.$
\item[(1)] \noindent{\bf Linear violation.} We take $g(z,x)=\sum_{l=1}^{p_z}\pi_l z_l+\phi(x)$ and $\mathcal{V}=\{z_1,\cdots,z_{p_z},\aw(x)\}.$ 
\item[(2)] \noindent{\bf Polynomial violation (for a single IV).} We consider $g(z,x)=h(z)+\phi(x)$, where the IV and baseline covariates affect the outcome in an additive manner. We set 
$\mathcal{V}=\{\vb_1(z),\cdots, \vb_{L}(z),\aw(x)\}$ and may take $\vb_1(z),\cdots, \vb_{L}(z)$ as (piecewise) polynomials of various orders.
\item[(3)] \noindent {\bf Interaction violation (for a single IV).} We consider $g(z,x) = \sum_{l=1}^{p_x} \alpha_l\cdot z \cdot x_l + \phi(x)$, allowing the IV to affect the outcome interactively with the base covariates. For such a violation form, we set $
\mathcal{V}=\{z\cdot x_1,\cdots,z\cdot x_{p_x}, \aw(x)\}.$
\end{enumerate}
We focus on the single IV setting for the polynomial and interaction violation to simplify the notations. It can be extended to the setting with multiple IVs. 

Two important remarks are in order for the choice of $g$ and $\mathcal{V}$. Firstly, the choice of $g$ and the corresponding basis set $\mathcal{V}$ is part of the outcome model specification \eqref{eq: outcome model}, reflecting the users' belief about how the IVs can potentially violate the assumptions (A2) and (A3). In applications, empirical researchers might apply domain knowledge and construct the pre-specified set of basis functions in \eqref{eq: basis function}. For example, \citet{angrist1999using} applied the Maimonides' rule to construct the IV as the transformation of {a variable ``enrollment"} 
and then adjust {with} the possible violation form generated by a linear, quadratic, or piecewise linear transformation of ``enrollment". 
Importantly, we do not have to assume the knowledge of $\mathcal{V}$ and will present in Section \ref{sec: vio sel} a data-dependent way to choose the best $\mathcal{V}$ from a collection of candidate basis sets. Secondly, the linear violation form with $g(z,x)=\sum_{l=1}^{p_z} \pi_l z_l+\phi(x)$ is the most commonly used violation form in the context of {the} multiple IV framework \citep{small2007sensitivity,kang2016instrumental,guo2018confidence,windmeijer2019use}. Most existing methods require at least a proportion of $Z_i$ to be valid, i.e., the corresponding $\pi_l$ being zero. In contrast, our framework allows all IVs to be invalid (i.e., all $\pi_l$ to be nonzero) by taking $\mathcal{V}=\{z_1,\cdots,z_{p_z},\aw(x)\}.$ 

With the set ${\cal V}$ of basis functions in \eqref{eq: basis function}, we define the matrix $V$ {which evaluates the  functions in ${\cal V}$ at the observed variables:}
\begin{equation}
V=\begin{pmatrix} V_{1}&\cdots&V_{n}\end{pmatrix}^{\intercal}\in \R^{n\times (L+p_w)} \quad \text{with} \quad  V_{i}=\left(\vb_1(Z_i, X_i),\cdots, \vb_{L}(Z_i, X_i), W_i^{\intercal}\right)^{\intercal}, 
\label{eq: violation matrix}
\end{equation}
for $1\leq i\leq n$. Note that we always use an intercept $W_{i,1} = 1$. {Instead of assuming $V_i$ perfectly generating $g(Z_i,X_i)$, we go with a broader framework by allowing {an}
approximation error of $g(X_i, Z_i)$ by the best linear {combination} of $V_i$,  defined as} 
\begin{equation}
R_i(V)\coloneqq g(Z_i, X_i)-V^{\intercal}_{i}\pi \quad \text{with} \quad \pi\coloneqq\argmin_{b} \E \left[g(Z_i, X_i)-V_i^{\intercal}b\right]^2. 
\label{eq: approx error}
\end{equation}
Define $R(V)=(R_1(V),\cdots,R_n(V))\in \R^{n}$. We shall omit the dependence on $V$ when there is no confusion.  With \eqref{eq: approx error}, we express the outcome model \eqref{eq: outcome model} as 
\begin{equation}
Y_i=D_i\beta+V_i^{\intercal} \pi+R_i(V)+\epsilon_i,\quad \text{for}\quad 1\leq i\leq n.
\label{eq: outcome model basis}
\end{equation}
If $g(Z_i, X_i)$ is well approximated by a linear combination of $V_i,$  $\|R(V)\|_2/\sqrt{n}$ is close to zero or even $\|R(V)\|_2=0$. %\Mengchu{added the plot. }Under the introduced framework, the outcome model we used in our proposal is robust to a general class of invalid IVs when certain conditions are satisfied, as shown in Figure \ref{fig:TSCI robustness}.

\begin{comment}
\begin{figure}[h]
    \centering
    \includegraphics[width=0.7\linewidth]{figures/TSCI robustness.png}
    \caption{\small Robustness of the outcome model in \eqref{eq: outcome model} against a general class of invalid IVs. $V_i$ is the evaluation of basis functions $\mathcal{V}$ at the $i$-th observation as in \eqref{eq: violation matrix}, and it only includes $W_i=\protect\aw(X_i)$ in valid IV settings. For identifiable violation, $V_i$ should only be able to span $g(Z_i,X_i)$ but not $f(Z_i,X_i)$. Any cases not satisfying any of these two conditions are unidentifiable. }
    \label{fig:TSCI robustness}
\end{figure}
\end{comment}
%%%%%%%%%%%%%%%%%%%%%%%%%%%%%%%%%%%%%%%%%%%%%%%%%%%%%%%%%%%%%%%%%%%%%%%%%%%%%%%%%%%%%%
\subsection{Causal interpretation: structural equation and potential outcome models} %\Mengchu{I think this part can be shortened because it is not the emphasis of our proposal}}
\label{sec: causal interp}
%%%%%%%%%%%%%%%%%%%%%%%%%%%%%%%%%%%%%%%%%%%%%%%%%%%%%%%%%%%%%%%%%%%%%%%%%%%%%%%%%%%%%%
%In the following, we interpret the outcome and treatment models \eqref{eq: outcome model} and \eqref{eq: treatment model} from both structural equation model (SEM) and the {potential outcome} perspectives. 
%{sec: causal interp}
%\vspace{2mm}
%\noindent {\bf Structural equation model interpretation.} 
%\Zijian{Peter, double check this paragraph as I add back all $X_i.$}
{We first provide the structural equation model (SEM) interpretation of the models \eqref{eq: outcome model} and \eqref{eq: treatment model}.} 
%For {expositional} simplicity, we focus on a special setting with no covariates $X_i$ and 
We consider the following SEM for $Y_i$ and $D_i$ and $X_i$, %with $1\leq i\leq n,$ 
\begin{equation}
Y_i \leftarrow a_0+ D_i\beta+g_1(Z_{i},X_i)+\nu_1(H_{i})+\epsilon^{0}_i, \quad \text{and}\quad D_i \leftarrow f_{1}(Z_i,X_i)+\nu_2(H_i)+\delta_i^0,
\label{eq: sem linear}
\end{equation}
where $H_i$ denotes some unmeasured confounders, $\epsilon^{0}_i$ and $\delta_i^0$ are random errors independent of $D_i, Z_{i},X_i, H_{i},$ and $a_0$ is the intercept such that $\E g_1(Z_i,X_i)=\E \nu_1(H_i)=0$. Define 
$g_{2}(Z_{i},X_i)=\E\left(\nu_1(H_{i}) \mid Z_{i},X_i\right)$ and $f_{2}(Z_{i},X_i)=\E\left(\nu_2(H_{i}) \mid Z_{i},X_i\right).$ In \eqref{eq: sem linear}, the unmeasured confounders $H_i$ might affect both the outcome and treatment, $g_1$ encodes a direct effect of the IVs on the outcome, and $g_2$ captures the association between the IVs and the unmeasured confounders. The SEM \eqref{eq: sem linear} together with the definitions of $f_2$ and $g_2$ imply the outcome model \eqref{eq: outcome model}
with  $\epsilon_i=\epsilon^0_i+\nu_1(H_{i})-g_2(Z_{i},X_i).$ The treatment model $D_i = f(Z_i,X_i)+\delta_i$ {arises} with $f(Z_{i},X_i)=f_1(Z_{i},X_i)+f_2(Z_{i},X_i)$ and $\delta_i=\delta^0_i+\nu_2(H_{i})-f_2(Z_{i},X_i).$

\begin{comment}
\tikzstyle{format} = [draw, thin,minimum height=1.35em, minimum width=1.35cm]
\tikzstyle{format1} = [draw, thin,minimum height=1.35em, minimum width=2.35cm]
\tikzstyle{medium} = [draw, thin,minimum height=1.35em]
\begin{figure}[htp]
\centering
%\begin{wrapfigure}{l}{0.58\textwidth}
\begin{tikzpicture}[node distance=3.5cm]
   % \path[use as bounding box] (-3.8,1) rectangle (10,-2);
    \node[format] at (0, 5) (treatment) {\footnotesize Treatment D};
    \node[format, right of=treatment, node distance=4.5cm] (outcome) {\footnotesize  Outcome Y };
    \draw[->]   (treatment) -- node[above] {\tiny Treatment Effect $\beta$} (outcome);
\node[medium, above of=outcome, left of=outcome, node distance=2.2cm](unmeasured){\footnotesize Unmeasured H};
\draw[->] (unmeasured) edge[] node {} (treatment);
\draw[->] (unmeasured) edge[]  node{} (outcome);
\node[format,left of=treatment, node distance=3cm] (IV) {\footnotesize Instrument Z};
\draw[->] (IV) edge node[swap,above]{}(treatment);
\draw[->] (IV) edge[bend right=20] node [swap, below=0.5mm]{\small $g_1\neq 0$ }(outcome);
\draw (IV) edge[bend left=20] node [swap, above=0.5mm]{\small $g_2\neq 0$}(unmeasured);
%\node [cross out,draw=black] at (-0.65, 6.78) {};
%\node [cross out, draw=black] at (1,4.1) {};
\end{tikzpicture}
\caption{Illustration of $g_1\neq 0$ and $g_2\neq 0$ in the SEM \eqref{eq: sem linear} with no covariates. } 
\label{fig: SEM vio}
\end{figure}
\end{comment}

{We now present how to interpret the invalid IV model with the potential outcome perspective.} 
For the $i$-th subject with the baseline covariates $X_i$, we use $Y^{(z,d)}(X_i)$ to denote the potential outcome with the IVs and the treatment assigned to $z\in \R^{p_z}$ and $d\in \R$, respectively. 
For $1\leq i\leq n,$ we consider the potential outcome model  \citep{splawa1990application,rubin1974estimating}, \begin{equation}
Y_i^{(z,d)}(X_i)=Y_i^{(0,0)}(X_i)+d\beta+g_1(z,X_i) \quad \text{and}\quad \E(Y_i^{(0,0)}(X_i)\mid Z_i, X_i)= g_2(Z_i,X_i),
\label{eq: potential}
\end{equation}
where $\beta\in \R$ denotes the treatment effect, $g_1: \R^{\pz+\px}\rightarrow \R$, and $g_2: \R^{\pz+\px}\rightarrow \R.$
%If the functions $g_1(z,x)$ and $g_2(z,x)$ change with the assigned IV value $z$, then the classical IV assumptions are violated. 
If $g_1(z,x)$ changes with  $z$, the IVs directly affect the outcome, violating the assumption (A3). If $g_2(z,x)$ changes with $z$, the IVs are associated with unmeasured confounders, violating the assumption (A2). The model \eqref{eq: potential} extends the Additive LInear, Constant Effects (ALICE) model of \citet{holland1988causal}
by allowing for a general class of invalid IVs. By the consistency assumption $Y_i=Y^{(Z_i,D_i)}_i(X_i),$ \eqref{eq: potential} implies \eqref{eq: outcome model} with $g(\cdot) = g_1(\cdot) + g_2(\cdot)$ and $\epsilon_i=Y_i^{(0,0)}(X_i)-g_2(Z_i,X_i).$ We can {easily} generalize \eqref{eq: potential} by considering $Y_i^{(z,d)}(X_i)=Y_i^{(0,0)}(X_i)+d\beta_i+g_1(z,X_i),$ where $\beta_i$ denotes the individual effect for the $i$-th individual. If we consider the random effect $\beta_i$ with $\E\beta_i=\beta$ and $\beta_i-\beta$ being independent of $(Z_i,X_i,D_i),$ then we obtain the model \eqref{eq: outcome model} with $\epsilon_i=Y_i^{(0,0)}(X_i)-g_2(Z_i,X_i)+(\beta_i-\beta)D_i$ and our proposed method can be applied to make inference for $\beta=\E \beta_i.$

%%%%%%%%%%%%%%%%%%%%%%%%%%%%%%%%%%%%%%%%%%%%%%%%%%%%%%%%%
%\subsection{Structural equation model interpretation}
%\label{sec: SEM}
%%%%%%%%%%%%%%%%%%%%%%%%%%%%%%%%%%%%

% The above SEM model \eqref{eq: sem linear} is illustrated in Figure \ref{fig: SEM vio}.

%%%%%%%%%%%%%%%%%%%%%%%%%
%\subsection{}
%%%%%%%%%%%%%%%%%%%%%%%%%
%\Zijian{Would it be fine that we significantly shorten this section? I wrote this section mainly to demonstrate the idea of ``making use of nonlinearity". Two issues now: the paper is a bit too long and then this might give people a feeling that the method is just about using nonlinearity and omitting the ``ML" part?}

%\vspace{-5mm}
%\paragraph{two-stage Curvature Identification of $\beta$.}

%When there are no covariates, we have $g(\cdot)=h(\cdot).$ 
%A combination of \eqref{eq: outcome model} and \eqref{eq: treatment model} leads to the following reduced form model,
%\begin{equation}
%\begin{aligned}
%Y_i=F(Z_{i})+\epsilon_i+\beta\delta_i \quad \text{with}\quad F(Z_{i})=\beta f(Z_i)+g(Z_{i}).
%\end{aligned}
%\label{eq: reduced form simple}
%\end{equation}

%%%%%%%%%%%%%%%%%%%%%%%%%%%%%%%%%%%%%%%%%%%%%%%%%%%%%
\subsection{Overview of {TSCI}} %\Mengchu{I think this part can be removed to shorten the length? We can write the transitions or summaries a little bit in the beginning of the next section. }}
\label{sec: overview}

We propose in Section \ref{sec: TSCI method} a novel methodology called two-stage curvature identification (\texttt{TSCI}) to identify $\beta$ under models \eqref{eq: outcome model} and \eqref{eq: treatment model}. {We first assume knowledge of the basis set $\mathcal{V}$ in \eqref{eq: basis function} that spans the function $g(\cdot).$} The proposal consists of two stages: in the first stage, we employ ML algorithms to learn the conditional mean $f(\cdot)$ in the treatment model \eqref{eq: treatment model}; in the second stage, we adjust for the IV invalidity form encoded by $\mathcal{V}$ and construct the confidence interval for $\beta$ by {leveraging the first-stage ML fits}. 

The success of \texttt{TSCI} relies on the following identification condition.
\begin{Condition} Define the best linear approximation of $f(Z_i,X_i)$ by $V_i$ in population as  $\gamma^*=\argmin_{\gamma}\E(f(Z_i,X_i)-V_i^{\intercal}\gamma)^2.$ We assume  $\E(f(Z_i,X_i)-V_i^{\intercal}\gamma^*)^2>0.$ \label{cond: identification}
\end{Condition} 
The condition states that the basis set $\mathcal{V}$, used for generating $g(\cdot)$, does not fully span the conditional mean function $f(\cdot)$ of the treatment model. 
For any user-specified $\mathcal{V}$ in \eqref{eq: basis function}, Condition \ref{cond: identification} can be {partially} examined by evaluating a generalized IV strength introduced in Section \ref{sec: IV strength}. When the generalized IV strength is sufficiently large, our proposed \texttt{TSCI} methodology leads to an estimator $\widehat{\beta}(\mathcal{V})$ in \eqref{eq: ML corrected hetero} such that 
\begin{equation}
(\widehat{\beta}(\mathcal{V})-\beta)/\widehat{\rm SE}(\mathcal{V})\cid N(0,1),
\label{eq: aimed results}
\end{equation}
with the estimated standard error $\widehat{\rm SE}(\mathcal{V})$ depending on the generalized IV strength.  Importantly, the basis set $\mathcal{V}$ used for generating $g(\cdot)$ does not need to be fixed in advance: we discuss in Section \ref{sec: vio sel} a data-driven strategy for choosing a ``good'' basis set $\mathcal{V}_{\widehat{q}}$ among a collection of candidate sets $\mathcal{V}_0\subset \mathcal{V}_1\subset \cdots \subset\mathcal{V}_{Q}$ for a positive integer $Q$, {where $\mathcal{V}_0$ corresponds to the valid IV setting. 
%We propose to choose the best basis set through comparing TSCI estimators given by valid IVs and a collection of invalid IVs generated in the form {of bases from the family} $\{\mathcal{V}_{q}\}_{1\leq q\leq Q}.$} 
%When there exists $\mathcal{V}_{q^*}$ such that $\mathcal{V}\subset \mathcal{V}_{q^*}$ and the {generalized} IV strength is large enough after adjusting {for} $\mathcal{V}_{q^*}$, we establish the asymptotic normality for  $\widehat{\beta}(\mathcal{V}_{\widehat{q}})$ as in \eqref{eq: aimed results}. 
{Our proposed \texttt{TSCI} compares estimators assuming valid IVs and adjusting for violation forms generated from the family $\{\mathcal{V}_{q}\}_{1\leq q\leq Q}.$ {Thus, we encompass the setting of valid IVs and through}
this comparison, the \texttt{TSCI} methodology provides more robust causal inference tools than directly assuming IVs' validity.} 

Intuitively, Condition \ref{cond: identification} requires $f(\cdot)$ to be more non-linear than $g(\cdot)$, which cannot be fully tested since $g(\cdot)$ is unknown. We shall provide two important remarks on the plausibility of Condition \ref{cond: identification}. Firstly, when the proposed IVs are valid with $g(z,x)$ being independent of $z$, Condition \ref{cond: identification} is satisfied as long as $f(z,x)$ depends on $z$. More interestingly, when $g(z,x)$ has a relatively weak dependence on $z$ (i.e., the IVs violate (A2) and (A3) locally), Condition \ref{cond: identification} holds if the machine learning algorithm captures a strong dependence of $f(z,x)$ on $z$. In practice, domain scientists identify IVs based on their knowledge, and these IVs, even {when} {violating assumptions (A2) and (A3), may have a relatively weak direct effect on the outcome or are weakly associated with the unmeasured confounders.} 
{In such a situation,} \texttt{TSCI} provides robust causal inference by allowing the proposed IV to be invalid. However, we are not suggesting the users take any covariate as an invalid IV to implement our proposal.  Secondly, even if Condition \ref{cond: identification} does not hold, our proposed \texttt{TSCI} estimator is reduced to {an estimator assuming valid IVs} 
which {then} has a similar performance to \texttt{TSLS} and \texttt{DML}. In Section \ref{sec: different nonlinearity sim}, we explore the performance of \texttt{TSCI} when Condition \ref{cond: identification} does not hold.

\section{{TSCI} with Machine Learning}
\label{sec: TSCI method}
%%%%%%%%%%%%%%%%%%%%%%%%%%%%%%%%%
We discuss the first and second stages of our \texttt{TSCI} methodology in Sections \ref{eq: split RF} and \ref{sec: second stage}, respectively. The main novelty of our proposal is to estimate {a bias}
%the overfitting bias 
caused by the first-stage ML and implement a follow-up bias correction. %We also introduce a generalized IV strength in Section \ref{sec: IV strength}, {providing empirical guidance on whether the identification condition for \texttt{TSCI} holds for a user-specified violation form.}%which indicates whether the association captured by the first-stage ML is sufficiently strong after adjusting for violation forms.
%We shall emphasize that a direct generalization of {the} two-stage least squares estimator by replacing its first stage {with ML does not work even when the ML method has an excellent prediction performance, see Section \ref{sec: Pitfalls} with an emphasis on random forests.}
%; see more details in Section \ref{sec: second stage}.}  
%\Peter{OK, sounds good!}
%{indicates whether
%the first-stage ML algorithm is under-fitted or not.}
%A key component of the TSCI is to estimate  $f(Z_i, X_i)=\E(D_i\mid X_i,Z_i)$. {We develop first 
%in 
%{the following of} this section 
%TSCI with estimation of the conditional mean function by random forests,}
%We extend our TSCI methodology to general ML methods in Section \ref{sec: TSML} and .} 
%Our proposed TSCI with random forests relies on viewing the split random forests as a weighting estimator, {a technique that}
%which 
%has been utilized in \citet{lin2006random,meinshausen2006quantile}. %{Based on such a weighting scheme,} we propose a novel TSCI estimator with random forests in Section \ref{sec: second stage}. 

%%%%%%%%%%%%%%%%%%%%%%%%%%%%%%
\subsection{First stage: machine learning models for the treatment model}
\label{eq: split RF}
%%%%%%%%%%%%%%%%%%%%%%%%%%%%%
We estimate the conditional mean $f(\cdot)$ in the treatment model \eqref{eq: treatment model} by a general ML fit. We randomly split the data into two disjoint subsets $\mathcal{A}_1$ and $\mathcal{A}_2$. Throughout the paper we use $\mathcal{A}_1=\{1, 2, \cdots, n_1\}$ with $n_1=\lfloor 2n/3\rfloor$ and set $\mathcal{A}_2=\{n_1+1,\cdots,n\}.$ {Our results can be extended to any other splitting with $|\mathcal{A}_1|\asymp |\mathcal{A}_2|.$ } Sample splitting is essential for removing the endogeneity of the ML predicted values.  
Without sample splitting, the ML predicted value for the treatment {can be close to the treatment (due to overfitting) and hence highly correlated with unmeasured confounders}, leading to a \texttt{TSCI} estimator suffering from a significant bias.

{The main step is to construct the ML prediction model with data belonging to $\mathcal{A}_2$ and express the ML estimator of $f_{\mathcal{A}_1}=({f}(Z_1, X_1),\cdots,{f}(Z_{n_1}, X_{n_1}))^{\intercal}$ as
\begin{equation}
\widehat{f}_{\mathcal{A}_1}=\Omega D_{\mathcal{A}_1}  \quad \text{for some matrix} \quad \Omega \in \R^{n_1\times n_1},
\label{eq: general trans matrix}
\end{equation}
where the data belonging to $\mathcal{A}_2$ is used to train the ML algorithm and construct the transformation matrix $\Omega$. 
Our proposed \texttt{TSCI} method mainly relies on expressing the first-stage prediction as the linear estimator in \eqref{eq: general trans matrix}. 
A wide range of first-stage ML algorithms are shown to have the expression in \eqref{eq: general trans matrix}. }
In the following, we follow the results in \citet{lin2006random,meinshausen2006quantile,wager2018estimation} and express the sample split random forests in the form \eqref{eq: general trans matrix}.
{In Sections \ref{sec: TS-boosting}, \ref{sec: TS-DNN}, and \ref{sec: TS-Bspline} in the supplement, we present the explicit definitions of $\Omega$ for boosting, deep neural network ({DNN}), and approximation with B-splines, respectively.} 

{Importantly, the transformation matrix $\Omega \in \R^{n_1\times n_1}$ {has a related meaning as the}
%serves as a similar role as the 
hat matrix in linear regression {procedures}: {however,} $\Omega$ is not a projection matrix for {nonlinear} 
ML algorithms, which creates additional challenges in adopting the first-stage ML fit; see Section \ref{sec: Pitfalls}. }

\noindent {\bf Split random forests in \eqref{eq: general trans matrix}.} {To avoid the overfitting of random forests,} {we adopt the construction of honest trees and forests as in \citet{wager2018estimation} and slightly modify it by excluding self-prediction for our purpose.} 
 We construct the random forests (RF) with the data $\{D_i, X_i, Z_i\}_{i\in \mathcal{A}_{2}}$ and estimate $f(Z_i, X_i)$ for $i\in \mathcal{A}_1$ by the constructed RF together with the data $\{X_i, Z_i, D_i\}_{i\in \mathcal{A}_1}$. 
RF aggregates $S\geq 1$ decision trees, with each decision tree being viewed as the partition of the whole covariate space $\R^{p_x+p_z}$ into disjoint subspaces $\{\mathcal{R}_l\}_{1\leq l\leq J}.$ Let $\theta$ denote the random parameter that determines how a tree is grown.
%where each tree is built with a bootstrapped sample from $\{X_i, Z_i, D_i\}_{i\in \mathcal{A}_{2}}$, and at each splitting, only a small number of randomly sampled covariates are considered to be split.  
 For any given $(z^{\intercal},x^{\intercal})^{\intercal}\in \R^{p_x+p_z}$ and a given tree with the parameter $\theta$, there exists an unique leaf $l(z,x,\theta)$ with $1\leq l(z,x,\theta)\leq J$ such that the subspace $\mathcal{R}_{l(z,x,\theta)}$ contains $(z^{\intercal},x^{\intercal})^{\intercal}.$ With the observations inside $\mathcal{R}_{l(z,x,\theta)}$, the decision tree {estimates}
$f(z,x)=\E(D\mid Z=z, X=x)$ by 
\begin{equation}
\widehat{f}_{\theta}(z,x)=\sum_{j\in \mathcal{A}_1} \omega_{j}(z,x,\theta) D_j \quad \text{with}\quad \omega_{j}(z,x,\theta)=\frac{{\bf{1}}\left[(Z_{j}^{\intercal}, X_{j}^{\intercal})^{\intercal}\in \mathcal{R}^0_{l(z,x,\theta)}\right]}{\sum_{k\in \mathcal{A}_1}{\bf{1}}\left[(Z_{k}^{\intercal}, X_{k}^{\intercal})^{\intercal}\in \mathcal{R}^0_{l(z,x,\theta)}\right]},
\label{eq: tree expression}
\end{equation}
where $\mathcal{R}^0_{l(z,x,\theta)}\coloneqq \mathcal{R}_{l(z,x,\theta)}/{(z,x)}$ is defined as the region excluding the point $(z,x).$ We use $\{\theta_{1},\cdots, \theta_{S}\}$ to denote the parameters corresponding to the $S$ trees. The estimator $\widehat{f}(z,x)=\frac{1}{S}\sum_{s=1}^{S}\widehat{f}_{\theta_s}(z, x)$ can be expressed as 
\begin{equation}
\widehat{f}(z, x)=\sum_{j\in \mathcal{A}_1}\omega_j(z, x) D_j \quad \text{where} \quad w_j(z, x)=\frac{1}{S}\sum_{s=1}^{S}\omega_{j}(z, x,\theta_s),
\label{eq: weight function}
\end{equation}
with $\omega_{j}(z, x,\theta_s)$ defined in \eqref{eq: tree expression}. {That is, the split RF estimator of  $f_{\mathcal{A}_1}$ attains the form \eqref{eq: general trans matrix} with $\Omega_{ij}=\omega_j(Z_i, X_i)$ for $i,j \in \mathcal{A}_1$.}

The construction of honest trees and removal of self-prediction help remove the endogeneity of the ML predicted values $\{\widehat{f}(Z_i,X_i)\}_{i\in \mathcal{A}_1}$. Firstly, due to the sample-splitting, the construction of $\Omega$ does not directly depend on $\{D_i\}_{i\in \mathcal{A}_{1}},$ which removes the endogeneity contained in the ML predicted values. Secondly, consider {an extreme setting} with $Z_i$ and $X_i$ not being predictive for $D_i$. In such a scenario, without removing the self-prediction, the estimator $\widehat{f}(Z_i, X_i)$ will be dominated by its own treatment observation $D_i$ and $\widehat{f}(Z_i, X_i)$ is still highly correlated with the unmeasured confounders, leading to a biased \texttt{TSCI} estimator. The removal of self-prediction can be viewed as an ML version of the ``leave-one-out" estimator for the treatment model, which was proposed in \cite{angrist1999jackknife} to reduce the bias of \texttt{TSLS} with many IVs. In the numerical studies, we have observed that the exclusion of self-prediction leads to a more accurate estimator than the corresponding \texttt{TSCI} estimator with self-prediction, regardless of the IV strength; see Table \ref{tab: ML Pitfall} for the detailed comparison. 
%is computed based on the covariate data $\{X_i, Z_i\}_{i\in \mathcal{A}_{1}}$ and the random forests  built with $\{X_i, Z_i, D_i\}_{i\in \mathcal{A}_{2}}$. Its construction  %{In Section \ref{sec: Pitfalls}, we provide further discussions on the necessity of sample splitting and removing self-prediction as in \eqref{eq: tree expression}.}  
%\Zijian{Why honest tree and modification?}
%\begin{Remark}[Removal of self-prediction]\rm
%{To allow for self-prediction, we simply replace {the notation} $\mathcal{R}^0_{l(z,x,\theta)}$ in \eqref{eq: tree expression} with $\mathcal{R}_{l(z,x,\theta)}$. } {To explain this, consider} 
%\end{Remark}

%{The above prediction model is slightly different from the split random forest since we remove the self-prediction. We explain in Section \ref{sec: Pitfalls} why the removal of self-prediction leads to more robust inference procedures in the presence of relatively weak IVs.}
%\noindent {\bf B-spline in the form of \eqref{eq: trans matrix}}

%%%%%%%%%%%%%%%%%%%%%%%%%%%%%%%%%%%%%%%%%%%%%%%%%%%%%%%%%%%%%%%%%%%%%%%%%%%%%%
\subsection{Second stage: adjusting for violation forms and correcting {bias}}
\label{sec: second stage}
%%%%%%%%%%%%%%%%%%%%%%%%%%%%%%%%%%%%%%%%%%%%%%%%%%%%%%%%%%%%
{In the second stage, we leverage the first-stage ML fits in the form of \eqref{eq: general trans matrix} and adjust for the possible IV invalidity forms.} 
{In the remaining of the construction, we {consider the outcome model in the form of \eqref{eq: outcome model basis}} and assume that {$V_i^{\intercal}\pi$ provides an accurate approximation of $g(Z_i,X_i)$}, {resulting in zero or sufficiently small $R_i(V)$}. Hence $R_i(V)$ does not enter in the following construction of the estimator for $\beta$. We quantify the effect of the error term $R_i(V)$ in the theoretical analysis in Section \ref{sec: theory} and propose in Section \ref{sec: vio sel} a data-dependent way of choosing $V$ such that $R_i(V)$ is sufficiently small.}

Applying the transformation  $\Omega$ in \eqref{eq: general trans matrix} to the model \eqref{eq: outcome model basis} with data in $\mathcal{A}_1$, we
obtain
\begin{equation}
\widehat{Y}_{\mathcal{A}_1}=\widehat{f}_{\mathcal{A}_1}\beta+\widehat{V}_{\mathcal{A}_1} \pi+\widehat{R}_{\mathcal{A}_1}+\widehat{\epsilon}_{\mathcal{A}_1}, %\quad \text{for} \quad 1\leq i\leq n
\label{eq: transformed subset}
\end{equation}
where $\widehat{Y}_{\mathcal{A}_1}=\Omega Y_{\mathcal{A}_1},\widehat{f}_{\mathcal{A}_1}=\Omega D_{\mathcal{A}_1}, \widehat{V}_{\mathcal{A}_1}=\Omega V_{\mathcal{A}_1}, \widehat{R}_{\mathcal{A}_1}=\Omega R_{\mathcal{A}_1},$ and  $\widehat{\epsilon}_{\mathcal{A}_1}=\Omega \epsilon_{\mathcal{A}_1}.$ {For a matrix $V$, we use $P_{V}$ and $P_{V}^{\perp}$ to denote the projection to the column spaces of $V$ and its orthogonal complement, respectively.} Based on \eqref{eq: transformed subset}, we {project out the columns of $\widehat{V}_{\mathcal{A}_1}$} and estimate the effect $\beta$ by
\begin{equation}
\widehat{\beta}_{\rm init}(V)\coloneqq\frac{\widehat{Y}_{\mathcal{A}_1}^{\intercal}P^{\perp}_{\widehat{V}_{\mathcal{A}_1}} \widehat{f}_{\mathcal{A}_1}}{\widehat{f}_{\mathcal{A}_1}^{\intercal}P^{\perp}_{\widehat{V}_{\mathcal{A}_1}} \widehat{f}_{\mathcal{A}_1}}=\frac{{Y}_{\mathcal{A}_1}^{\intercal}\T(V) {D}_{\mathcal{A}_1}}{{D}_{\mathcal{A}_1}^{\intercal}\T(V) {D}_{\mathcal{A}_1}} \quad \text{with}\quad \T(V)=\Omega^{\intercal}P^{\perp}_{\widehat{V}_{\mathcal{A}_1}} \Omega.
\label{eq: RF init}
\end{equation}

%$\Omega_{ij}=\omega_j(Z_i, X_i)$ for $i,j \in \mathcal{A}_1$.}
%\begin{equation}{eq: weight function}
%{where $V$ is defined in \eqref{eq: violation matrix}.

%\Peter{Does this matrix $V$ appear in the equation above? Isn't it just saying that the expression depends on $V$ -- and it doesn't matter whether it is a matrix or not?} \Zijian{The matrix $V$ appears in  $P^{\perp}_{\widehat{V}_{\mathcal{A}_1}},$ where $\widehat{V}_{\mathcal{A}_1}=\Omega V_{\mathcal{A}_1}.$} \Peter{NEW: OK, all fine. The terminology "violation matrix" is used many times and pretty clear from the context, I think -- although it is (perhaps?) nowhere really defined...}
When $R_i(V)=0$, then we decompose the error of the above estimator as
\begin{equation}
\widehat{\beta}_{\rm init}(V)-\beta=\frac{{{\epsilon}_{\mathcal{A}_1}^{\intercal}\T(V) {\delta}_{\mathcal{A}_1}}}{{{D}_{\mathcal{A}_1}^{\intercal}\T(V) {D}_{\mathcal{A}_1}}}+\frac{{\epsilon}_{\mathcal{A}_1}^{\intercal}\T(V) {f}_{\mathcal{A}_1}}{{D}_{\mathcal{A}_1}^{\intercal}\T(V) {D}_{\mathcal{A}_1}}.
\label{eq: init decomposition special}
\end{equation}
In the above decomposition, the first term on the right-hand side is a bias component appearing 
{due to the use of the first-stage ML. To explain this, we consider the homoscedastic correlation 
${\rm Cov}(\epsilon_i,\delta_i\mid Z_i, X_i)={\rm Cov}(\epsilon_i,\delta_i)$ and obtain the explicit bias expression,}
\begin{equation}
\frac{{\epsilon}_{\mathcal{A}_1}^{\intercal}\T(V) {\delta}_{\mathcal{A}_1}}{{D}_{\mathcal{A}_1}^{\intercal}\T(V) {D}_{\mathcal{A}_1}}\approx \frac{{\rm Cov}(\delta_i,\epsilon_i)\cdot {\rm Tr}[\T(V) ]}{{D}_{\mathcal{A}_1}^{\intercal}\T(V) {D}_{\mathcal{A}_1}},
\label{eq: bias identification}
\end{equation}
where ${\rm Tr}[\T(V) ]$ denotes the trace of $\T(V)$ defined in \eqref{eq: RF init}. {Note that ${\rm Tr}[\T(V)]$ can be viewed as degrees of freedom {or complexity measure} for the ML algorithm. {It {may} become particularly large in the overfitting regime.}
%{The ML algorithms, such as random forests and deep neural network,} might overfit the data and lead to a large degree of freedom, 
{The bias in \eqref{eq: bias identification} also becomes large when the denominator 
${D}_{\mathcal{A}_1}^{\intercal}\T(V) {D}_{\mathcal{A}_1}$ is small, which means a weak generalized IV strength in the following \eqref{eq: IV strength homo}. Thus, both the numerator and denominator in \eqref{eq: bias identification} can lead to a large bias of $\widehat{\beta}_{\rm init}(V)$.} In Figure \ref{fig:bias correction}, we illustrate the bias of $\widehat{\beta}_{\rm init}(V)$ in settings {with different values of the IV strength scaled to ${D}_{\mathcal{A}_1}^{\intercal}\T(V) {D}_{\mathcal{A}_1}$.} %When {the} IV strength increases from 28.77 to 111.70, the bias of $\widehat{\beta}_{\rm init}(V)$ shrinks from 0.060 to 0.016. }%\Mengchu{When $a$ or $n$ increases, the denominator ${D}_{\mathcal{A}_1}^{\intercal}\T(V) {D}_{\mathcal{A}_1}$ becomes larger. Although ${\rm Tr}[\T(V) ]$ enlarges with the increase of $n$ as well, its increasing speed is much lower than the denominator. }\Zijian{Okay, but may you give some rough numbers to have a more accurate sense?}\Mengchu{For example, }

\begin{figure}[H]
    \centering
    \includegraphics[width=0.9\linewidth]{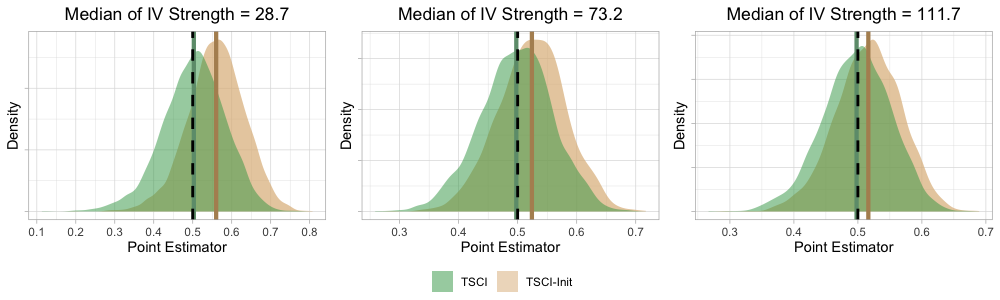}
    \caption{\small Density plot of \texttt{TSCI} and \texttt{TSCI-Init} estimators (with 500 simulations), which are after and before bias correction, respectively. The three panels from left to right correspond to settings with increasing IV strengths; see Section \ref{sec: thol 40} in the supplement for details. The black dashed line represents the true value $\beta=0.5$. The green and brown solid lines indicate the means of \texttt{TSCI} and \texttt{TSCI-Init} estimators. }
    \label{fig:bias correction}
\end{figure}

To address this {finite-sample bias} of $\widehat{\beta}_{\rm init}(V)$, we propose the following bias-corrected estimator,
\begin{equation}
\widehat{\beta}(V)=\frac{{Y}_{\mathcal{A}_1}^{\intercal}\T(V) {D}_{\mathcal{A}_1}}{{D}_{\mathcal{A}_1}^{\intercal}\T(V) {D}_{\mathcal{A}_1}}-\frac{\sum_{i=1}^{n_1}[\T(V)]_{ii}  \widehat{\delta}_i[\widehat{\epsilon}(V)]_i}{{{D}_{\mathcal{A}_1}^{\intercal}\T(V) {D}_{\mathcal{A}_1}}}.
\label{eq: ML corrected hetero}
\end{equation}
with $\widehat{\delta}_{i}=D_{i}-\widehat{f}_{i}$ for $1\leq i\leq n_1$ and $\widehat{\epsilon}(V)=P^{\perp}_{V}[Y-D\widehat{\beta}_{\rm init}(V)]$. We show {in Figure \ref{fig:bias correction} that our proposal effectively corrects the bias of $\widehat{\beta}_{\rm init}(V)$. }
%\begin{comment}
%\Zijian{To improve the above estimator of $\epsilon,$ we construct \begin{equation}\widehat{\epsilon}(V)=P^{\perp}_{V_{\mathcal{A}_1}^{\intercal},\widehat{f}_{\mathcal{A}_1}}[Y_{\mathcal{A}_1}-D_{\mathcal{A}_1}\widehat{\beta}_{\rm init}(V)]
%\end{equation}
%Alternatively, we may conduct the bias correction as 
%\begin{equation}
%\betaHe(V)=\widehat{\beta}_{\rm init}(V)-\frac{(D_{\mathcal{A}_1}-\widehat{f}_{\mathcal{A}_1})^{\intercal}\Omega^{\intercal}P^{\perp}_{\widehat{V}_{\mathcal{A}_1},\widehat{f}_{\mathcal{A}_1}}\Omega Y_{\mathcal{A}_1}}{{{D}_{\mathcal{A}_1}^{\intercal}\T(V) {D}_{\mathcal{A}_1}}},  
%\label{eq: rf corrected alter}
%\end{equation}
%} 
%\end{comment}
Importantly, our proposed bias correction is effective for both homoscedastic and {heteroscedastic} correlations. We present a simplified bias correction assuming homoscedastic correlation in Section \ref{sec: homo extra} in the supplement.

Centering at $\betaHe(V)$ defined in \eqref{eq: ML corrected hetero}, we construct the confidence interval
{
\begin{equation}
{\rm CI}(V)=
\left(\widehat{\beta}(V)-z_{\alpha/2}\widehat{\rm SE}(V),\widehat{\beta}(V)+z_{\alpha/2}\widehat{\rm SE}(V)\right), 
\label{eq: ML CI}
\end{equation}}
with {$z_{\alpha/2}$ denoting the upper $\alpha/2$ quantile of the standard normal distribution and} 
\begin{equation}
\widehat{\rm SE}(V)=\frac{\sqrt{\sum_{i=1}^{n_1}[\widehat{\epsilon}(V)]^2_i[\T(V) {D}_{\mathcal{A}_1}]_i^2}
}{{D}_{\mathcal{A}_1}^{\intercal}\T(V) {D}_{\mathcal{A}_1}} \quad \text{with}\quad \widehat{\epsilon}(V)=P^{\perp}_{V}\left[Y-D\widehat{\beta}_{\rm init}(V)\right].
\label{eq: TSRF SE hetero}
\end{equation}
\begin{Remark}\rm The bias component in \eqref{eq: bias identification} results from the correlation between $\epsilon_i$ and $\delta_i$ for $i\in \mathcal{A}_1$. In the regime of many IVs, \cite{angrist1999jackknife} proposed the ``leave-one-out" jackknife-type IV estimator, that is, instead of fitting the first stage model with all the data, the treatment model for the $i$-th observation is fitted without using itself. Such an estimator effectively removes the bias {for \texttt{TSLS}} due to the correlation between $\epsilon_i$ and $\delta_i$.
%for \texttt{TSLS}. 
Our proposed removal of self-prediction is of a similar spirit to the jackknife. However, even after removing the self-prediction, the diagonal of $\T(V)$ is not zero, and hence the correlation between $\epsilon_i$ and $\delta_i$ still leads to a bias component, which requires the bias correction as in \eqref{eq: ML corrected hetero}. %\Zijian{Stop here.}
\end{Remark}

\vspace{-3mm}
%%%%%%%%%%%%%%%%%%%%%%%%%%%%%%%%%
%%%%%%%%%%%%%%%%%%%%%%%%%%%%%%%%%
\subsection{Generalized IV strength: {detection of under-fitted machine learning}} 
\label{sec: IV strength}
%%%%%%%%%%%%%%%%%%%%%%%%%%%%%%%%%
%%%%%%%%%%%%%%%%%%%%%%%%%%%%%%%%%

The IV strength is particularly important for identifying the treatment effect stably, and the weak IV is a major concern in practical applications of IV-based methods \citep{stock2002survey,hansen2008estimation}. {With a larger basis set $\mathcal{V}$, the IV strength will generally decrease as the information contained in $\mathcal{V}$ is projected out from the first-stage ML fit. It is crucial to introduce a generalized IV strength measure in consideration of invalid IV forms and the ML algorithm.} {Similarly to the classical setup, good performance 
of our proposed \texttt{TSCI} estimator also requires {a relatively large generalized} IV strength.} We introduce a generalized IV strength measure as, \begin{equation}
\conpara\coloneqq {{f}_{\mathcal{A}_1}^{\intercal}\T(V) {f}_{\mathcal{A}_1}}/{\left[\sum_{i\in \mathcal{A}_1} {\rm Var}(\delta_i\mid X_i,Z_i)/n_1\right]}.
\label{eq: IV strength homo}
\end{equation}
If ${\rm Var}(\delta_i\mid X_i,Z_i)=\sigma_{\delta}^2,$ then $\conpara$ is reduced to ${f}_{\mathcal{A}_1}^{\intercal}\T(V) {f}_{\mathcal{A}_1}/\sigma_{\delta}^2.$ A sufficiently large strength $\conpara$ will guarantee stable point and interval estimators defined in \eqref{eq: ML corrected hetero} and \eqref{eq: ML CI}. Hence,  we need to check whether $\conpara$ is sufficiently large. Since $f$ is unknown, we estimate ${f}_{\mathcal{A}_1}^{\intercal}\T(V) {f}_{\mathcal{A}_1}$ by its sample version ${D}_{\mathcal{A}_1}^{\intercal}\T(V){D}_{\mathcal{A}_1}$ and estimate $\conpara$ by 
$\widehat{\conpara}\coloneqq {{D}_{\mathcal{A}_1}^{\intercal}\T(V){D}_{\mathcal{A}_1}}/\left[{\|D_{\mathcal{A}_1}-\widehat{f}_{\mathcal{A}_1}\|_2^2/n_1}\right].
$

{In Section \ref{sec: theory}, we show that our point estimator in \eqref{eq: ML corrected hetero} is consistent when the IV strength $\conpara$ is much larger than ${\rm Tr}\left[\T(V)\right]$; see Condition {\rm (R2)}. We now develop a bootstrap test to provide an empirical assessment of this IV strength requirement.}
{Since
\citet{rothenberg1984approximating} and \citet{stock2002survey} suggested the concentration parameter {being larger than}
%above 
$10$ as {being} ``adequate",  we develop a bootstrap test for $\conpara\geq \max\{2{\rm Tr}\left[\T(V)\right],10\}.$} 
We apply the wild bootstrap method and construct a probabilistic upper bound for the estimation error $
{D}_{\mathcal{A}_1}^{\intercal}\T(V){D}_{\mathcal{A}_1}-{f}_{\mathcal{A}_1}^{\intercal}\T(V) {f}_{\mathcal{A}_1}=2 {f}_{\mathcal{A}_1}^{\intercal}\T(V) {\delta}_{\mathcal{A}_1}+{\delta}_{\mathcal{A}_1}^{\intercal}\T(V) {\delta}_{\mathcal{A}_1}.
$ For $1\leq i\leq n_1$, we define $\widehat{\delta}_i=D_i-\widehat{f}_i$ and compute  $\widetilde{\delta}_i=\widehat{\delta}_i-\bar{\mu}_{\delta}$ with $\bar{\mu}_{\delta}=\frac{1}{n_1}\sum_{i=1}^{n_1}\widehat{\delta}_i.$ For $1\leq l\leq  L,$ we generate $\delta^{[l]}_i=U^{[l]}_i\cdot \widetilde{\delta}_i$ for $1\leq i\leq n_1,$ with $\{U^{[l]}_i\}_{1\leq i\leq n_1}$ generated as i.i.d. standard normal random variables, and compute $S^{[l]}=\left[{2 {f}_{\mathcal{A}_1}^{\intercal}\T(V) {\delta}^{[l]}+({\delta}^{[l]})^{\intercal}\T(V) {\delta}^{[l]}}\right]/\left[{{{\|D_{\mathcal{A}_1}-\widehat{f}_{\mathcal{A}_1}\|_2^2/n_1}}}\right].$ We use  $\mathcal{S}_{\alpha_0}(V)$ to denote the upper $\alpha_0$ empirical quantile of $\{|S^{[l]}|\}_{1\leq l\leq L}$. We conduct the {generalized} IV strength test 
$
\widehat{\conpara}\geq \max\{2{\rm Tr}\left[\T(V)\right],10\}+{\mathcal{S}_{\alpha_0}(V)},
$ with ${\mathcal{S}_{\alpha_0}(V)}$ being a high probability upper bound for $|\widehat{\conpara}-\conpara|.$
{We use $\alpha_0 = 0.025$ throughout this paper.}
If the above {generalized} IV strength test is passed, the IV is  claimed to be strong after adjusting for the matrix $V$ defined in \eqref{eq: violation matrix}; 
otherwise, the IV is claimed to be weak after adjusting for the matrix $V.$ {Empirically, we observe reliable inference properties when the estimated {generalized} IV strength $\widehat{\mu(V)}$ is above 40}; see Figure \ref{fig: thol 40} for details. %\Mengchu{plot added} \Zijian{More settings.}

\vspace{-2mm}
%%%%%%%%%%%%%%%%%%%%%%%%%%%%%%%%%
%%%%%%%%%%%%%%%%%%%%%%%%%%%%%%%%%%
\subsection{Data-dependent selection of $\mathcal{V}$} 
\label{sec: vio sel}
%%%%%%%%%%%%%%%%%%%%%%%%%%%%%%%%%%
Our proposed \texttt{TSCI} estimator in \eqref{eq: ML corrected hetero} requires 
prior knowledge of {the basis set $\mathcal{V}$, which 
%is used to 
generates the function $g(\cdot)$.} 
In the following, we consider the nested sets of basis functions $\V_0\subset \V_1\subset \cdots \subset \V_{Q},$ where $Q$ is a positive integer. We devise a data-dependent way to choose the best one among $\{\V_{q}\}_{0\leq q\leq Q}$. 
% The extension to the non-nested collection $\{\V_{q}\}_{0\leq q\leq Q}$ is discussed in the following Remark \ref{rem: non-nested}.

%\Mengchu{This part is a bit repeated from Section \ref{sec: general}. Maybe we can just refer to that to save the space. }
We define $\V_0\coloneqq\{\aw(x)\}$ as the set of basis functions for the valid IV setting. For $q\geq 1$, define $\V_{q}\coloneqq \{v_1(\cdot),\cdots,v_{L_q}(\cdot),\aw(x)\}$ {as the basis set for different invalid IV forms}, where $L_q\geq 1$ is the number of basis functions. We present two examples of $\{\mathcal{V}_q\}_{1\leq q\leq Q}$ for the single IV setting. 
\begin{enumerate}
\item[(1)] Polynomial violation: $\mathcal{V}_q=\{z,z^2,\cdots,z^q,\aw(x)\},$ for $1\leq q\leq Q.$ 
\item[(2)] Interaction violation: $\mathcal{V}_1=\left\{z, z\cdot x_1,z\cdot x_2,\cdots,z\cdot x_{p_x},\aw(x)\right\}.$
\end{enumerate}
%{For the multiple IV setting $z\in \R^{p_z}$, we may extend the above specification as
%\begin{enumerate}
%\item[(3)] $\mathcal{V}_q=\left\{z_1,\cdots, z_1^{q}, \cdots, z_{p_z},\cdots, z_{p_z}^{q},\aw(x)\right\}$  for $1\leq q\leq Q$;
%\item[(4)] $\mathcal{V}_1=\left\{z_1,\cdots, z_{p_z},\aw(x)\right\}$ and $\mathcal{V}_2=\left\{\{z_j, z_j\cdot x_1,z_j\cdot x_2,\cdots,z_j\cdot x_{p_x}\}_{1\leq j\leq p_z},\aw(x)\right\}$ 
%\end{enumerate}}
{For $0\leq q\leq Q$, we define the matrix $V_{q}\in \R^{n\times (L_q+p_{w})}$ with its $i$-th row defined as $
(V_{q})_{i}=\left(\vb_1(Z_i,X_i),\cdots, \vb_{L_q}(Z_i,X_i),W_i^{\intercal}\right)^{\intercal}$ {with} $W_i=\aw(X_i)$ for $1\leq i\leq n$.} %\Mengchu{Already appeared in \eqref{eq: violation matrix}. Can we write $W_i$ as $\aw(X_i)$?}

%\Peter{All what follows is only for polynomial violation? I mean: with (b), there is nothing to choose from, there is only ${\cal V}_1$?} \Zijian{For (b), we choose between ${\cal V}_0$ and ${\cal V}_1.$} \Peter{OK. Perhaps we can then write:\\
%\Peter{The blue stuff looks good to me!} 
For estimating $q \in \{0,1,\ldots \}$ from data, we proceed as follows.
We implement the {generalized} IV strength test in Section \ref{sec: IV strength} and define $Q_{\max}$ as, \begin{equation}
Q_{\max}\coloneqq \max_{q\geq 0} \left\{q:\widehat{{\mu}(V_q)}\geq \max\{2{\rm Tr}\left[\T(V_q)\right],10\}+{\mathcal{S}_{\alpha_0}(V_q)}\right\},
\label{eq: max index RF}
\end{equation}
where $\alpha_0$ is set at $0.025$ by default. For a larger $q$, we tend to adjust out more information and have relatively weaker IVs. Intuitively, ${Q_{\max}}$ denotes the largest index such that the IVs still have enough strength after adjusting for $V_q$.  With $Q_{\max},$ we shall choose among $\{V_{q}\}_{0\leq q\leq Q_{\max}}.$ {As a remark, when there is no $q\geq 0$ satisfying \eqref{eq: max index RF}, this corresponds to the weak IV regime. In such a scenario, we shall implement the valid IV estimator and output a warning of weak IV.} %our proposal will report a warning of weak IV and output the \texttt{TSCI} estimator assuming valid IVs.} 

For any given $0\leq q\leq Q_{\max},$ we apply the generalized estimator in \eqref{eq: ML corrected hetero} and construct 
\begin{equation}
\widehat{\beta}(V_q)=\frac{{Y}_{\mathcal{A}_1}^{\intercal}\T(V_q) {D}_{\mathcal{A}_1}}{{D}_{\mathcal{A}_1}^{\intercal}\T(V_q) {D}_{\mathcal{A}_1}}-\frac{\sum_{i=1}^{n_1}[\T(V_{q})]_{ii}  \widehat{\delta}_i[\widehat{\epsilon}(V_{Q_{\max}})]_i}{{{D}_{\mathcal{A}_1}^{\intercal}\T(V_q) {D}_{\mathcal{A}_1}}},
\label{eq: ML corrected hetero seq}
\end{equation}
with $\widehat{\delta}_{\mathcal{A}_1}=D_{\mathcal{A}_1}-\widehat{f}_{\mathcal{A}_1},$ $\T(\cdot)$ defined in \eqref{eq: RF init}, and $\widehat{\epsilon}(\cdot)$ defined in \eqref{eq: TSRF SE hetero}. 

{The selection of the best $V_q$ among $\{V_q\}_{0\leq q\leq Q_{\max}}$ relies on comparing the estimators $\{\widehat{\beta}(V_q)\}_{0\leq q\leq Q_{\max}}.$} 
We start with comparing the difference between the estimators $\widehat{\beta}_{{V}_{q}}$ and $\widehat{\beta}_{{V}_{q'}}$ with $0\leq q< q'\leq Q_{\max}.$ 
When $\{g(X_i,Z_i)\}_{1\leq i\leq n}$ are well approximated by both $V_q$ and $V_{q'}$, the approximation errors $R({V}_{q})$ and $R({V}_{q'})$ defined in \eqref{eq: approx error} are small and the main difference $\widehat{\beta}(V_{q'})-\widehat{\beta}(V_q)$ is approximately due to the randomness of the errors $\epsilon_{\mathcal{A}_1}.$ {In such cases.} we {then} have $\widehat{\beta}(V_{q'})-\widehat{\beta}(V_q)$ is approximated centered at zero with the following conditional variance, 
%$$
%\widehat{\beta}(V_{q'})-\widehat{\beta}(V_q)\approx \frac{f_{\mathcal{A}_1}^{\intercal} \T(V_{q'}) \epsilon_{\mathcal{A}_1}}{{f}_{\mathcal{A}_1}^{\intercal} \T(V_{q'}) f_{\mathcal{A}_1}}-\frac{f_{\mathcal{A}_1}^{\intercal} \T(V_{q}) \epsilon_{\mathcal{A}_1}}{f_{\mathcal{A}_1}^{\intercal} \T(V_{q}) f_{\mathcal{A}_1}},
%$$
%where the right-hand side has the following conditional variance, 
\begin{equation}
\begin{aligned}
\widehat{H}(V_{q},V_{q'})&=\frac{\sum_{i=1}^{n_1}[\widehat{\epsilon}(V_{Q_{\max}})]^2_i[\T(V_{q'}) {D}_{\mathcal{A}_1}]_i^2}{[{D}_{\mathcal{A}_1}^{\intercal} \T(V_{q'}) D_{\mathcal{A}_1}]^2}+\frac{\sum_{i=1}^{n_1}[\widehat{\epsilon}(V_{Q_{\max}})]^2_i[\T(V_{q}) {D}_{\mathcal{A}_1}]_i^2}{[{D}_{\mathcal{A}_1}^{\intercal} \T(V_{q}) D_{\mathcal{A}_1}]^2}\\
&-2\frac{\sum_{i=1}^{n_1}[\widehat{\epsilon}(V_{Q_{\max}})]^2_i [\T(V_{q'}) {D}_{\mathcal{A}_1}]_i[\T(V_{q}) {D}_{\mathcal{A}_1}]_i}{[{D}_{\mathcal{A}_1}^{\intercal} \T(V_{q'}) D_{\mathcal{A}_1}]\cdot[{D}_{\mathcal{A}_1}^{\intercal} \T(V_{q}) D_{\mathcal{A}_1}]}.
\end{aligned}
\label{eq: diff normalization hetero}
\end{equation}
Based on the above approximation, we further conduct the following test for whether $\widehat{\beta}(V_q)$ is significantly different from $\widehat{\beta}(V_{q'})$,
\begin{equation}
\mathcal{C}(V_q, V_{q'})= {\bf 1}\left({\left|\widehat{\beta}(V_q)-\widehat{\beta}(V_{q'})\right|}/{\sqrt{\widehat{H}(V_{q},V_{q'})}}\geq z_{\alpha_0}\right), 
%\begin{cases} 0& \text{if}\;  \\1& \text{otherwise}\end{cases},
\label{eq: comparison test}
\end{equation}
where $\widehat{\beta}(V_q)$ and $\widehat{\beta}(V_{q'})$ are defined in
\eqref{eq: ML corrected hetero seq}. %and $z_{\alpha_0}$ is the upper $\alpha_0$ quantile of the standard normal random variable. 
{On the other hand, if the smaller matrix $V_{q}$ does not provide a good approximation of $\{g(Z_i,X_i)\}_{1\leq i\leq n}$, the difference $\left|\widehat{\beta}(V_q)-\widehat{\beta}(V_{q'})\right|$ tends to be much larger than the threshold in \eqref{eq: comparison test} and hence $\mathcal{C}(V_q, V_{q'})=1$, indicating that $V_{q}$ does not fully generate $\{g(Z_i,X_i)\}_{1\leq i\leq n}.$} %the estimator $\widehat{\beta}(V_q)$ is not reliable.}

For $Q_{\max}\geq 2,$ we generalize the pairwise comparison to multiple comparisons. For $0\leq q\leq Q_{\max}-1,$  we compare $\widehat{\beta}(V_q)$ to any $\widehat{\beta}(V_{q'})$ with ${q+1\leq q'\leq Q_{\max}}$.  Particularly, for $0\leq q\leq Q_{\max}-1,$ we define the test  
\begin{equation}
\mathcal{C}(V_q)={\bf 1}\left(\max_{q+1\leq q'\leq Q_{\max}}\left[{\left|\widehat{\beta}(V_q)-\widehat{\beta}(V_{q'})\right|}/{\sqrt{\widehat{H}(V_{q},V_{q'})}}\right]\geq \widehat{\rho}\right), 
\label{eq: lawyer test RF}
\end{equation}
{where the thresholding $\widehat{\rho}$ is chosen by the wild bootstrap; see more details in Section \ref{eq: sample rho} in the supplement.} We define $\mathcal{C}(V_{Q_{\max}})=0$ as there is no index larger than ${Q_{\max}}$ that we might compare to. {We interpret $\mathcal{C}(V_q)=0$ as follows: none of the differences $\{|\widehat{\beta}(V_{q})-\widehat{\beta}(V_{q'})|\}_{q+1\leq q'\leq Q_{\max}}$ is large, indicating that $V_q$  (approximately) generates the function $g(\cdot)$ and $\widehat{\beta}(V_{q})$ is a reliable estimator.}

We choose the index $\widehat{q}_c \in [0, Q_{\max}]$ as
$\widehat{q}_c=\min_{0\leq q\leq Q_{\max}}\left\{q: \mathcal{C}(V_q)=0\right\},
$
which is the smallest $q$ such that  $\mathcal{C}(V_q)$ is zero. The index $\widehat{q}_c$ is interpreted as follows: any matrix containing ${V}_{\widehat{q}_c}$ as a submatrix does not lead to a substantially different estimator from that by ${V}_{\widehat{q}_c}.$ This provides evidence for ${V}_{\widehat{q}_c}$ forming a good approximation of $\{g(Z_i,X_i)\}_{1\leq i\leq n}.$}  Here, the sub-index ``c" represents the comparison as we compare different $\{V_{q}\}_{0\leq q\leq Q_{\max}}$ to choose the best one. 
Note that $\widehat{q}_c$ must exist since $\mathcal{C}(V_{Q_{\max}})=0$ by definition. In finite samples, there are chances that certain violations cannot be detected, especially if the \texttt{TSCI} estimators given by $V_{\widehat{q}_c}$ and  $V_{\widehat{q}_c+1}$ are not significantly different. We propose a more robust choice of the index as 
$
\widehat{q}_r=\min\{\widehat{q}_c+1,Q_{\max}\},
$
where the sub index ``r" denotes the robust selection; see the discussion after Theorem \ref{thm: post-selection RF}. We summarize our proposed \texttt{TSCI} estimator in Algorithm \ref{algo: TSCI selection}.

%\vspace{-3mm}

\begin{algorithm}[htp!]
\caption{\texttt{TSCI} with machine learning}
\begin{flushleft}
%\hspace*{\algorithmicindent} 
\textbf{Input:} Data $X\in \R^{n\times p_{x}}, Z,D,Y\in \R^{n}$; Sets of basis $\{\mathcal{V}_{q}\}_{q\geq 0}$ for approximating $g(\cdot)$; \\
%\hspace*{\algorithmicindent} 
\textbf{Output:} 
$\widehat{q}_c$ and $\widehat{q}_r$;  $\widehat{\beta}(V_{\widehat{q}_c})$ and $\widehat{\beta}(V_{\widehat{q}_r})$; ${\rm CI}(V_{\widehat{q}_c})$ and ${\rm CI}(V_{\widehat{q}_r})$. 
\end{flushleft}
\begin{algorithmic}[1]
   \State Generate matrices $V_q$ based on $\mathcal{V}_q$ for $q\geq 0$ as in \eqref{eq: violation matrix}; 
    \State Compute $Q_{\max}$ as in \eqref{eq: max index RF} and $\widehat{\epsilon}(V_{Q_{\max}})$ as in \eqref{eq: TSRF SE hetero}; %\Comment{Estimate the noise level}
 %   \State $\widehat{q}_{c}=0$ \Comment{Set the initial $q$ value as zero}
  \State Compute $\{\widehat{\beta}(V_q)\}_{0\leq q\leq Q_{\max}}$ as in \eqref{eq: ML corrected hetero seq} and $\{\widehat{H}(V_{q},V_{q'})\}_{0\leq q<q'\leq Q_{\max}}$ as in \eqref{eq: diff normalization hetero};
%  \State Compute $\widehat{\rho}$ as in \eqref{eq: sample quantile}; 
  \If{$Q_{\max}\geq 1$}   
    \State Compute $\{\mathcal{C}(V_q)\}_{0\leq q \leq Q_{\max}-1}$ as in  \eqref{eq: lawyer test RF};
   \EndIf    
    \State Set $\mathcal{C}(V_{Q_{\max}})=0$;
  %  \State Compute $\widehat{\beta}(V_{\widehat{q}_c})$ and ${\rm CI}(V_{\widehat{q}_c})$ by {\bf TSCI}  Algorithm \ref{algo: TSCI} with $V_{\widehat{q}_c}$;
% \State Set ${\rm I}_{\rm IV}=1$ if $\mathcal{C}(V_0)=1$ and ${\rm I}_{\rm IV}=0$ otherwise; \Comment{IV Invalidity test.} 
\State Compute $\widehat{q}_c=\min\left\{0\leq q\leq Q_{\max}: \mathcal{C}(V_q)=0\right\}$; \Comment{Comparison selection}
% \If {$\mathcal{C}_{2\alpha/\log n}(V_{q}, V_{q'})=1$}
 \State   Compute $\widehat{q}_r=\min\{\widehat{q}_c+1,Q_{\max}\}$;  \Comment{Robust selection}
     \State Compute $\betaHe(V_{\widehat{q}_c})$ and $\betaHe(V_{\widehat{q}_r})$ as in \eqref{eq: ML corrected hetero} with $V=V_{\widehat{q}_c},V_{\widehat{q}_r},$ respectively; 
     \State Compute ${\rm CI}_{\rm RF}(V_{\widehat{q}_c})$ and ${\rm CI}_{\rm RF}(V_{\widehat{q}_r})$ as in \eqref{eq: ML CI} with $V=V_{\widehat{q}_c},V_{\widehat{q}_r},$ respectively.  
%   \Else
%   \State Compute ${\rm CI}_{\rm RF}(V_{0})$ and ${\rm CI}_{\rm RF}(V_{0})$ as in \eqref{eq: ML CI} Implement the valid IV analysis; report ${\rm I}_{\rm IV}=0$.

  %  \vspace*{0.25cm}
 \end{algorithmic}
\label{algo: TSCI selection}
\end{algorithm}

%\vspace{-3mm}

%\begin{Remark}\rm
%{For a single IV, it is straightforward to generate the nested sets $\mathcal{V}_0\subset \mathcal{V}_1 \subset \cdots \mathcal{V}_{Q}$; e.g., $\mathcal{V}_q=\{z,\cdots,z^q\}$.
%\Mengchu{this remark is reflected in the multiple IV simulation. We can move the dicussion there.} \Zijian{You can move this to the supplement.} 
%\label{rem: non-nested}
%\end{Remark}
As a remark, $\mathcal{C}(V_q, V_{q'})$ in \eqref{eq: comparison test} can be used to test the IV validity. 
%{If we set $q=0$ in \eqref{eq: comparison test}, we are comparing the valid IV estimator and an estimator adjusting for $V_{q'}.$ 
For any $q'\geq 1$, $\mathcal{C}(V_0, V_{q'})=1$ represents that the estimator assuming valid IV is {sufficiently} different from an estimator allowing for the violation form generated by $V_{q'}$, {indicating that} the valid IV assumption is violated.}
%\label{rem: IV validity test}
%\end{Remark}
%%%%%%%%%%%%%%%%%%%%%%%%%%%%%%%%%%%%%%%%%%%%%%%%%%%%%%%%%%%%%%%%%%%%%%%%%%%%%%%%%%%%%%%%%%%%%%%%
%\subsection{}
%\label{sec: multi-splitting}
%%%%%%%%%%%%%%%%%%%%%%%%%%%%%%%%%%%%%%%%%%%%%%%%%%%%%%%%%%%%%%%%%%%%%%%%%%%%%%%%%%%%%%%%%%%%%%%%%%%%%%%%%%%%%%%%%%%%%%%%%%%%%
%Our proposed TSCI estimator  randomly splits the data into two subsamples. 

%%%%%%%%%%%%%%%%%%%%%%%%%%%%%%%%%%%%%%%%%%%%%%%%%%%%%%%%%%%%
%\subsection{Falsification of the identification assumption}
%\label{sec: falsification}
%%%%%%%%%%%%%%%%%%%%%%%%%%%%%%%%%%%%%%%%%%%%%%%%%%%%%%%%%%%%

\vspace{-2mm}
%%%%%%%%%%%%%%%%%%%%%%%%%%%%%%%%%%%%%%%%%%%%%%%%%%%%%%%%%%%%
\subsection{Comparison to Double Machine Learning and Machine Learning IV}
\label{sec: dml comp}
%%%%%%%%%%%%%%%%%%%%%%%%%%%%%%%%%%%%%%%%%%%%%%%%%%%%%%%%%%%%
We compare our proposed \texttt{TSCI} with the double machine learning (\texttt{DML}) estimator \citep{chernozhukov2018double} and {the} machine learning IV (\texttt{MLIV}) estimator \citep{chen2020mostly,liu2020deep}. As the most significant difference, \texttt{TSCI} provides a valid inference that is robust to a certain class of invalid IVs while \texttt{DML} and \texttt{MLIV} are designed for valid IV regimes {only}. We focus on the valid IV setting in the following and compare our proposal to \texttt{DML} and \texttt{MLIV}. In the DML framework, \citet{chernozhukov2018double} considered the outcome model $Y_i=D_i\beta+g(X_i)+\epsilon_i$, which is a special case of our outcome model \eqref{eq: outcome model} by assuming valid IVs. %that is, the unknown $g(\cdot)$ does not depend on $Z_i$. %For the treatment model, DML only considers the linear association between $D_i$ and $Z_i$ while TSCI applies $D_i=f(Z_i,X_i)+\delta_i$ to formulate the nonlinear association. 
The population parameter $\beta$ in \texttt{DML} can be identified through the following expression \citep{chernozhukov2018double,emmenegger2021regularizing}  
\begin{equation}\label{eq: dml identification cond}
\beta=\frac{\E\left[Y_i-\E(Y_i\mid X_i)\right][Z_i-\E(Z_i\mid X_i)]}{\E\left[D_i-\E(D_i\mid X_i)\right][Z_i-\E(Z_i\mid X_i)]},
\end{equation}
where $\E(Y_i\mid X_i)$, $\E(Z_i\mid X_i),$ and $\E(D_i\mid X_i)$ are fitted by machine learning algorithms. 
%carrying the interpretation as the \texttt{TSLS} estimator after adjusting the outcome, treatment, and IV by baseline covariates. 
%using the residuals $Y_i-\E(Y_i\mid X_i)$, $D_i-\E(D_i\mid X_i),$ and $Y_i-\E(Y_i\mid X_i)$. 

As a fundamental difference, we apply machine learning algorithms to capture the nonlinear relation between $D_i$ and $Z_i,X_i$ while \eqref{eq: dml identification cond} identifies $\beta$ based on the linear association $\E\left[D_i-\E(D_i\mid X_i)\right][Z_i-\E(Z_i\mid X_i)]$. Due to the first-stage ML, our proposed \texttt{TSCI} estimator is generally more efficient than the \texttt{DML} estimator. On the other hand, we approximate $g(z,x)$ by a set of basis functions, which might provide an inaccurate estimator when the basis functions are misspecified. In contrast, \texttt{DML} learns the conditional mean model by general machine learning algorithms and does not particularly require such a specification of basis functions. We compare the performance of \texttt{DML} and \texttt{TSCI} with valid IVs in Section \ref{sec: dml valid IV sim} and with invalid IVs in Section \ref{sec: different nonlinearity sim}.

%When the basis functions are mis-specified advantage over \texttt{TSCI} in terms of estimating $g(x)$ more accurately

%According to \eqref{eq: dml identification cond}, the DML estimator only takes advantage of the linear association between the IV and the treatment as in TSLS, while TSCI uses ML algorithms to explore and capture the non-linearity in the treatment model, which helps increase the IV strength and improve the efficiency of the estimator. Moreover, TSCI considers the possible violation in the outcome model and proposes a data-dependent way to select the appropriate violation form, while DML only makes inference by assuming the IV is valid. 

%In valid IV settings, . For \texttt{TSCI}, we approximate the nonlinear function $g(x)$ or $g(z,x)$ using a set of basis functions. However, this set of basis functions might be misspecified, leading to the bias of estimating the effect $\beta$. 

%Although TSCI is able to accurately span , it might misspecify it when there are interactions among covariates and cause the estimation bias. Yet DML can capture the most of the nonlinear associations including interactions well using ML algorithms. 

\citet{chen2020mostly,liu2020deep} proposed to use the ML prediction values $\widehat{f}_{\mathcal{A}_1}$ in \eqref{eq: general trans matrix} as the IV, referred to as the \texttt{MLIV} in the current paper. With this \texttt{MLIV}, the standard \texttt{TSLS} estimator can be implemented: in the first stage, use
%implement 
OLS regression of $D_{\mathcal{A}_1}$ on the \texttt{MLIV} $\widehat{f}_{\mathcal{A}_1}$ and the baseline covariates $V_{\mathcal{A}_1}$ and construct the predicted value $\widehat{D}_{\mathcal{A}_1}=c_f \widehat{f}_{\mathcal{A}_1}+{V}_{\mathcal{A}_1}c_v$ with $c_f\in \R$ and the vector $c_v$ denoting the \texttt{OLS} regression coefficients; in the second stage, use
%implement 
the outcome regression by 
replacing 
${D}_{\mathcal{A}_1}$ with $\widehat{D}_{\mathcal{A}_1}$. The
%and the 
TSLS estimator with \texttt{MLIV} can then be expressed as 
%$\widehat{f}_{\mathcal{A}_1}$ and the adjusted covariates  $V_{\mathcal{A}_1}$, 
\begin{equation}
\widehat{\beta}_{\rm \texttt{MLIV}}\coloneqq {Y_{\mathcal{A}_1}^{\intercal} P_{V_{\mathcal{A}_1}}^{\perp}\widehat{D}_{\mathcal{A}_1}}/{\widehat{D}_{\mathcal{A}_1}^{\intercal} P_{V_{\mathcal{A}_1}}^{\perp}\widehat{D}_{\mathcal{A}_1}}=\beta+{\epsilon_{\mathcal{A}_1}^{\intercal}P_{V_{\mathcal{A}_1}}^{\perp}\widehat{f}_{\mathcal{A}_1}}/[{c_f\cdot \widehat{f}_{\mathcal{A}_1}^{\intercal}P_{V_{\mathcal{A}_1}}^{\perp}\widehat{f}_{\mathcal{A}_1}}].
\label{eq: MLIV}
\end{equation}
The last equality of \eqref{eq: MLIV} holds by plugging in the outcome model $Y_i=D_i\beta+V_i^{\intercal}\pi+\epsilon_i$ and $\widehat{D}_{\mathcal{A}_1}=c_f \widehat{f}_{\mathcal{A}_1}+{V}_{\mathcal{A}_1}c_v$ and applying the orthogonality between  $D_{\mathcal{A}_1}-\widehat{D}_{\mathcal{A}_1}$ and $\{\widehat{f}_{\mathcal{A}_1},{V}_{\mathcal{A}_1}\}.$ 
%Equivalently, $\widehat{\beta}_{\rm \texttt{MLIV}}$ is the solution to the estimating equation $\widehat{f}_{\mathcal{A}_1}^{\intercal}P^{\perp}_{{V}_{\mathcal{A}_1}}(Y_{\mathcal{A}_1}-\beta D_{\mathcal{A}_1})=0$ as in \citet{liu2020deep},  which takes $\widehat{f}_{\mathcal{A}_1}$ as IV and  ${V}_{\mathcal{A}_1}$ as the adjusted covariates. 
{Note that a dominating term of $\widehat{\beta}_{\rm init}(V)-\beta$ in \eqref{eq: init decomposition special} is ${\widehat{\epsilon}_{\mathcal{A}_1}^{\intercal}P^{\perp}_{\widehat{V}_{\mathcal{A}_1}} \widehat{f}_{\mathcal{A}_1}}/{\widehat{f}_{\mathcal{A}_1}^{\intercal}P^{\perp}_{\widehat{V}_{\mathcal{A}_1}} \widehat{f}_{\mathcal{A}_1}}.$ 
In comparison, the last term in \eqref{eq: MLIV} is roughly inflated by a factor of $1/c_f$}, appearing due to the additional first-stage regression using the \texttt{MLIV}. When the IV is relatively strong, then this factor $c_f$ is around 1, and \texttt{MLIV} and $\widehat{\beta}_{\rm init}(V)$ are of similar performance for valid IV settings. However, in the presence of relatively weak IVs, the coefficient $c_f$ in front of $\widehat{f}_{\mathcal{A}_1}$ has a large fluctuation due to the weak association between $D_{\mathcal{A}_1}$ and $\widehat{f}_{\mathcal{A}_1}.$ The weak IV problem deteriorates with the extra first stage regression using the \texttt{MLIV}.  This explains why the estimator $\widehat{\beta}_{\rm \texttt{MLIV}}$ in \eqref{eq: MLIV} has a much larger bias 
%\Peter{Why bias? isn't it just the large fluctuation -- and hence variance?} \Zijian{It is hard to say whether it is the bias or variance. It is like the weak IV problem, which leads to both bias and variance. In Table \ref{tab: ML Pitfall}, we observe that MLIV has both large bias and variance. and hence large MSE.} 
and standard error than our proposed \texttt{TSCI} estimator in \eqref{eq: ML corrected hetero} when the IVs are relatively weak; see Section \ref{sec: ML comparison} in the supplementary for the detailed comparison. %We also present a {histogram}
\section{Theoretical justification}
\label{sec: theory}
We first establish the asymptotic normality of $\widehat{\beta}(V)$ for a given $V$. To begin with, we present the required conditions. The first assumption is imposed on the regression errors in models \eqref{eq: outcome model} and \eqref{eq: treatment model} and the data $\{V_i,f_i\}_{1\leq i\leq n}$ with $f_i=f(Z_i,X_i)$.
%\vspace{3mm}
\begin{enumerate}
\item[(R1)] Conditioning on $Z_i,X_i$, $\epsilon_i$ and  $\delta_i$ are sub-gaussian random variables, that is, there exists a positive constant $K>0$ such that 
$$\sup_{Z_i, X_i}\max\left\{\PP(|\epsilon_i|>t\mid Z_i, X_i),\PP(|\delta_i|>t\mid Z_i, X_i)\right\}\leq \exp(-{K^2 t^2}/{2}),$$
with $\sup_{Z_i, X_i}$ denoting the supremum over the support of the density of $Z_i, X_i$. The {random variables} 
$\{V_i, f_i\}_{1\leq i\leq n_1}$ satisfy $\lambda_{\min}\left(\sum_{i=1}^{n_1}V_i V_i^{\intercal}/n_1\right)\geq c$,  $\left\|\sum_{i=1}^{n_1}V_i f_{i}/n_1\right\|_2\leq C,$ $\max_{1\leq i\leq n_1}\{|f_i|,\|V_i\|_2\} \leq C \sqrt{\log n_1},$ and $\left\|\sum_{i=1}^{n_1}V_i[R(V)]_{i}/n_1\right\|_2\leq C \|R(V)\|_{\infty}$, where $C>0$ and $c>0$ are constants independent of $n$ and $p.$ The matrix $\Omega$ defined in \eqref{eq: general trans matrix} satisfies $\lambda_{\max}(\Omega)\leq C$ for some positive constant $C>0.$
\end{enumerate}
The conditional sub-gaussian assumption is required to establish some concentration results. For the special case {where}
%that 
$\epsilon_i$ and $\delta_i$ are independent of $Z_i, X_i,$ {it is sufficient}
%this is required 
to assuming sub-gaussian errors $\epsilon_i$ and $\delta_i.$ The sub-gaussian conditions on regression errors may be relaxed to moment conditions. The conditions on $V_i$ and $f_i$ will be automatically satisfied with high probability if $\E V_iV_i^{\intercal}$ is positive definite and $V_i$ and $f_i$  are sub-gaussian random variables, where the sub-gaussianality conditions on $\{V_i,f_i\}_{1\leq i\leq n}$ may be relaxed to moment conditions. 
The proof of Lemma \ref{lem: general transform} in the supplement shows that $\lambda_{\max}(\Omega)\leq 1$ for 
%the 
random forests and deep neural networks.

The second assumption is imposed on the {generalized} IV strength $\conpara$ defined in \eqref{eq: IV strength homo}. {Throughout the paper, the asymptotics is taken as $n \to \infty.$} 

%\vspace{3mm}

\begin{enumerate}
\item[(R2)]  ${f}_{\mathcal{A}_1}^{\intercal}\T(V) {f}_{\mathcal{A}_1}$ satisfies 
$
{f}_{\mathcal{A}_1}^{\intercal}\T(V) {f}_{\mathcal{A}_1}\rightarrow \infty$ and $ {f}_{\mathcal{A}_1}^{\intercal}\T(V) {f}_{\mathcal{A}_1}\gg {\rm Tr}[\T({V})]$, with $\T({V})$ defined in \eqref{eq: RF init}.
 \end{enumerate}
{The above condition is closely related to Condition \ref{cond: identification}, where ${f}_{\mathcal{A}_1}^{\intercal}\T(V) {f}_{\mathcal{A}_1}$  measures the variability of the estimated $\widehat{f}$ after adjusting for $V$ 
used to approximate $\{g(Z_i,X_i)\}_{1\leq i\leq n}$. Intuitively, a larger value of ${f}_{\mathcal{A}_1}^{\intercal}\T(V) {f}_{\mathcal{A}_1}$ indicates that $\E(f(Z_i,X_i)-V_i^{\intercal}\gamma^*)^2>0$ in Condition \ref{cond: identification} holds more plausibly. The generalized IV strength $\conpara$ introduced in \eqref{eq: IV strength homo} is proportional to  ${f}_{\mathcal{A}_1}^{\intercal}\T(V) {f}_{\mathcal{A}_1}$ when ${\rm Var}(\delta_i\mid X_i,Z_i)$ is of a constant scale. Our proposed test in Section
\ref{sec: IV strength} is designed to test whether $\conpara$ is sufficiently large for  \texttt{TSCI}. For the setting with $\E(f(Z_i,X_i)-V_i^{\intercal}\gamma^*)^2$ in Condition \ref{cond: identification} being a positive constant, ${f}_{\mathcal{A}_1}^{\intercal}\T(V) {f}_{\mathcal{A}_1}$ can be of the {order}
%scale 
$n$, making Condition {\rm (R2)} a mild assumption.}  %which is closely related to the above condition.
%Importantly, 

%Two important remarks are in order. Firstly, 
%If we assume that the hat matrix $\Omega$ leads to an accurate estimator of $f$, then {$\Omega f_{\mathcal{A}_1}\approx \Omega D_{\mathcal{A}_1}\approx f_{\mathcal{A}_1}$} and hence
%${f}_{\mathcal{A}_1}^{\intercal}\T(V) {f}_{\mathcal{A}_1}\approx {f}_{\mathcal{A}_1}^{\intercal}P^{\perp}_{\widehat{V}_{\mathcal{A}_1}} {f}_{\mathcal{A}_1}= \sum_{i=1}^{n_1}\left[ f_i-\tau^{\intercal}(\widehat{V}_i^{\intercal},\widehat{W}_i^{\intercal})\right]^2$
%where $\widehat{V}$ and $\widehat{W}$ are defined in \eqref{eq: RF init} and $\tau$ is the regression coefficient of regressing $f_{\mathcal{A}_1}$ on $\widehat{V}_{\mathcal{A}_1}$ and $\widehat{W}_{\mathcal{A}_1}.$ {{\rm (R2)} essentially requires} $$\min_{1\leq i\leq n_1}\E \left[ f_i-\tau^{\intercal}(\widehat{V}_i^{\intercal},\widehat{W}_i^{\intercal})\right]^2 \gg \frac{\max\{1,{\rm Tr}[\T({V})]\}}{n_1}, \quad \text{with} \quad n_1=\lceil 2n/3\rceil,$$
%where the expectation is taken conditioning on the data in $\mathcal{A}_2.$ This condition intuitively requires every of $\{f_i=f(Z_i,X_i)\}_{1\leq i\leq n_1}$ to have enough variability after adjusting for the covariates and the violation form. More importantly, 

The following proposition establishes the consistency of $\widehat{\beta}_{\rm init}(V)$ if {Conditions (R1) and (R2) hold} and the approximation errors $\{R_i(V)=g(Z_i, X_i)-V^{\intercal}_{i}\pi\}_{1\leq i\leq n}$ are small.

\begin{Proposition}
Consider the models \eqref{eq: outcome model} and \eqref{eq: treatment model}. If Conditions {\rm (R1)} and {\rm (R2)} hold and $\|R_{\mathcal{A}_1}(V)\|_2^2\ll {f}_{\mathcal{A}_1}^{\intercal}\T(V) {f}_{\mathcal{A}_1}$, then $\widehat{\beta}_{\rm init}(V)$ defined in \eqref{eq: RF init} satisfies 
$
\widehat{\beta}_{\rm init}(V)\cip \beta.
$
\label{prop: initial consistency}
\end{Proposition}
%{The condition 
%$\|R_{\mathcal{A}_1}(V)\|_2^2\ll {f}_{\mathcal{A}_1}^{\intercal}\T(V) {f}_{\mathcal{A}_1}$
%is satisfied if $\{g(Z_i, X_i)\}_{1\leq i\leq n}$ are well approximated by $V$.} %In the extreme case, $R(V)=0$ and this condition is automatically satisfied. %\Zijian{Stops here.}
\subsection{Improvement with bias correction}
%\noindent {\bf Improvement with bias correction.} 
In the following, we present the distributional properties of the bias-corrected {\texttt{TSCI}} estimator $\widehat{\beta}(V)$ defined in \eqref{eq: ML corrected hetero} {and demonstrate the advantage of this extra bias correction step.} 
For establishing the asymptotic normality, we further impose a stronger {generalized} IV strength condition than (R2). 
\begin{enumerate}
\item[(R2-I)]  
%${f}_{\mathcal{A}_1}^{\intercal}[\T(V)]^2 {f}_{\mathcal{A}_1}$ satisfies 
%\Peter{I deleted the expression}
${f}_{\mathcal{A}_1}^{\intercal}[\T(V)]^2 {f}_{\mathcal{A}_1}\rightarrow \infty,$ ${f}_{\mathcal{A}_1}^{\intercal}[\T(V)]^2 {f}_{\mathcal{A}_1}\gg \|R(V)\|_2^2,$ and  \begin{equation*}
{f}_{\mathcal{A}_1}^{\intercal}[\T(V)]^2 {f}_{\mathcal{A}_1}\gg {\max\left\{({\rm Tr}[\T(V)])^{c}, \log n\cdot \eta_n(V)^2\cdot ({\rm Tr}[\T(V)])^2\right\}},
\end{equation*} 
where $c>1$ is some positive constant and $\eta_n(\cdot)$ is defined as
{\small
\begin{equation}
\eta_n(V)=\|f_{\mathcal{A}_1}-\widehat{f}_{\mathcal{A}_1}\|_{\infty}+\left(|\beta-\widehat{\beta}_{\rm init}(V)|+\|R(V)\|_{\infty}+{{\frac{\log n}{\sqrt{n}}}}\right)(\sqrt{\log n}+\|f_{\mathcal{A}_1}-\widehat{f}_{\mathcal{A}_1}\|_{\infty}).
\label{eq: approximation rate hetero}
\end{equation}}
\end{enumerate}

The above condition can be viewed as requiring the IV to be sufficiently strong, where ${f}_{\mathcal{A}_1}^{\intercal}[\T(V)]^2 {f}_{\mathcal{A}_1}$ can be viewed as another measure of IV strength. 
If the treatment model is fitted with neural network, we have ${f}_{\mathcal{A}_1}^{\intercal}[\T(V)]^2 {f}_{\mathcal{A}_1}= {f}_{\mathcal{A}_1}^{\intercal}\T(V) {f}_{\mathcal{A}_1}$. For random forests, we only have ${f}_{\mathcal{A}_1}^{\intercal}[\T(V)]^2 {f}_{\mathcal{A}_1}\leq {f}_{\mathcal{A}_1}^{\intercal}\T(V) {f}_{\mathcal{A}_1}$. $\eta_{n}(V)$ defined in \eqref{eq: approximation rate hetero} depends on the accuracy of the ML prediction model $\widehat{f}.$  We have $\eta_n(V)\rightarrow 0$ if $\|f_{\mathcal{A}_1}-\widehat{f}_{\mathcal{A}_1}\|_{\infty}\rightarrow 0$, $\widehat{\beta}_{\rm init}(V)\cip \beta$, and $\|R(V)\|_{\infty}\rightarrow 0$.} However, even for inconsistent $\widehat{f}$, {$\eta_{n}(V)$ is of a smaller order of magnitude than $\sqrt{\log n}$ as long as $\|f_{\mathcal{A}_1}-\widehat{f}_{\mathcal{A}_1}\|_{\infty}$ is of a constant scale.
%regardless whether $\widehat{f}$ is consistent or not.}  %for a consistent estimator  and a small approximation error $\|R(V)\|_{\infty}.$} % Hence, in comparison to (R2), Condition (R2-I) is a stronger condition on the {generalized} IV strength. 

{The following theorem establishes the asymptotic normality of the \texttt{TSCI} estimator.}
%of our proposed bias-corrected estimator $\widehat{\beta}(V)$ defined in \eqref{eq: ML corrected hetero} and further justify the validity of the confidence interval.} 
\begin{Theorem}
Consider the models \eqref{eq: outcome model} and \eqref{eq: treatment model}. Suppose that Condition {\rm (R1)}, {\rm (R2-I)} hold and
\begin{equation}
\frac{\max_{1\leq i\leq n_1} \sigma_i^2\cdot\left[\T(V) {f}_{\mathcal{A}_1}\right]_i^2}{\sum_{i=1}^{n_1}\sigma_i^2\cdot\left[\T(V) {f}_{\mathcal{A}_1}\right]_i^2}\rightarrow 0 \quad \text{with}\quad \sigma_i^2=\E(\epsilon_i^2\mid Z_i,X_i).
\label{eq: linder condition}
\end{equation}
Then $\widehat{\beta}(V)$ defined in \eqref{eq: ML corrected hetero}
 satisfies 
$$
\frac{1}{{\rm SE}(V)} \left(\widehat{\beta}(V)-\beta\right)\cid N(0,1), \quad \text{with}\quad {\rm SE}(V)=\frac{\sqrt{\sum_{i=1}^{n_1}\sigma^2_i[\T(V) {f}_{\mathcal{A}_1}]_i^2}
}{{f}_{\mathcal{A}_1}^{\intercal}\T(V) {f}_{\mathcal{A}_1}}.
$$
If $\widehat{\rm SE}(V)$ used in \eqref{eq: ML CI} satisfies $\widehat{\rm SE}(V)/{\rm SE}(V)\cip 1,$ then the confidence interval ${\rm CI}(V)$ in \eqref{eq: ML CI}
 satisfies 
 $\liminf_{n\rightarrow \infty}\PP(\beta\in {\rm CI}(V))= 1-\alpha.$
\label{thm: limiting TSRF hetero}
\end{Theorem}
{We emphasize that the validity of our proposed confidence interval does not require the ML prediction model $\widehat{f}$ to be a consistent estimator of $f$. Particularly, Condition {\rm (R2)} and {\rm (R2-I)} can be plausibly satisfied as long as the ML algorithms capture enough association between the treatment and the IVs.}  %Even if the difference $\|\widehat{f}_{\mathcal{A}_1}-f_{\mathcal{A}_1}\|_2$ is large, Condition (R2-I) can still hold {when the generalized IV strength is sufficiently large}. 
 The condition \eqref{eq: linder condition} is imposed such that not a single entry of the vector $\T(V) {f}_{\mathcal{A}_1}$ dominates all other entries, which is needed for verifying the Linderberg condition. The standard error ${\rm SE}(V)$ relies on the {generalized} IV strength for {a given matrix $V$}. If ${f}_{\mathcal{A}_1}^{\intercal}\T(V) {f}_{\mathcal{A}_1}/n_1$ is of constant order, then ${\rm SE}(V)\lesssim {1}/{\sqrt{n}}.$ A larger  matrix $V$ will generally lead to a larger ${\rm SE}(V)$ because  ${f}_{\mathcal{A}_1}^{\intercal}\T(V) {f}_{\mathcal{A}_1}$ decreases {after adjusting {for} more information contained in $V$.} The consistency of $\widehat{\rm SE}(V)$ is presented in Lemma \ref{lem: variance consistency} in the supplement.

%still holds 

{We now explain the effectiveness of the bias correction for the homoscedastic setting with ${\rm Cov}(\epsilon_i,\delta_i\mid Z_i, X_i)={\rm Cov}(\epsilon_i,\delta_i)$. {For the initial estimator $\widehat{\beta}_{\rm init}(V),$} we can establish that $\frac{1}{{\rm SE}(V)}\left(\widehat{\beta}_{\rm init}(V)-\beta\right)={\mathcal{G}}(V)+\widetilde{\mathcal{E}}(V),$
where ${\mathcal{G}}(V)\cid N(0,1)$ and
\begin{equation}
\left|\widetilde{\mathcal{E}}(V)\right|\leq \frac{{\rm Cov}(\epsilon_i,\delta_i)\cdot {\rm Tr}[\T(V)]+\|R(V)\|_2}{\sqrt{{f}_{\mathcal{A}_1}[\T(V)]^2 {f}_{\mathcal{A}_1}}}+\frac{\sqrt{{\rm Tr}([\T({V})]^2)}}{({f}_{\mathcal{A}_1}[\T(V)]^2 {f}_{\mathcal{A}_1})^{c_0}},
\label{eq: error bound no correction}
\end{equation}
for some positive constant $c_0>0.$ 
Our proposed bias-corrected estimator $\widehat{\beta}(V)$ is effective in reducing the bias component $\widetilde{\mathcal{E}}(V).$ Particularly, Theorem 
\ref{thm: homo bias correction effect}
 in Section \ref{sec: homo extra} in the supplement establishes that { the term ${\rm Cov}(\epsilon_i,\delta_i)\cdot {\rm Tr}[\T(V)]/\sqrt{{f}_{\mathcal{A}_1}[\T(V)]^2 {f}_{\mathcal{A}_1}}$ in \eqref{eq: error bound no correction} is reduced to $\eta_n(V)\cdot {\rm Tr}[\T(V)]/\sqrt{{f}_{\mathcal{A}_1}[\T(V)]^2 {f}_{\mathcal{A}_1}}$}. If $\eta_n(V)\rightarrow 0$, which can be achieved for a consistent ML prediction 
$\widehat{f}$, the bias-corrected \texttt{TSCI} estimator effectively reduces the finite-sample {or higher-order} bias. However, even if $\widehat{f}_{\mathcal{A}_1}$ is inconsistent with $\|f_{\mathcal{A}_1}-\widehat{f}_{\mathcal{A}_1}\|_{2}\lesssim \sqrt{n}$, the bias correction will not lead to a worse estimator. %We demonstrate the finite-sample performance of the bias correction in Tables \ref{tab: Model 2 Error 3} and \ref{tab: binary IV model 3}.} %\Peter{OK, reads good!}

\subsection{Guarantee for Algorithm \ref{algo: TSCI selection}}
%\noindent{\bf Guarantee for Algorithm \ref{algo: TSCI selection}.} 
{We now justify the validity of the confidence interval {from the \texttt{TSCI}}
%outputed by 
Algorithm \ref{algo: TSCI selection}, which is based on the asymptotic normality of $\widehat{\beta}(V_{\widehat{q}})$ with a  data-dependent index $\widehat{q}$. %The key is to analyze
% multiple normalized differences $\left\{\left(\widehat{\beta}(V_q)-\widehat{\beta}(V_{q'})\right)\right\}_{0\leq q<q'\leq Q_{\max}}$.} 
 {The property of $\widehat{q}$ relies on a careful analysis of the difference $\widehat{\beta}(V_q)-\widehat{\beta}(V_{q'})$ for $0\leq q<q'\leq Q_{\max}$.
 
 We start with the setting where both $V_{q}$ and $V_{q'}$ provide good approximations to $\{g(Z_i,X_i)\}_{1\leq i\leq n},$ leading to sufficiently small $\|R(V_{q})\|_2$ and $\|R(V_{q'})\|_2$. In this case, the dominating component of $\widehat{\beta}(V_q)-\widehat{\beta}(V_{q'})$} is $S^{\intercal}\epsilon_{\mathcal{A}_1}$ with
\begin{equation}
S=\left(\frac{1}{f_{\mathcal{A}_1}^{\intercal} \T(V_{q'}) f_{\mathcal{A}_1}}\T(V_{q'})-\frac{1}{f_{\mathcal{A}_1}^{\intercal} \T(V_{q}) f_{\mathcal{A}_1}}\T(V_{q})\right)f_{\mathcal{A}_1}\in \R^{n_1}.
\label{eq: vector diff}
\end{equation}
Conditioning on the data in $\mathcal{A}_2$ and $\{X_i, Z_i\}_{i\in \mathcal{A}_1},$ $S^{\intercal}\epsilon_{\mathcal{A}_1}$ is of zero mean and variance
{\small
\begin{equation}
\begin{aligned}
{H}(V_{q},V_{q'})=\frac{\sum_{i=1}^{n_1}\sigma_i^2[\T(V_{q'}) {f}_{\mathcal{A}_1}]_i^2}{[{f}_{\mathcal{A}_1}^{\intercal} \T(V_{q'}) f_{\mathcal{A}_1}]^2}+\frac{\sum_{i=1}^{n_1}\sigma_i^2[\T(V_{q}) {f}_{\mathcal{A}_1}]_i^2}{[{f}_{\mathcal{A}_1}^{\intercal} \T(V_{q}) f_{\mathcal{A}_1}]^2}
-2\frac{\sum_{i=1}^{n_1}\sigma_i^2 [\T(V_{q'}) {f}_{\mathcal{A}_1}]_i[\T(V_{q}) {f}_{\mathcal{A}_1}]_i}{[{f}_{\mathcal{A}_1}^{\intercal} \T(V_{q'}) f_{\mathcal{A}_1}]\cdot[{f}_{\mathcal{A}_1}^{\intercal} \T(V_{q}) f_{\mathcal{A}_1}]},
\end{aligned}
\label{eq: difference variance}
\end{equation}
}
with $\sigma_i^2=\E(\epsilon_i^2\mid Z_i,X_i).$
The following condition  requires that the variance of $S^{\intercal}f_{\mathcal{A}_1}$ dominates other components of  $\widehat{\beta}(V_{q})-\widehat{\beta}(V_{q'}),$ which are mainly finite-sample approximation errors.
%\vspace{3mm}

\begin{enumerate}
\item[(R3)] The variance $H(V_q, V_{q'})$ in \eqref{eq: difference variance} satisfies 
\begin{equation*}
\sqrt{H(V_q, V_{q'})} \gg \max_{V\in\{V_{q}, V_{q'}\}}\left\{\frac{1}{\mu(V)}\left[1+(1+\sqrt{\log n}\cdot\eta_n(V_{Q_{\max}}))\cdot {\rm Tr}[\T(V)]\right]\right\},
\label{eq: approx error assump 2}
\end{equation*}
with $\eta_n(\cdot)$ defined in \eqref{eq: approximation rate hetero}.
There exists $c>0$ such that ${\rm Var}(\delta_i\mid Z_i,X_i)\geq c.$
\end{enumerate}
%{As discussed after Condition (R2-I), $\eta_n(V_{Q_{\max}})\leq C 
%\log n$ no matter whether the ML prediction model $\widehat{f}$ is consistent or not.}
%If $\widehat{f}_{\mathcal{A}_1}$ accurately estimate $f_{\mathcal{A}_1}$ and $g$ is well approximated by $V_{Q_{\max}}$, $\sqrt{\log n}\cdot\eta_n(V_{Q_{\max}})\leq c$ for some positive constant $c>0.$ 
When the first stage is fitted with the basis method or neural network  and ${\rm Var}(\epsilon_i\mid Z_i,X_i)=\sigma_{\epsilon}^2, {\rm Var}(\delta_i\mid Z_i,X_i)=\sigma_{\delta}^2,$ we have
$H(V_q, V_{q'})=\sigma_{\epsilon}^2\left({1}/{f^{\intercal} \T({V}_{q'}) f}-{1}/{f^{\intercal} \T({V}_{q}) f}\right)$ and $ {\mu(V_q)}={f^{\intercal} \T({V}_{q}) f}/\sigma_{\delta}^2$ for $q\leq q'.$ 
If we assume that $f^{\intercal} \T({V}_{q'}) f=c_*f^{\intercal} \T({V}_{q}) f$ for some $0<c_*<1,$ we have $H(V_q, V_{q'})=\frac{1-c_*}{c_*}\frac{\sigma_{\epsilon}^2}{\sigma_{\delta}^2}\mu(V_q).$ In this case, Condition {\rm (R3)} is satisfied if $\mu(V_q)\gg ({\rm Tr}[\T(V_q)])^2$ and  $\mu(V_{q'})\gg ({\rm Tr}[\T(V_{q'})])^2$ up to some polynomial order of $\log n$. In this case, Condition (R3) is slightly stronger than Condition (R2-I).

%When both $V_q$ and $V_{q'}$ accurately approximate
%$
%\left(\widehat{\beta}(V_q)-\widehat{\beta}(V_{q'})\right)/{\sqrt{H(V_q,V_{q'})}}\cid N(0,1),
%$

The following theorem establishes the asymptotic normality of 
$\widehat{\beta}(V_q)-\widehat{\beta}(V_{q'})$ under the null setting where both $R(V_q)$ and $R(V_{q'})$ are small. %More theoretical results about comparing $\widehat{\beta}(V_q)$ and $\widehat{\beta}(V_{q'})$ can be found in Section \ref{sec: comparison two} in the supplement.
\begin{Theorem} 
Consider the model \eqref{eq: outcome model} and \eqref{eq: treatment model}. Suppose that {\rm (R1)} and {\rm (R3)}  hold, {\rm (R2)} 
holds for $V\in\{V_{q}, V_{q'}\}$, and $S$ defined in \eqref{eq: vector diff} satisfies $\max_{i\in \mathcal{A}_1}{S_i^2}/(\sum_{i\in \mathcal{A}_1}S_i^2)\rightarrow 0$.  If $
\sqrt{H(V_q, V_{q'})} \gg \max_{V\in\{V_{q}, V_{q'}\}}{\|R(V)\|_2}/{\sqrt{\mu(V)}},$
then 
$
\left(\widehat{\beta}(V_q)-\widehat{\beta}(V_{q'})\right)/{\sqrt{H(V_q,V_{q'})}}\cid N(0,1),
$
with $\widehat{\beta}(\cdot)$ and $H(V_{q},V_{q'})$ defined in \eqref{eq: ML corrected hetero seq} and  \eqref{eq: difference variance}, respectively. 
\label{thm: comparison RF}
\end{Theorem}
{For small approximation errors $\|R(V_q)\|_2$ and $\|R(V_{q'})\|_2$, the condition $\sqrt{H(V_q, V_{q'})} \gg \max_{V\in\{V_{q}, V_{q'}\}} {\|R(V)\|_2}/{\sqrt{\mu(V)}} $ holds. In this case, Theorem \ref{thm: comparison RF} establishes that the difference $\widehat{\beta}(V_{q})-\widehat{\beta}(V_{q'})$ is centered at zero and has an asymptotic normal distribution. 

We now move on to the case that at least one of $V_{q}$ and $V_{q'}$ does not approximate $\{g(Z_i,X_i)\}_{1\leq i\leq n}$ well. In this case,  Theorem \ref{thm: comparison RF} does not hold and we establish Theorem \ref{thm: power RF} in the supplement that $(\widehat{\beta}(V_{q})-\widehat{\beta}(V_{q'}))/{\sqrt{H(V_q,V_{q'})}}$ is centered at}  
\begin{equation}
\mathcal{L}_n(V_{q}, V_{q'})=\frac{1}{\sqrt{H(V_q,V_{q'})}}\left(\frac{D_{\mathcal{A}_1}^{\intercal} \T(V_{q}) [R(V_{q})]_{\mathcal{A}_1} }{{D}^{\intercal}_{\mathcal{A}_1} \T(V_{q})D_{\mathcal{A}_1}}-\frac{D_{\mathcal{A}_1}^{\intercal} \T(V_{q'}) [R(V_{q'})]_{\mathcal{A}_1}}{{D}^{\intercal}_{\mathcal{A}_1} \T(V_{q'})D_{\mathcal{A}_1}}\right).
\label{eq: sig difference RF}
\end{equation}
%\Peter{Does this diverge? Do we have power?}\Zijian{We have added back the following theorem.} \Peter{OK, fine now!}

%The limiting distribution in Theorem \ref{thm: comparison RF} implies the size and power of the comparison test $\mathstadlercal{C}(V_q, V_{q'})$ in \eqref{eq: comparison test}. We state this result
% in . 

%\Peter{Do we have equla to  "=" or smaller or equal to $\le$?}

%$\mathcal{L}_n(V_{q}, V_{q'})$ means the difference between the bias terms of $\widehat{\beta}(V_q)$ and $\widehat{\beta}(V_{q'})$. In a special case, if $h$ is exactly spanned by $V_q'$, then $\mathcal{L}_n(V_{q}, V_{q'})$ measures the bias of $\widehat{\beta}(V_q).$ 

%

To simultaneously quantify the errors of $\{\widehat{\beta}(V_q)-\widehat{\beta}(V_{q'})\}_{0\leq q< q'\leq Q_{\max}},$ we define the $\alpha_0$ quantile for the maximum of multiple random error components in \eqref{eq: vector diff},
%To facilitate the theoretical discussion, we define,
\begin{equation}
\PP\left(\max_{0\leq q< q'\leq Q_{\max}}\frac{1}{\sqrt{{H}(V_{q},V_{q'})}}\left|\frac{f_{\mathcal{A}_1}^{\intercal} \T(V_{q'}) \epsilon_{\mathcal{A}_1}}{{f}_{\mathcal{A}_1}^{\intercal} \T(V_{q'}) f_{\mathcal{A}_1}}-\frac{f_{\mathcal{A}_1}^{\intercal} \T(V_{q}) \epsilon_{\mathcal{A}_1}}{f_{\mathcal{A}_1}^{\intercal} \T(V_{q}) f_{\mathcal{A}_1}}\right|\geq \rho(\alpha_0)\right)=\alpha_0.
\label{eq: quantile}
\end{equation}

{We introduce the following condition for accurate selection among $\{\mathcal{V}_q\}_{0\leq q\leq Q}.$
%$\mathcal{V}_{\widehat{q}}$ provides a good approximation for $g(\cdot)$ in the outcome model. %accurate selection of  %among $\{\mathcal{V}_q\}_{0\leq q\leq Q}$.}
\begin{enumerate}
\item[(R4)] For $\mathcal{V}_{0} \subset \mathcal{V}_{1}\subset \cdots \subset \mathcal{V}_{Q}$ and corresponding matrices $\{{V}_{q}\}_{0\leq q\leq Q}$, there exists $q^{*} \in \{0,1,2,\cdots, Q\}$ such that $q^{*}\leq Q_{\max}$ and $R(V_{q^{*}})=0$ with $Q_{\max}$ defined in \eqref{eq: max index RF}. For any integer $q \in[0, q^{*}-1],$ there exists an integer $q' \in [q+1, q^{*}]$ such that  
\begin{equation}
\mathcal{L}_n(V_{q}, V_{q'})\geq A\rho({\alpha_0}) \quad \text{with}\quad A>2,
\label{eq: condition optimal Q}
\end{equation}
where $\mathcal{L}_n(V_{q}, V_{q'})$ is defined in \eqref{eq: sig difference RF} and $\rho({\alpha_0})$ is defined in \eqref{eq: quantile}.
\end{enumerate} 

{The above condition ensures that there exists $\mathcal{V}_{q^{*}}$ such that the function $g$ is well approximated by the column space of $V_{q^{*}}$.} The well-separation condition \eqref{eq: condition optimal Q} is  interpreted as follows: {if $g$ is not well approximated by the column space of $V_q$}, then the estimation bias of $\widehat{\beta}(V_q)$ is larger than the uncertainty $\rho({\alpha_0})$ due to the multiple random error components. Note that $\mathcal{L}_n(V_{q}, V_{q^{*}})$ defined in \eqref{eq: sig difference RF} is a measure of the bias of $\widehat{\beta}(V_q).$

The following theorem guarantees the coverage property for the CIs corresponding to $V_{\widehat{q}_c}$ and $V_{\widehat{q}_r}$ in Algorithm \ref{algo: TSCI selection}.
%We now state the validity of our proposed confidence interval .

\begin{Theorem}
Consider the model \eqref{eq: outcome model} and \eqref{eq: treatment model}. Suppose that Conditions {\rm (R1)} and {\rm (R4)} hold, Condition {\rm (R2-I)} holds for $V\in\{V_{q} \}_{0\leq q\leq Q_{\max}}$, Condition {\rm (R3)} holds for any $0\leq q<q'\leq Q_{\max}$, $\widehat{H}(V_{q},V_{q'})/H(V_{q},V_{q'})\cip 1$, and $\widehat{\rho}/\rho(\alpha_0)\cip 1$
with $\widehat{\rho}$ used in \eqref{eq: lawyer test RF} and $\rho(\alpha_0)$ defined in 
\eqref{eq: quantile} respectively. Our proposed CI in Algorithm \ref{algo: TSCI selection}  satisfies 
$
\liminf_{n\rightarrow \infty} \PP\left[\beta \in {\rm CI}(V_{\widehat{q}})\right]\geq 1-\alpha-2\alpha_0$ with  $\widehat{q}=\widehat{q}_c$ {or}  $\widehat{q}_r$
{where $\alpha_0$ is used in \eqref{eq: quantile}.}
\label{thm: post-selection RF}
\end{Theorem}

%\begin{Remark}
%\label{rem: uniform} 
%\rm  %\Zijian{Peter, double check.}
The condition \eqref{eq: condition optimal Q} of (R4) is critical to guarantee the selection consistency $\widehat{q}_c=q^*$, ensuring that ${\rm CI}(V_{\widehat{q}_c})$ 
%to 
achieves the desired coverage. However, in finite samples, we may make mistakes in conducting the selection among $\{V_q\}_{0\leq q\leq Q_{\max}}.$ The selection method $\widehat{q}_{r}$ is more robust in the sense that statistical inference based on $V_{\widehat{q}_r}$ is still valid even if we cannot separate $V_{q^{*}-1}$ and $V_{q^{*}}.$ To achieve the uniform inference without requiring the well-separation condition \eqref{eq: condition optimal Q}, we may simply apply \texttt{TSCI} with $\mathcal{V}_{Q_{\max}}$. However, such a confidence interval might be conservative by adjusting for a large $V_{Q_{\max}}.$ %\Peter{COMMENT: is $V_{Q_{max}}$ a matrix or rather $M(V_{Q_{max}})$? We could just write "... by adjusting for a large $V_{Q_{max}}$?} %requires that at least one of the bias measures $\left\{\mathcal{L}_n(V_{q}, V_{q'})\right\}_{q+1\leq q'\leq q^{*}}$ The CI based on $V_{\widehat{q}_r}$ is more robust to the violation space selection error. 
%In particular, the well-separation condition \eqref{eq: condition optimal Q} can be relaxed as follows: for any integer $q \in[0, q^{*}-2],$ there exists an integer $q' \in [q+1, q^{*}]$ such that  \eqref{eq: condition optimal Q} holds. That is, 
%\end{Remark}
%\Peter{Should thsi remark be "displayed" more clearly?}\Zijian{I have put this as a separate remark. Let me know if you have other suggestions.} \Peter{OK, all good.}

%%%%%%%%%%%%%%%%%%%%%%%%%%%%%%%%%%%%%%%%%%%%%%%%%%%%%%%%%%
\section{Simulation studies}
\label{sec: sim}
%%%%%%%%%%%%%%%%%%%%%%%%%%%%%%%%%%%%%%%%%%%%%%%%%%%%%%%%%%
%\Zijian{Peter, double check this section.}
%\Zijian{Shorten the paper to 31-32 pages and highlight whatever moved to the supplement.}
%\Mengchu{added new introductions}
In Section \ref{sec: dml valid IV sim}, we consider the valid IV settings and compare our proposal to the \texttt{DML} estimator proposed in \citet{chernozhukov2018double}. In Sections \ref{sec: invalid IV sim} and \ref{sec: different nonlinearity sim}, we demonstrate our proposal for general invalid IV settings and also compare it with the \texttt{DML} estimator. In Section \ref{sec: multiple IV} in the supplement, we further consider multiple IVs settings and compare \texttt{TSCI} with existing methods based on the majority rule. We compute all measures using 500 simulations throughout the numerical studies. The code for replicating the numerical results is available at \url{https://github.com/zijguo/TSCI-Replication}. %including \texttt{TSHT} and \texttt{CIIV} proposed in \citet{guo2018confidence,windmeijer2019confidence}, respectively.  %\Zijian{Mengchu, update the code.}

%\Zijian{List what else we have in the supplement?}

%%%%%%%%%%%%%%%%%%%%%%%%%%%%%%%%%%%%%%%%%%%%%%%%%%
%%%%%%%%%%%%%%%%%%%%%%%%%%%%%%%%%%%%%%%%%%%%%%%%%%
\subsection{Comparison with DML in valid IV settings}
\label{sec: dml valid IV sim}
%%%%%%%%%%%%%%%%%%%%%%%%%%%%%%%%%%%%%%%%%%%%%%%%%%
%%%%%%%%%%%%%%%%%%%%%%%%%%%%%%%%%%%%%%%%%%%%%%%%%%
%We compare DML and TSCI  show TSCI's capability of increasing the IV strength by capturing the nonlinear relationship for the treatment model. 
We focus on the valid IV settings and illustrate the advantages and limitations of both  \texttt{TSCI} and \texttt{DML}. %We consider the valid IV setting here and move on to the invalid IV setting in Section \ref{sec: different nonlinearity sim}. 
We adopt the data generative setting used in the R package \texttt{dmlalg} \citep{dmlalg_emmenegger_2021} and implement \texttt{DML} by the R package \texttt{DoubleML} \citep{bach2021doubleml}. We set the sample size as $n=3000$ and generate the baseline covariate $X_i\in \R$ following the uniform distribution on $[-\pi,\pi]$. Conditioning on $X_i$, we generate the IV $Z_i$ and the hidden confounder $H_i$ following $Z_i\mid X_i\sim N(3\cdot\tanh(2X_i-1),1)$ and $H_i\mid X_i \sim N(2\cdot\sin(X_i),1)$, respectively. We generate the outcome $Y_i$ and treatment $D_i$ using the SEM in \eqref{eq: sem linear}. Particularly, we generate the outcome $Y_i$ following $Y_i=D_i+ X_i^2/2 - 3\cdot\cos(\pi H_i/4)  + e_{i,2}$ and consider the following two treatment models,
\begin{itemize}
    \item Setting S1: $D_i=-a\cdot |Z_i|  - 2\cdot\tanh(X_i)- H_i + e_{i,1},$
    \item Setting S2: $D_i = a\cdot Z_i^2/2-2\cdot\tanh(X_i)-H_i+e_{i,1},$
\end{itemize}
where $a$ controls the nonlinearity of the treatment model {and strength of the IV} and $e_{i,1}\sim N(0,1)$ and $e_{i,2}\sim N(0,1)$ are independent noises.  Note that a larger value of $a$ in the treatment model leads to a more nonlinear dependence between $D_i$ and $Z_i$. The treatment model and the outcome model are consistent to \eqref{eq: treatment model} and \eqref{eq: outcome model} with $\delta_i$ and $\epsilon_i$ being composed of a hidden confounder term and an independent random noise. 

In addition to the above two settings, we consider another setting where $g(X_i)$ in the outcome model is more complicated. We set $p_x=5$ and for $1\leq i\leq n$, we generate $\{X_{i,j}\}_{1\leq j\leq p_{x}}$ which are independently distributed, and each of them follows a uniform distribution on $[-1,1]$. We generate $Z_i$ and $H_i$ following $Z_i\mid X_i\sim N(3\cdot\tanh(2X_{i,1}-1),1)$ and $H_i\mid X_i \sim N(2\cdot\sin(X_{i,1}),1)$. We generate $\{D_i,Y_i\}_{1\leq i\leq n}$ following 

\begin{itemize}
    \item Setting S3: $D_i=b\cdot Z_i+\sin(2\pi Z_i)+3/2\cdot\cos(2\pi Z_i)-1/5\cdot\sum_{j=1}^{p_x}X_{i,j}- 1/2\cdot H_i+e_{i,1}$ and $Y_i=D_i/2+g(X_i)-\cos(\pi H_i/4)+e_{i,2}$ with $g(X_i)=\sum_{1\leq j\neq k \leq p_x}X_{i,j}X_{i,k}$,
\end{itemize}
where the value of $b$ controls the linear association strength in Setting S3. We implement Algorithm \ref{algo: TSCI selection} with random forests by specifying a collection of basis functions for $g(x)$. Since the IV is valid here, we use the set of basis functions $\mathcal{V}_0=\{1,\overrightarrow{\bf b}_1(x_1),\cdots,\overrightarrow{\bf b}_{p_x}(x_{p_x})\}$ with $\overrightarrow{\bf b}_j(x_j)$ denoting the B-spline basis with 5 knots of $x_j$ for $1\leq j\leq p_x$. 
\begin{figure}[ht!]
    \centering
    \includegraphics[width=0.8\linewidth]{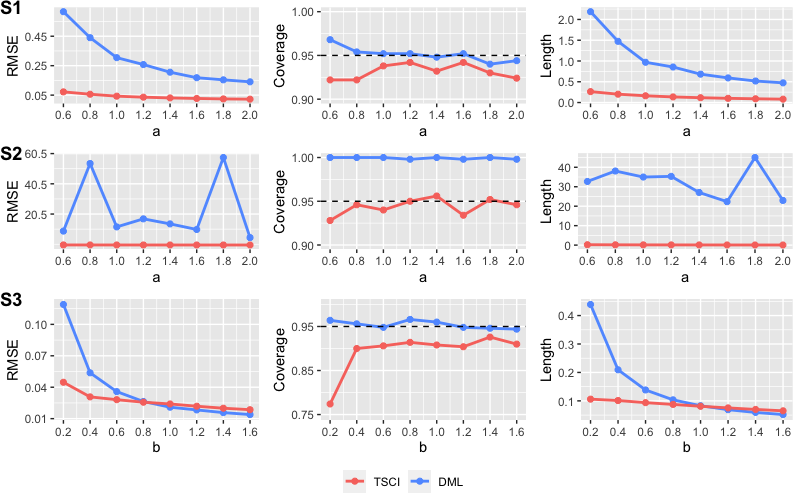}
    \caption{\small Comparison of \texttt{DML} and \texttt{TSCI} in terms of RMSE, CI coverage, and length under Settings S1, S2, and S3. The larger value of the constant $a$ in Settings S1 and S2 stands for a higher nonlinearity level in the treatment model, and the larger value of $b$ in Setting S3 for a higher linearity level. {{In addition,} a larger value of $a$ and $b$ indicate larger generalized IV strength.}}%``\texttt{TSCI}'' is our proposal in Algorithm \ref{algo: TSCI selection} with random forests. ``\texttt{DML}'' is the method proposed in \citet{chernozhukov2018double} implemented by  \texttt{DoubleML} \citep{bach2021doubleml}.} %\Peter{Is a small value of $a$ also indicating that the generalized IV stregth is low?}\Zijian{Yes, please check my added sentence.}}
    \label{fig: dml valid}
\end{figure}

In Figure \ref{fig: dml valid}, we compare \texttt{DML} and \texttt{TSCI} under Settings S1, S2, and S3. In the top two panels, \texttt{TSCI} achieves the desired coverage level in Settings S1 and S2. Even though \texttt{DML} achieves desired coverage in setting S1, the RMSE and CI length of \texttt{DML} is uniformly larger than those of our proposed \texttt{TSCI}. As discussed in Section \ref{sec: dml comp}, this happens since \texttt{DML} only uses the linear association in the treatment model while \texttt{TSCI} can increase the IV strength by applying ML algorithms to fit nonlinear treatment models. We have designed setting S2 as a less favorable setting for \texttt{DML}, where the linear association between the treatment and IV is nearly zero but the nonlinear association is strong. Consequently, the CI length and RMSE of the \texttt{DML} procedure is much larger than our proposed \texttt{TSCI}. In Setting S3, \texttt{TSCI} does not show the advantage over \texttt{DML} and has an under-coverage problem. This is caused by the estimation bias due to the misspecification of $g(x)$ in the outcome model. \texttt{DML} is not exposed to such bias because it uses ML algorithms to fit $E(Y_i \mid X_i)$. We also implement \texttt{TSLS} in Settings S1, S2 and S3, but we do not include it in the figures because it has low coverage due to misspecification of nonlinear $g(x)$.

%DML estimator in Setting S2 are too long to make meaningful inference. In Setting 1 with moderate nonlinearity level, DML is able to generate valid CIs with normal length, but it is still less efficient than TSCI.  

%%%%%%%%%%%%%%%%%%%%%%%%%%%%%%%%%%%%%%%%%%%%%%%%%%
%%%%%%%%%%%%%%%%%%%%%%%%%%%%%%%%%%%%%%%%%%%%%%%%%%
\subsection{\texttt{TSCI} with invalid IVs}
\label{sec: invalid IV sim}
%%%%%%%%%%%%%%%%%%%%%%%%%%%%%%%%%%%%%%%%%%%%%%%%%%
%%%%%%%%%%%%%%%%%%%%%%%%%%%%%%%%%%%%%%%%%%%%%%%%%%
{We now demonstrate our \texttt{TSCI} method under the general setting with possibly invalid IVs.}
{In the following, we focus on the continuous treatment and will investigate the performance of \texttt{TSCI} for binary treatment in Section \ref{sec: binary treatment} in the supplement.} We generate $X^{*}_{i}\in \R^{p_x+1}$ following a multivariate normal distribution with zero mean and covariance matrix $\Sigma$ where $\Sigma_{ij}=0.5^{|i-j|}$ for $1\leq i,j\leq p_x+1$. With $\Phi$ {denoting} 
the standard normal cumulative distribution function, we define $X_{ij}=\Phi(X^{*}_{ij})$ for $1\leq j\leq p_x.$ We generate a continuous IV as $Z_{i}=4(\Phi(X^{*}_{i,p_x+1})-0.5)\in (-2,2)$. 

We consider the following two conditional mean models {for the treatment}, 
\begin{itemize}
    \item Setting B1: 
    $f(Z_{i},X_{i})= -\frac{25}{12}+Z_{i}+Z_i^3/3+Z_{i}\cdot(a\cdot\sum_{j=1}^{5}X_{ij})-\frac{3}{10}\cdot\sum_{j=1}^{p}X_{ij},$
    
    \item Setting B2:
    $f(Z_{i},X_{i})= \sin(2\pi Z_{i})+\frac{3}{2}\cos(2\pi Z_{i})+Z_{i}\cdot(a\cdot\sum_{j=1}^{5}X_{ij})-\frac{3}{10}\cdot\sum_{j=1}^{p}X_{ij}.$
\end{itemize}
The value of the constant $a$ controls the interaction strength between $Z_i$ and the first five variables of $X_{i}$, and when $a=0$, the interaction term disappears. We vary $a$ across $\{0,0.5,1\}$ and $n$ across $\{1000,3000,5000\}$ to observe the performance. 

{We consider the outcome model \eqref{eq: outcome model} with two forms of $g(Z_i,X_i)$: (a)Vio=1: $g(Z_i,X_i) = Z_i+1/5\cdot \sum_{j=1}^{p_x}X_{ij}$;
(b)Vio=2: $g(Z_i,X_i) = Z_i+Z_i^2-1+1/5\cdot \sum_{j=1}^{p_x}X_{ij}$. We set the errors $\{(\delta_i, \epsilon_i)^\intercal\}_{1\leq i\leq n}$ as heteroscedastic following \citet{bekker2015jackknife}: for $1\leq i\leq n,$ generate $\delta_i \sim N(0,Z_i^2+0.25)$ and  
$
\epsilon_i=0.6 \delta_i+\sqrt{{[1-0.6^2]}/[0.86^4+1.38072^2]}(1.38072\cdot \tau_{1,i}+0.86^2\cdot \tau_{2,i}),
$
where conditioning on $Z_i$, $\tau_{1,i}$ and $\tau_{2,i}$ are generated to be independent of $\delta_i$, with $\tau_{1,i}\sim N(0, Z_i^2+0.25)$ and $\tau_{2,i} \sim N(0,1).$

% We consider two different distributions for the errors $\{(\delta_i,\epsilon_i)^{\intercal}\}_{1\leq i\leq n}$ in \eqref{eq: outcome model} and \eqref{eq: treatment model}:\\

% \vspace{-2mm}
% \noindent {\bf Error distribution 1 (homoscedastic error).} Generate $\{(\delta_i,\epsilon_i)^{\intercal}\}_{1\leq i\leq n}$ following i.i.d. bivariate normal with zero mean, unit variance, and covariance 0.5.\\ 

% \vspace{-2mm}
% \noindent {\bf Error distribution 2 (heteroscedastic error).} Following  \citet{bekker2015jackknife}, we   generate $\delta_i \sim N(0,Z_i^2+0.25)$ and  
% $
% \epsilon_i=0.6 \delta_i+\sqrt{{[1-0.6^2]}/[0.86^4+1.38072^2]}(1.38072\cdot \tau_{1,i}+0.86^2\cdot \tau_{2,i}),
% $ for $1\leq i\leq n,$
% where conditioning on $Z_i$, $\tau_{1,i}\sim N(0, Z_i^2+0.25),$ $\tau_{2,i} \sim N(0,1),$ and $\tau_{1,i}$ and $\tau_{2,i}$ are independent of $\delta_i.$  
% \vspace{2mm}

%Error distributions 1 and 2 respectively correspond to  and s, with the generation of heteroscedastic errors .

We shall implement \texttt{TSCI} with random forests as detailed in Algorithm \ref{algo: TSCI selection}. Since the IV is possibly invalid, we consider four possible basis sets $\{\mathcal{V}_q\}_{0\leq q\leq 3}$ to approximate $g(z,x)$, where $\mathcal{V}_0=\{\overrightarrow{\bf w}(x)\}$ and $\mathcal{V}_q=\{z,\cdots,z^q,\aw(x)\}$ for $1\leq q\leq 3$ with $\overrightarrow{\bf w}(x)=\{1,x_1,\cdots,x_{p_x}\}$. Our proposed \texttt{TSCI} is designed to choose the best $\mathcal{V}_{q}$.

%\Zijian{Mengchu, move the following to the supplement?}For Setting B3 with a binary IV, we consider the interaction violation and two sets of basis functions to approximate $g(z,x)$ where   
%$\mathcal{V}_0=\{\aw(x)\}$ and $\mathcal{V}_1=\{z,z\cdot x_1,\cdots, z\cdot x_5,\aw(x)\}$ with $\overrightarrow{\bf w}(x)=\{1,x_1,\cdots,x_{p_x}\}$.  

%with specified basis sets and choose the best $\mathcal{V}_q$ by the comparison and robust selection methods. 
% {In addition, we investigate the TSCI estimator with replacing the first stage random forests by B-spline basis expansion, where the number of basis element is chWould you have time today to chat? I’d be curious to hear ;)osen by cross validation; see details in Section \ref{sec: TS-Bspline} in the supplement.} 
%{The code for implementing our proposed TSCI method is submitted together with the paper.} 

{As ``naive'' benchmarks,} we compare \texttt{TSCI} with three different random forests based methods in the oracle setting, where the best $\mathcal{V}_q$ is known a priori.} %The ``naive'' benchmark methods include RF-Init, RF-Plug and RF-Full. 
In particular, RF-Init denotes the \texttt{TSCI} estimator without bias correction; RF-Full is to implement \texttt{TSCI} without data-splitting; RF-Plug is the estimator of directly plugging the ML fitted treatment in the outcome model, as a naive generalization of \texttt{TSLS}. We give the detailed expressions of these three estimators and their CIs in Section \ref{sec: Pitfalls} in the supplement. %and demonstrate the numerical evidence for the problems of RF-Plug and RF-Full . 

\begin{table}[h]
\centering
\resizebox{\linewidth}{!}{
\begin{tabular}{|c|c|c|cccc|cccc|c|ccc|}
  \hline
\multicolumn{3}{|c|}{} & \multicolumn{4}{c|}{\texttt{TSCI-RF}} & \multicolumn{4}{c|}{Proportions of selection} & \texttt{TSLS} & \multicolumn{3}{c|}{Other RF(oracle)} \\
  \hline
vio & a & n & Oracle & Comp & Robust & Invalidity & $\widehat{q}_c=0$ & $\widehat{q}_c=1$ & $\widehat{q}_c=2$ & $\widehat{q}_c=3$ &  & Init & Plug & Full \\ 
  \hline
  \multirow{9}{*}{1} & \multirow{3}{*}{0.0} & 1000 & 0.91 & 0.01 & 0.01 & 0.01 & 0.99 & 0.01 & 0.00 & 0.00 & 0.00 & 0.80 & 0.38 & 0.01 \\ 
  &  & 3000 & 0.92 & 0.92 & 0.92 & 1.00 & 0.00 & 0.84 & 0.16 & 0.00 & 0.00 & 0.91 & 0.64 & 0.00 \\ 
  &  & 5000 & 0.91 & 0.92 & 0.93 & 1.00 & 0.00 & 0.85 & 0.15 & 0.00 & 0.00 & 0.89 & 0.74 & 0.00 \\ 
  \cline{2-15}
  & \multirow{3}{*}{0.5} & 1000 & 0.91 & 0.23 & 0.25 & 0.24 & 0.76 & 0.24 & 0.01 & 0.00 & 0.00 & 0.84 & 0.56 & 0.02 \\ 
  &  & 3000 & 0.95 & 0.94 & 0.94 & 1.00 & 0.00 & 0.97 & 0.02 & 0.01 & 0.00 & 0.91 & 0.43 & 0.00 \\ 
  &  & 5000 & 0.92 & 0.92 & 0.91 & 1.00 & 0.00 & 0.98 & 0.01 & 0.01 & 0.00 & 0.88 & 0.09 & 0.00 \\ 
  \cline{2-15}
  & \multirow{3}{*}{1.0} & 1000 & 0.96 & 0.92 & 0.92 & 0.95 & 0.05 & 0.93 & 0.01 & 0.00 & 0.00 & 0.91 & 0.52 & 0.08 \\ 
  &  & 3000 & 0.94 & 0.94 & 0.95 & 1.00 & 0.00 & 0.99 & 0.01 & 0.00 & 0.00 & 0.92 & 0.00 & 0.00 \\ 
  &  & 5000 & 0.94 & 0.94 & 0.94 & 1.00 & 0.00 & 0.98 & 0.02 & 0.01 & 0.00 & 0.92 & 0.00 & 0.00 \\ 
  \hline
  % \multirow{9}{*}{2} & \multirow{3}{*}{0.0} & 1000 & 0.92 & 0.01 & 0.01 & 0.00 & 1.00 & 0.00 & 0.00 & 0.00 & 0.00 & 0.82 & 0.30 & 0.01 \\ 
  % &  & 3000 & 0.94 & 0.94 & 0.94 & 1.00 & 0.00 & 0.00 & 1.00 & 0.00 & 0.00 & 0.91 & 0.63 & 0.00 \\ 
  % &  & 5000 & 0.94 & 0.94 & 0.94 & 1.00 & 0.00 & 0.00 & 1.00 & 0.00 & 0.00 & 0.88 & 0.77 & 0.00 \\ 
  % \cline{2-15}
  % & \multirow{3}{*}{0.5} & 1000 & 0.92 & 0.11 & 0.12 & 0.21 & 0.79 & 0.08 & 0.13 & 0.00 & 0.00 & 0.86 & 0.49 & 0.01 \\ 
  % &  & 3000 & 0.94 & 0.93 & 0.92 & 1.00 & 0.00 & 0.00 & 0.97 & 0.03 & 0.00 & 0.90 & 0.37 & 0.00 \\ 
  % &  & 5000 & 0.94 & 0.94 & 0.94 & 1.00 & 0.00 & 0.00 & 0.99 & 0.01 & 0.00 & 0.88 & 0.03 & 0.00 \\ 
  % \cline{2-15}
  % & \multirow{3}{*}{1.0} & 1000 & 0.95 & 0.89 & 0.89 & 0.96 & 0.04 & 0.02 & 0.93 & 0.01 & 0.00 & 0.89 & 0.40 & 0.01 \\ 
  % &  & 3000 & 0.93 & 0.93 & 0.93 & 1.00 & 0.00 & 0.00 & 0.99 & 0.01 & 0.00 & 0.92 & 0.00 & 0.00 \\ 
  % &  & 5000 & 0.93 & 0.93 & 0.93 & 1.00 & 0.00 & 0.00 & 1.00 & 0.00 & 0.00 & 0.90 & 0.00 & 0.00 \\ 
  %  \hline
\end{tabular}}
\caption{\small Coverage for Setting B1 with Vio=1. The columns indexed with ``\texttt{TSCI-RF}" correspond to our proposed \texttt{TSCI} with random forests, where the columns indexed with ``Oracle", ``Comp" and ``Robust" correspond to {the estimators with $\mathcal{V}_q$ selected by the oracle knowledge, the comparison method,} and the robust method. The column indexed with ``Invalidity" reports the proportion of detecting the proposed IV as invalid. The columns indexed with ``Proportions of selection'' reports the proportions of basis sets $\mathcal{V}_q$ for $0\leq q\leq 3$ selected by \texttt{TSCI-RF}. The column indexed with ``\texttt{TSLS}" corresponds to the \texttt{TSLS} estimator. The columns indexed with ``Init", ``Plug", and ``Full" correspond to the RF estimators without bias correction, the plug-in RF estimator and the no data-splitting RF estimator, {with the oracle knowledge of the best $\mathcal{V}_q$}.}
\label{tab: cover and V selctn invalid IV model1}
\end{table}

In Table \ref{tab: cover and V selctn invalid IV model1}, we compare the coverage properties of our proposed \texttt{TSCI}, \texttt{TSLS}, and the above three random forests based methods with Vio=1 in the outcome model.
We observe that \texttt{TSLS} fails to have desired coverage {probability 95\%} due to the existence of invalid IV; furthermore, {RF-Full} and {RF-Plug} have coverages far below 95\% and RF-Init has slightly lower coverage due to the finite-sample bias. In contrast, our proposed \texttt{TSCI} achieves the desired coverage with a relatively strong interaction or large sample size. This is mainly due to its correct selection of $\mathcal{V}_q$ in those settings. In some settings with small interaction and small sample size (like $a=0$, $n=1000$), \texttt{TSCI} fails to select the correct basis set, and the coverage is below 95\%. For the outcome model with Vio=2, \texttt{TSCI} can have desired coverage as well with a relatively strong interaction or large sample size; see Table \ref{tab: Coverage Model 1 vio2} in the supplement.

%We highlight two interesting observations. Firstly, for smaller interaction strengths and sample sizes, $Q_{\max}$ is taken as $0$ due to small IV strengths after adjusting any basis set. This forces \texttt{TSCI} to make the selection $\widehat{q}_c=0$ and hence the desired coverage is not achieved. However, the IV validity test still reports invalid IV in some challenging settings (like $a=0.5, n=1000$). Secondly, {RF-Init} suffers from a larger bias than \texttt{TSCI} as it does not have bias correction, which leads to its undercoverage in all settings.
For Setting B1, we further report the absolute bias and CI length in Tables \ref{tab: bias and V selctn invalid IV model1} and \ref{tab: length and V selctn invalid IV model1} in the supplement, respectively. In Section \ref{sec: additional invalid IV setting}, we consider a binary IV setting, denoted as B3, and observe that Settings B2 and B3 exhibit a similar pattern to those of B1. 
Setting B2 is easier than B1 in the sense that the generalized IV strength remains relatively large even after adjusting for quadratic violation forms in the basis set $\mathcal{V}_2$. %We report the empirical coverage of Settings B2 Table \ref{tab: Coverage Model 2} in the supplement. We also consider a setting with binary IV denoted as Setting B3, in which the observations are generally similar to those in Setting B1; see Table \ref{tab: binary IV model 3} for details. 

%%%%%%%%%%%%%%%%%%%%%%%%%%%%%%%%%%%%%%%%%%%%%%%%%%%%%%%%%%%%
\subsection{\texttt{TSCI} with various nonlinearity levels}
\label{sec: different nonlinearity sim}
%%%%%%%%%%%%%%%%%%%%%%%%%%%%%%%%%%%%%%%%%%%%%%%%%%%%%%%%%%%%
%\Mengchu{rewritten the setting description. need to make a choice. }
%In the last two sections, we have shown that \texttt{TSCI} has valid and efficient inference when the treatment model has more significant nonlinearity than the function $g(z,x)$ in the outcome model. 
In the following, we explore the performance of our proposed \texttt{TSCI} when this identification condition (Condition \ref{cond: identification}) fails to hold, that is, the conditional mean model $f(z,x)$ is not more complex than $g(z,x).$ To approximate such a regime, we consider the outcome model $Y_i=D_i/2+Z_i+\sum_{i=1}^{p_x}X_{i,j}^2-\cos(\pi H_i/4)+e_{i,2}$ and the treatment model $D_i=f(Z_i,X_i)-H_i/2+e_{i,1}$ with different specification of $f$ detailed in Table \ref{tab:settings nonlinearity} where $e_{i,1}$,$e_{i,2}$ are independent random noises following the standard normal distribution. In such a generating model, when $a$ gets close to zero, $f(z,x)$ becomes a linear function in $z$, and Condition \ref{cond: identification} fails since $g(z,x)$ is also linear in $z.$

\begin{table}[h]
    \centering
    \resizebox{\linewidth}{!}{
    \renewcommand{\arraystretch}{1.2}
    \begin{tabular}{|c|c|c|}
        \hline
        Settings & Distribution $Z_i$ and $X_i$ & Treatment model \\
        \hline
        C1 & Same as Setting B1 in Section \ref{sec: invalid IV sim} & \multirow{2}{*}{$f(Z_i,X_i)=Z_i+1/2\cdot Z_i\cdot(a\cdot\sum_{i=1}^{p_x}X_{i,j})-25/12 - 3/10\cdot\sum_{i=1}^{p_x}X_{i,j}$} \\
        \cline{1-2}
        C2 & Same as Setting S3 in Section \ref{sec: dml valid IV sim} & \\ 
        \hline
        C3 & Same as Setting B1 in Section \ref{sec: invalid IV sim} & $f(Z_i,X_i)=Z_i+a\cdot(\sin(2\pi Z_i)+3/2\cdot\cos(2\pi Z_i) - 3/10\cdot\sum_{i=1}^{p_x}X_{i,j}$ \\
        \hline
    \end{tabular}
    \renewcommand{\arraystretch}{1}}
    \caption{\small Different simulation settings, where $f(z,x)$ becomes a linear function of $z$ with $a\rightarrow 0.$} 
    %Three settings with different nonlinearity levels used in Section \ref{sec: different nonlinearity sim}. \Mengchu{move the description into the caption.}
    \label{tab:settings nonlinearity}
\end{table}

We fix $n=3000$ and $p_x=5$, generate $Z_i$ and $X_i$ as detailed in Table \ref{tab:settings nonlinearity}, and generate $H_i$ as $H_i\mid X_{i}\sim N(2\cdot\sin(X_{i,1}),1).$ We implement \texttt{TSCI} detailed in Algorithm \ref{algo: TSCI selection} by specifying $\{\mathcal{V}_q\}_{0\leq q\leq 3}$ as $\mathcal{V}_q=\{z, \dots, z^q,\overrightarrow{\bf b}(x)\}$ with $\overrightarrow{\bf b}(x)=\{1,\overrightarrow{\bf b}_1(x_1),\cdots,\overrightarrow{\bf b}_{p_x}(x_{p_x})\}$, where $\overrightarrow{\bf b}_j(x_j)$ denotes the B-spline basis functions defined in Section \ref{sec: dml valid IV sim}. 

In Figure \ref{fig:diff_nonlinear_levels}, we compare \texttt{TSCI} to \texttt{DML} and \texttt{TSLS} when treatment conditional mean models change from linear ($a=0$) to nonlinear ($a>0$); {Note that in addition, a larger value of $a$ increases the generalized IV strength.}
%\Peter{But isn't a large value of $a$ also increasing IV strength? Just an internal queestion...} \Zijian{Yes, the IV strength increases with $a$. The key point here is that we use $a$ to control nonlinear dependence on $Z_i$ in the $f(Z_i,X_i)$ and a larger value of $a$ means that the more nonlinear of $f$ and hence our identification condition 1 is more likely to hold.} \Peter{NEW: Yes, I understand -- see my additional small half sentence above,}
We vary the value of $a$ in $\{0,0.1,0.2,\cdots,1.2\}$ where larger values of $a$ represent more nonlinear dependence between $D_i$ and $Z_i$. When $a$ is 0, Condition \ref{cond: identification} fails to hold, and hence \texttt{TSCI} cannot identify the treatment effect when there are invalid IVs. In such settings, \texttt{TSCI} has similar performance to \texttt{DML} and \texttt{TSLS}. 
However, when $a$ gets slightly larger than $0$, \texttt{TSCI} has a better RMSE than \texttt{DML} and \texttt{TSLS}, especially for settings C2 and C3. When $a$ is sufficiently large (e.g., $a\geq 0.1$ in Setting C2), \texttt{TSCI} detects the invalid IV and achieves the desired coverage. However, \texttt{DML} and \texttt{TSLS} do not have coverage since they are designed for valid IV settings.  

\begin{figure}[htb!]
    \centering
    \includegraphics[width=0.95\linewidth]{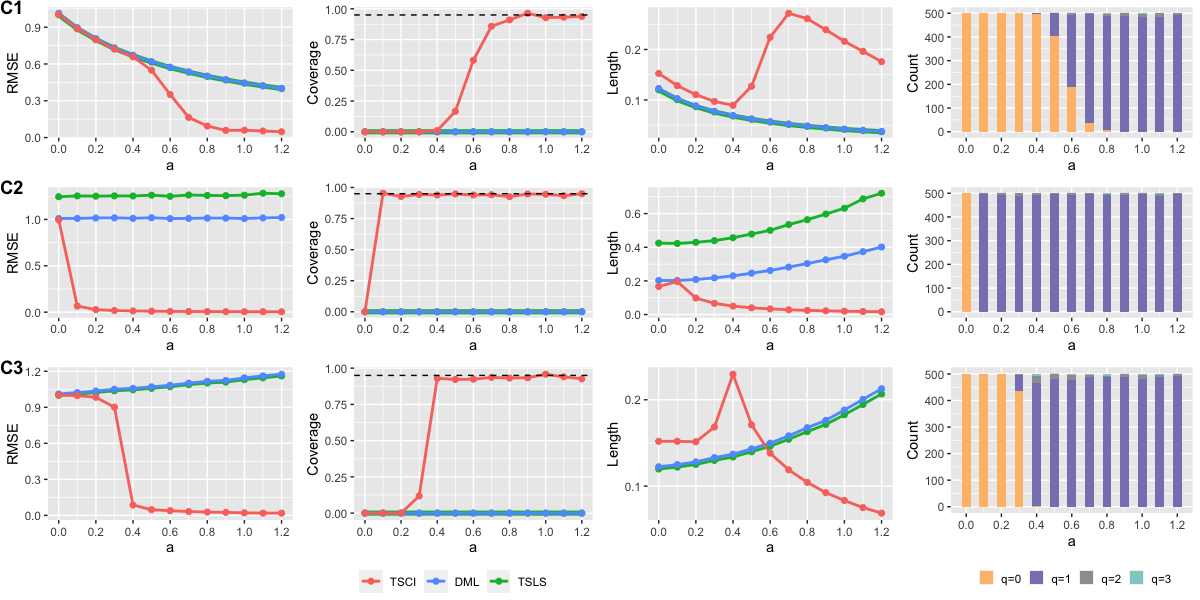}
    \caption{\small Comparison of \texttt{TSCI}, \texttt{DML} and \texttt{TSLS} in terms of RMSE, CI coverage, and length in Settings C1, C2 and C3, where $a$ controls the nonlinearity level in the treatment model. The stacked bar charts show the basis selection of \texttt{TSCI}, where {$q = 1$ corresponds to the correct selection.}} %``\texttt{TSCI}'' is our proposal and ``\texttt{DML}'' is the method proposed in \citet{chernozhukov2018double}. }
    \label{fig:diff_nonlinear_levels}
\end{figure}

\vspace{-2pt}
%\Mengchu{revised the basis description. }

We shall point out that, in Setting C2, the confidence intervals by TSCI are shorter than that of DML, even after adjusting for the linear invalid IV forms. This happens since the remaining nonlinear IV strength after adjusting for the invalidity form is even larger than the IV strength due to the linear association.
\section{Real data {analysis}}
\label{sec: real} 
%%%%%%%%%%%%%%%%%%%%%%%%%%%%%%%%%%%%%%%
We revisit the question on the effect of education on income \citep{card1993using}. We follow \citet{card1993using} and analyze the same data set from the National Longitudinal Survey of Young Men. The outcome is the log wages (\texttt{lwage}), and the treatment is the years of schooling (\texttt{educ}). As argued in \citet{card1993using}, there are various reasons that the treatment is endogenous. For example, the unobserved confounder ``ability bias" may affect both the schooling years and wages, leading to the \texttt{OLS} estimator having a positive bias. Following \citet{card1993using}, we use the indicator for a nearby 4-year college in 1966 {(\texttt{nearc4})} as the IV and include the following covariates: a quadratic function of potential experience {(\texttt{exper} and \texttt{expersq})}, a race indicator {(\texttt{black})}, and dummy variables for residence in a standard metropolitan statistical area in 1976 {(\texttt{smsa})} and in 1966 {(\texttt{smsa66})}, and the dummy variable for residence in the south in 1976 {(\texttt{south})}, {and a full set of 8 regional dummy variables}(\texttt{reg1} to \texttt{reg8}). The data set consists of $n=3010$ observations, which is made available by the R package \texttt{ivmodel}
\citep{kang2021ivmodel}. 

We implement random forests for the treatment model and report the variable importance in Figure \ref{fig: var_importance} in the supplement, where the IV \texttt{nearc4} ranks the seventh in terms of variable importance, after the covariates \texttt{exper,expersq}, \texttt{black,south,smsa,smsa66.}   

We allow for different IV violation forms by approximating $g(z,x)$ with various basis functions $\{\mathcal{V}_q\}_{q=0,1,2}$ detailed in Table \ref{tab:basis real data}. Particularly, since $\mathcal{V}_0$ does not involve the IV \texttt{nearc4}, $\mathcal{V}_0$ corresponds to the valid IV setting as assumed in the analysis of \cite{card1993using}; moreover, $\mathcal{V}_1$ and $\mathcal{V}_2$ correspond to invalid IV settings by allowing the IV \texttt{nearc4} to affect the outcome directly or interactively with the baseline covariates. The main difference is that $\mathcal{V}_1$ includes the interaction with the six most important covariates while $\mathcal{V}_2$ includes all fourteen covariates.  
%Since $\mathcal{V}_0$ assumes that \texttt{nearc4} is a valid IV, the \texttt{TSCI} estimator based on $\mathcal{V}_0$ \citet{card1993using} but we are using a nonlinear treatment model. For $\mathcal{V}_1$ and $\mathcal{V}_2$, we construct them by interacting the IV with the covariates which are grouped into the six most important one and the remaining.  We abuse the notation $\mathcal{V}_q$ here to denote both the basis functions and their evaluations at the observed variables. 

\begin{table}[h]
    \centering
    \renewcommand\arraystretch{1.5}
    \resizebox{\linewidth}{!}{
    \begin{tabular}{|c|c|}
        % \hline
       %  & Basis functions for $g(\cdot)$ function \\
        \hline
        $\mathcal{V}_0$ & \makecell{$\{\texttt{exper,expersq,black,south,smsa,smsa66,} \texttt{reg1,reg2,reg3,reg4,reg5,reg6,reg7,reg8}\}$} \\
        \hline
        $\mathcal{V}_1$ & \makecell{$\mathcal{V}_0\cup\texttt{nearc4}\cdot\{1, \texttt{exper}, \texttt{expersq}, \texttt{black},\texttt{south}, \texttt{smsa}, \texttt{smsa66}\}$} \\
        \hline
        $\mathcal{V}_2$ & \makecell{$\mathcal{V}_0\cup\texttt{nearc4}\cdot\{1, \texttt{exper}, \texttt{expersq}, \texttt{black},\texttt{south}, \texttt{smsa}, \texttt{smsa66}, \texttt{reg1}, \texttt{reg2}, \texttt{reg3}, \texttt{reg4},\texttt{reg5}, \texttt{reg6}, \texttt{reg7}, \texttt{reg8}\}$} \\
        \hline
    \end{tabular}}
    \caption{\small Definitions of $\mathcal{V}_0, \mathcal{V}_1$ and  $\mathcal{V}_2$, which are used to approximate $g(\cdot)$.} %and required as inputs for Algorithm \ref{algo: TSCI selection}.} %$\mathcal{V}_0$ includes all baseline covariates; $\mathcal{V}_1$ includes $\mathcal{V}_0$ and the interaction of the IV \texttt{nearc4} and six most important covariates in the first stage. For $\mathcal{V}_2$, we further include the interactions of the IV with the remaining covariates which are the full set of regional dummies. }
    \label{tab:basis real data}
\end{table}

%After adjusting for $g(z,x)$ spanned by $\mathcal{V}_1$ and $\mathcal{V}_2$, our proposed \texttt{TSCI} method is robust even if the IV \texttt{nearc4} affects the outcome directly or through the interaction with baseline covariates. We report the point estimates and IV strength in Figure \ref{fig: hist}. 

We implement Algorithm \ref{algo: TSCI selection} to choose the best from $\{\mathcal{V}_0, \mathcal{V}_1,\mathcal{V}_2\},$ and construct the \texttt{TSCI} estimator. Since the \texttt{TSCI} estimates depend on the specific sample splitting, we report 500 \texttt{TSCI} estimates due to 500 different splittings.
On the leftmost panel of Figure \ref{fig: hist}, we compare estimates of \texttt{OLS}, \texttt{TSLS}, and \texttt{TSCI}, which uses $\mathcal{V}_{\widehat{q}_c}$ reported from Algorithm \ref{algo: TSCI selection}. The median of these 500 \texttt{TSCI} estimators is $0.0604$, 87.2\% of these 500 estimates are smaller than the \texttt{OLS} estimate (0.0747), and 100\% of \texttt{TSCI} estimates are smaller than the \texttt{TSLS} estimate (0.1315). In contrast to the \texttt{TSLS} estimator, the \texttt{TSCI} estimators tend to be smaller than the \texttt{OLS} estimator, which helps correct the positive ``ability bias". }

\begin{figure}[htp!]
\centering
\includegraphics[width=\linewidth]{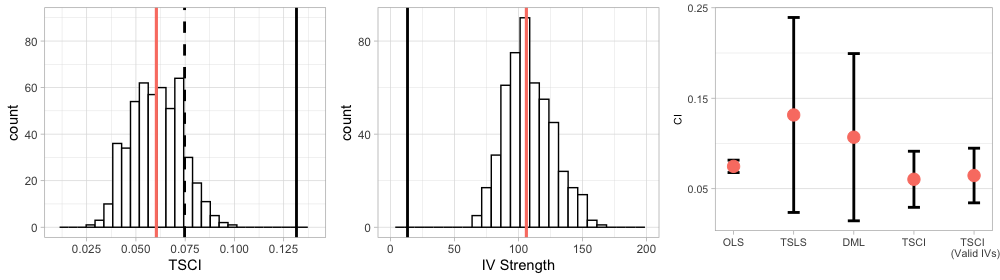}
\caption{\small The leftmost panel reports the histograms of the \texttt{TSCI} (random forests) estimates with the comparison method, where the estimates {differ}
due to the randomness of {different 500 realized} sample splittings; the solid red line corresponds to the median of {the} \texttt{TSCI} estimates, while the solid and dashed black lines correspond to the \texttt{TSLS} and \texttt{OLS} estimates, respectively. The middle {panel displays} 
the histogram of the generalized IV strength (after adjusting for $\mathcal{V}_{\widehat{q}_c}$) {over the different 500 realized sample splittings}; the solid red line denotes the median of all IV strength for \texttt{TSCI} while the solid black line denotes the IV strength of \texttt{TSLS}. The rightmost panel compares different confidence intervals (CIs) produced by \texttt{OLS}, \texttt{TSLS}, \texttt{DML}, and our proposed \texttt{TSCI} with $\mathcal{V}_{\widehat{q}_c}$, and \texttt{TSCI} assuming valid IVs. The CIs of TSCI are adjusted by the multi-splitting method due to finite samples; see Appendix \ref{sec:finite-sample adjmt}. }
\label{fig: hist}
\end{figure}

 On the rightmost panel of Figure \ref{fig: hist}, we compare different confidence intervals (CIs). The \texttt{TSLS} CI is $(0.0238,0.2393)$, and the \texttt{DML} CI is $(0.0061,0.1907)$, which are both much wider than the CIs by \texttt{TSCI}. This wide interval results from the relatively weak IV because \texttt{TSLS} and \texttt{DML} only consider the linear association between IV and the treatment. %We can refer the weak IV problem to the middle panel in Figure \ref{fig: hist}. 
The CI by \texttt{TSCI} with random forests varies with the specific sample splitting. We follow \citet{meinshausen2009p} and implement the multi-split method to adjust the finite-sample variation due to sample splitting; see Section \ref{sec:finite-sample adjmt} in the supplement. The multi-split CI is (0.0294, 0.0914). The CIs based on the \texttt{TSCI} estimator with random forests are pushed to the lower part of the wide CI by \texttt{TSLS}.  The CIs by our proposed \texttt{TSCI} in Algorithm \ref{algo: TSCI selection} tend to shift down in comparison to the \texttt{TSCI} assuming valid IVs.

We shall point out that the IV strengths for the \texttt{TSCI} estimators, even after adjusting for the selected basis functions, are typically much larger than the \texttt{TSLS} (the concentration parameter is $13.33$), which is illustrated {in} the middle panel of Figure \ref{fig: hist}. {This happens since the first-stage ML captures much more associations than the linear model in \texttt{TSLS}, even after adjusting for the possible IV violations captured by $\mathcal{V}_{\widehat{q}_c}$.

Among 500 sample splittings, the proportion of choosing $\mathcal{V}_{0}, \mathcal{V}_{1},$ and $\mathcal{V}_{2}$ are $59.2\%$, $38.2\%$, and $2.6\%$, respectively, where around $41\%$ of splittings report $\texttt{nearc4}$ as invalid IVs.  
Importantly, $93.6\%$ out of the 500 splittings report $Q_{\text{max}}>\widehat{q}_c$, which represents that $\widehat{\beta}_{\mathcal{V}_{\widehat{q}_c}}$ is not statistically different from $\widehat{\beta}_{\mathcal{V}_{Q_{\text{max}}}}$ produced by the largest $\mathcal{V}_{Q_{\text{max}}}$. This indicates that $\mathcal{V}_{\widehat{q}_c}$ already provides a reasonably good approximation of $g(Z_i,X_i)$ in the outcome model. In Section \ref{sec: add real} of the supplement, we consider alternative choices of $\mathcal{V}_2$ and observe that the results are consistent with that reported above. In Section \ref{sec: fal}, we further propose a falsification argument regarding Condition \ref{cond: identification} for the data analysis.

\section{Conclusion and discussion}
%%%%%%%%%%%%%%%%%%%%%%%%%%%%%%%%%%%%%%%%%%%%%%
We integrate modern ML {algorithms} into the {framework of} instrumental variable analysis. 
{We devise a} novel \texttt{TSCI} methodology which provides 
reliable causal conclusions even with {a wide range of violation forms}. Our proposed generalized IV strength {measure} helps {to} understand when our proposed method is reliable and {supports}
%facilitate 
the basis functions' selection in approximating the violation form {of possibly invalid IVs.} The current methodology is focused on inference for a linear and constant {treatment} effect.
{An} interesting {future} research {direction is about} inference for heterogeneous treatment effects \citep{athey2016recursive,wager2018estimation} {but in presence of potentially invalid instruments}.

%%%%%%%%%%%%%%%%%%%%%%%%%%%%%%%%%%%%%
\section*{Acknowledgement}
%%%%%%%%%%%%%%%%%%%%%%%%%%%%%%%%%%%%%
We acknowledge valuable suggestions from Xu Cheng, Tirthankar Dasgupta, Qingliang Fan, Michal Koles\'{a}r, Greg Lewis, Yuan Liao, Molei Liu, Zhonghua Liu, Zuofeng Shang, and Frank Windmeijer.
%The authors acknowledge {feedback from} the participants at the Center for Causal Inference Seminar at Upenn and the Online Causal Inference Seminar. In particular, we thank Tirthankar Dasgupta, Qingliang Fan, Michal Koles\'{a}r, Greg Lewis, Yuan Liao, Molei Liu, Zhonghua Liu, Zuofeng Shang, and Frank Windmeijer for their suggestions about an earlier version of the manuscript. 
Z. Guo gratefully acknowledges financial support from NSF DMS 1811857, 2015373, NIH R01GM140463, and financial support for visiting the Institute of Mathematical Research (FIM) at ETH Zurich. P. B\"{u}hlmann received funding from the European Research Council (ERC) under the European Union’s Horizon 2020 research and innovation program (grant agreement No. 786461)
%\bibliographystyle{chicago}

%\Peter{The Bach et al. reference appeared here\\
%https://www.jmlr.org/papers/volume23/21-0862/21-0862.pdf\\
%Card (1993) has no journal pr preprint name.\\
%Perhaps check whether other arxiv papers have been published by now}

%\Mengchu{Peter, thanks for your comments! The published Bach et al paper is a Python package while the paper needed to cite here is R package and it is not published yet. Other comments are all addressed.} \Peter{Great -- thank you!}

%\Peter{Reference Kang et al. Write "R" instead of "r" in R-package.}

\bibliographystyle{chicago.bst}
\bibliography{IVRef}
\restoregeometry

\newpage

\appendix

\setcounter{page}{1}
\setcounter{table}{0}
\setcounter{figure}{0}

% https://www.overleaf.com/project/5f54333526a4f100013935eb
%\renewcommand{\thesection}{S\arabic{section}}  
\renewcommand{\thetable}{S\arabic{table}}  
\renewcommand{\thefigure}{S\arabic{figure}}
%%%%%%%%%%%%%%%%%%%%%%%%%%%%%%%%%%%%%%%%%%%%%%%%%%%%%%%%%%%%%%%%%%%%%%%%%%%%%%%%%%%%%%%%%%%%%%%%%%%%%%%%%%%%%%%%%%
\section{Additional Methods and Theories}
%%%%%%%%%%%%%%%%%%%%%%%%%%%%%%%%%%%%%%%%%%%%%%%%%%%%%%%%%
%%%%%%%%%%%%%%%%%%%%%%%%%%%%%%%%%%%%%%%%%%%%%%%%%%%%%%%%%
\subsection{Identification in random experiments with non-compliance and invalid IVs}
\label{sec: noncompliance}
In the following, we explain how a nonlinear treatment model helps identify the causal effect in randomized experiments with non-compliance and possibly invalid IVs. For the $i$-th subject, we use $Z_i\in\{0,1\}$ to denote the random treatment assignment (which serves as an instrument) and $D_i\in \{0,1\}$ to denote whether the subject actually takes the treatment or not. We use $X_i\in \{0,1\}$ to denote a binary covariate and  $Y^{(z,d)}_i(x)$ to denote the potential outcome for a subject with the assignment $z$, the treatment value $d$, and the covariate $x$.
%\Peter{Question: why is this invalid? If it is randomized...? Is the idea that it may not be randomized?} \Zijian{The main reason of being invalid is due to having a direct effect on the outcome model.}  

A primary concern is that $Z_i$ may be an invalid IV even if it is randomly assigned.  To facilitate the discussion, we follow the discussion in \citetsupp{imbens2014instrumental} and adopt the example from \citetsupp{mcdonald1992effects} studying the effect of influenza vaccination on being hospitalized for flu-related reasons. The researchers randomly sent letters to physicians encouraging them to vaccinate their patients. In this example, $Z_i$ denotes the indicator for the physician receiving the encouragement letter, while $D_i$ stands for the patient being vaccinated or not. The outcome $Y_i$ denotes whether the patient is hospitalized for a flu-related reason. 
As pointed out in  \citetsupp{imbens2014instrumental}, the physician receiving the letter may ``take actions that affect the likelihood of the patient getting infected with the flu other than simply administering the flu vaccine", indicating {that} $Z_i$ may have a direct effect on the outcome and violate the classical IV assumptions.
%\Peter{So here is the explanation why the instrument may be invalid, correct? It is hard to grasp when you introduce the setting above -- at least to me. Should we write above that the example here says why the instrument can ba invalid?}\Zijian{I have implemented the suggestion.} 

For a covariate $x$, we define the intention-to-treat (ITT) on the outcome as ${\rm ITT}_{Y}(x)=\E(Y_i\mid Z_i=1, X_i=x)-\E(Y_i\mid Z_i=0, X_i=x)$ and the ITT on the treatment as ${\rm ITT}_{D}(x)=\E(D_i\mid Z_i=1, X_i=x)-\E(D_i\mid Z_i=0, X_i=x).$ %\Peter{Twice the same quantities... should it be a potential outcome?}\Zijian{I have fixed the typo and please double check.} 
In the following proposition, we assume the constant effect and demonstrate the treatment effect identification even if the IV $Z_i$ directly affects the outcome (violating assumption (A2)). 

\begin{Proposition}
Suppose that {\rm (I1)} ${\rm ITT}_{D}(x=1)-{\rm ITT}_{D}(x=0)\neq 0$, {\rm (I2)} $Z_i$ is randomly assigned, {\rm (I3)} $Y^{(z',d)}_i(x)-Y^{(z,d)}_i(x)=(z'-z)\pi$ for $x\in \{0,1\}$, and {\rm (I4)} $Y^{(z,d')}_i(x)-Y^{(z,d)}_i(x)=(d'-d)\beta.$   Then we identify $\beta$ as
$
\frac{{\rm ITT}_{Y}(x=1)-{\rm ITT}_{Y}(x=0)}{{\rm ITT}_{D}(x=1)-{\rm ITT}_{D}(x=0)}.
$  
\label{prop: special identification}
\end{Proposition}
The proof of the above proposition is presented in Section \ref{sec: identification}. 
Condition (I4) assumes a constant effect while (I1)-(I3) are relaxations of the classical IV assumptions (A1)-(A3), respectively. Particularly, the random assignment required by (I2) implies (A2) and (I2) is satisfied in randomized experiments with non-compliance. 
{Condition (I3) allows the IV to affect the outcome directly but assumes that the IV's direct effect does not interact with the baseline covariate $X_i$. When there exists a direct effect, the identification of $\beta$ relies on the 
Condition (I1). Condition (I1) essentially requires that the treatment assignment is interactively determined by $Z_i$ and $X_i$: {a special example is given by} 
%with a special example 
$D_i={\bf 1}\left(\gamma_0+\gamma_z Z_i+\gamma_{x} X_i+\gamma Z_i\cdot X_i+ \delta_i>0\right)$ where $\delta_i$ denotes a random variable independent of $Z_i$ and $X_i.$ Such identification conditions have been proposed in \cite{shardell2016instrumental,ten2007causal}. Particularly, \cite{shardell2016instrumental} has pointed out that ${\rm ITT}_{D}(x=1)-{\rm ITT}_{D}(x=0)$ is referred to as the compliance score and Condition (I1) requires that the compliance score changes with the covariate value $x$.} 

{We provide {now} some discussions on the plausibility of Condition (I1).} We still use the vaccination example from \citetsupp{mcdonald1992effects} and imagine that we may have access to the covariate $X_i$, an indicator for whether the patients took regular flu shots in the past few years. Patients who regularly take flu shots (with $X_i=1$) are more likely to follow the physician's encouragement than those who do not (with $X_i=0$), indicating ${\rm ITT}_{D}(x=1)-{\rm ITT}_{D}(x=0)\neq 0$. However, regarding the concern ``take actions that affect the likelihood of the patient getting infected with the flu other than simply administering the flu vaccine" \citepsupp{imbens2014instrumental}, such a direct effect of $Z_i$ might not interact with $X_i$.

\subsection{Pitfalls with the naive machine learning}
\label{sec: Pitfalls}

%We give the detailed equations for point estimators and standard errors for $\widehat{\beta}_{\text{plug}}$ and $\widehat{\beta}_{\text{full}}$. 
As the benchmarks in Section \ref{sec: invalid IV sim}, we also include the point estimators $\widehat\beta_{\text{init}}$, $\widehat{\beta}_{\text{plug}},$ and $\widehat{\beta}_{\text{full}}$ together with their corresponding standard errors. We construct the 95\% confidence interval as $(\widehat{\beta}-1.96*\widehat{\rm SE},\widehat{\beta}+1.96*\widehat{\rm SE}).$

\noindent {\bf RF-Init.} We compute $\widehat{\beta}$ as $\widehat{\beta}_{\rm init}$ in \eqref{eq: RF init} and $\widehat{\sigma}_{\epsilon}(V)=\sqrt{\|P^{\perp}_{V}[Y-D\widehat{\beta}_{\rm init}(V)]\|_2^2/(n-r)}.$
Calculate 
$\widehat{\rm SE}={\widehat{\sigma}_{\epsilon}\sqrt{{D}_{\mathcal{A}_1}^{\intercal}[\T(V)]^2 {D}_{\mathcal{A}_1}}}/{{D}_{\mathcal{A}_1}^{\intercal}\T(V) {D}_{\mathcal{A}_1}}.$ 

\vspace{3mm}

\noindent {\bf RF-Plug and its pitfall.} {A simple idea of combining \texttt{TSLS} with ML {is to directly use $\widehat{f}_{\mathcal{A}_1}$ in the second stage}. We may calculate such a two-stage estimator as
$
\widehat{\beta}_{\rm plug}(V)={{Y}_{\mathcal{A}_1}^{\intercal}P^{\perp}_{{V}_{\mathcal{A}_1}} \widehat{f}_{\mathcal{A}_1}}/{\widehat{f}_{\mathcal{A}_1}^{\intercal}P^{\perp}_{{V}_{\mathcal{A}_1}} \widehat{f}_{\mathcal{A}_1}},
$
where the second-stage regression is implemented by regressing $Y_{\mathcal{A}_1}$ on $\widehat{f}_{\mathcal{A}_1}=\Omega D_{\mathcal{A}_1}$ and  $V_{\mathcal{A}_1}$. We compute the standard error $\widehat{\rm SE}=\sqrt{\frac{\|P^{\perp}_{V}(Y-\widehat{\beta}_{\rm plug}(V)D)\|_2^2}{|\mathcal{A}_1|\cdot {\widehat{D}_{\mathcal{A}_1}^{\intercal}P^{\perp}_{{V}_{\mathcal{A}_1}}}}}.$  \citetsupp{angrist2009mostly}[Sec.4.6.1] criticized such use of the nonlinear first-stage model under the name of ``forbidden regression'' and \citetsupp{chen2020mostly} pointed out that $\widehat{\beta}_{\rm plug}(V)$ is biased since $\Omega$ in \eqref{eq: general trans matrix} is not a projection matrix for random forests. }

\vspace{3mm}

\noindent {\bf RF-Full and its pitfall.} 
Sample splitting is essential for removing the endogeneity of the ML-predicted values.  
Without sample splitting, the ML predicted value for the treatment {can be close to the treatment (due to overfitting) and hence highly correlated with unmeasured confounders}, leading to a \texttt{TSCI} estimator suffering from a significant bias. 
{{In the following, we}
%We 
take the non-split Random Forests as an example.} Similarly to $\Omega$ defined in \eqref{eq: general trans matrix}, we define the transformation matrix $\Omega^{\rm full}$ for random forests constructed with the full data. 
As a modification of \eqref{eq: RF init}, we consider
the corresponding \texttt{TSCI} estimator,
\begin{equation}
\widehat{\beta}_{\rm full}(V)={Y^{\intercal}\T^{\rm full}(V) {D}}/{{D}^{\intercal}\T^{\rm full}(V) {D}} \quad \text{with}\quad \T^{\rm full}_{\rm RF}(V)=(\Omega^{\rm full})^{\intercal}P^{\perp}_{\widetilde{V}} \Omega^{\rm full},
\label{eq: full RF}
\end{equation}
where $\widetilde{V}=\Omega^{\rm full} V$. We calculate its standard error as 
$$\widehat{\rm SE}=\frac{\widehat{\sigma}_{\rm full}(V)\sqrt{ {{D}^{\intercal}[\T^{\rm full}(V)]^2 {D}}}}{{D}^{\intercal}\T^{\rm full}(V) {D}}\quad \text{with}\quad \widehat{\sigma}_{\rm full}(V)=\sqrt{\|P^{\perp}_{V}(Y-\widehat{\beta}_{\rm RF}^{\rm full}(V)D)\|_2^2/{n}}.$$ The simulation results in Section \ref{sec: sim} show that the estimator $\widehat{\beta}_{\rm full}(V)$ in \eqref{eq: full RF} suffers from a large bias and the resulting confidence interval does not achieve the desired coverage. %This happens since the endogeneity of the treatment is not completely removed: the predicted values $\Omega^{\rm full}D$ might be highly correlated with the errors $\delta$ due to the overfitting nature of the ML algorithm.

\subsection{Choice of $\widehat{\rho}$ in \eqref{eq: lawyer test RF}}
\label{eq: sample rho}
{In the following, we present {the choice of}
%choosing 
$\widehat{\rho}=\widehat{\rho}(\alpha_0)>0$ used in \eqref{eq: lawyer test RF} by the wild bootstrap.} For $1\leq i\leq n_1,$ we compute $\widetilde{\epsilon}_i=[\widehat{\epsilon}(V_{Q_{\max}})]_i-\bar{\mu}_{\epsilon}$ where $\widehat{\epsilon}(V_{Q_{\max}})$ is defined in \eqref{eq: TSRF SE hetero} with $V=V_{Q_{\max}}$ and  $\bar{\mu}_{\epsilon}=\frac{1}{n_1}\sum_{i=1}^{n_1}[\widehat{\epsilon}(V_{Q_{\max}})]_i$. We generate $\epsilon^{[l]}_i=U^{[l]}_i\cdot \widetilde{\epsilon}_i$ for $1\leq i\leq n_1$ and $1\leq l\leq  L,$ where $\{U^{[l]}_i\}_{1\leq i\leq n_1}$ are i.i.d. standard normal random variables. We define \begin{equation*}
T^{[l]}=\max_{0\leq q<q'\leq Q_{\max}}\frac{1}{{\sqrt{\widehat{H}(V_{q},V_{q'})}}}\left[\frac{f_{\mathcal{A}_1}^{\intercal} \T(V_{q'}) \epsilon^{[l]}}{{f}_{\mathcal{A}_1}^{\intercal} \T(V_{q'}) f_{\mathcal{A}_1}}-\frac{f_{\mathcal{A}_1}^{\intercal} \T(V_{q}) \epsilon^{[l]}}{f_{\mathcal{A}_1}^{\intercal} \T(V_{q}) f_{\mathcal{A}_1}}\right]\; \text{for}\; 1\leq l\leq L.
\label{eq: bootstrap sample}
\end{equation*}
We set $\widehat{\rho}=\widehat{\rho}(\alpha_0)$ to be the upper $\alpha_0$ empirical quantile of $\{|T^{[l]}|\}_{1\leq l\leq L}$, that is, 
\begin{equation}
\widehat{\rho}(\alpha_0)=\min\left\{\rho\in \R: \frac{1}{L}\sum_{l=1}^{L} {\bf 1}\left(|{T^{[l]}}|\leq \rho\right)\geq 1-\alpha_0 \right\},
\label{eq: sample quantile}
\end{equation}
{where $\alpha_0$ is set as $0.025$ by default.}
\subsection{Finite-sample adjustment of uncertainty from data splitting}
\label{sec:finite-sample adjmt}
Even though our asymptotic theory in Section \ref{sec: theory} is valid for any random sample splitting, the constructed point estimators and confidence intervals do vary with different sample splittings in finite samples. This randomness due to sample splittings has been {also} observed in 
%the 
double machine learning  \citepsupp{chernozhukov2018double} and multi-splitting \citepsupp{meinshausen2009p}. Following \citetsupp{chernozhukov2018double} and \citetsupp{meinshausen2009p}, we introduce a confidence interval 
%by adjusting the additional uncertainty due to the data splitting. The proposed interval 
{which} aggregates multiple confidence intervals due to different splittings. Consider $S$ random splittings and for the $s$-th splitting, we use $\widehat{\beta}^{s}$ and $\widehat{\rm SE}^{s}$ to denote 
%all 
the {corresponding} \texttt{TSCI} point and standard error estimators, respectively. Following Section 3.4 of \citetsupp{chernozhukov2018double}, we introduce the median estimator $\widehat{\beta}^{\rm med}={\rm median}\{\widehat{\beta}^{s}\}_{1\leq s\leq S}$ together with 
$\widehat{\rm SE}^{\rm med}={\rm median}\left\{\sqrt{[\widehat{\rm SE}^{s}]^2+(\widehat{\beta}^{s}-\widehat{\beta}^{\rm med})^2}\right\}_{1\leq s\leq S},$
and construct the median confidence interval as $(\widehat{\beta}^{\rm med}-z_{\alpha/2}\widehat{\rm SE}^{\rm med},\widehat{\beta}^{\rm med}+z_{\alpha/2}\widehat{\rm SE}^{\rm med}).$ Alternatively, following  \citetsupp{meinshausen2009p}, we construct the p values $p^s(\beta_0)=2(1-\Phi(|\widehat{\beta}^{s}-\beta_0|/\widehat{\rm SE}^{s}))$ for $\beta_0\in \R$ and $1\leq s\leq S,$ where $\Phi$ is the CDF of the standard normal. We define the multi-splitting confidence interval as 
$\{\beta_0\in \R: 2\cdot {\rm median}\{p^s(\beta_0)\}_{s=1}^{S}\geq \alpha\}.$ See equation (2.2) of \citetsupp{meinshausen2009p}. %for more details.

\subsection{Choices of Violation Basis Functions for Multiple IVs}
When there are multiple IVs, there are more choices to specify the violation form, and $\{\mathcal{V}_q\}_{1\leq q\leq Q}$ are not necessarily nested. For example, when we have two IVs $z_{1}$ and $z_{2}$, we may set $\mathcal{V}_{q_1,q_2}= \{1,z_1,\cdots,z_{1}^{q_1},z_2,\cdots,z_2^{q_2}\},$ and then $\{\mathcal{V}_{q_1,q_2}\}_{0\leq q_1,q_2\leq Q}$ are not nested. However, even in such a case, our proposed selection method is still applicable. Specifically, for any $0\leq q_1,q_2\leq Q$, we compare the estimator generated by $\mathcal{V}_{q_1,q_2}$ to that by $\mathcal{V}_{q'_1,q'_2}$ with $q'_1\geq q_1$ and $q'_2\geq q_2.$ 
%%%%%%%%%%%%%%%%%%%%%%%%%%%%
\subsection{\texttt{TSCI} with boosting}
\label{sec: TS-boosting}
%%%%%%%%%%%%%%%%%%%%%%%%%%%%
%%%%%%%%%%%%%%%%%%%%%%%%%%%%
%\Zijian{Move to the supplement?} \Peter{Yes. And if there is not enough space, perhaps even move the entire Section 4 to the supplement?}
In the following, we demonstrate how to express the $L_2$ boosting estimator  \citepsupp{buhlmann2007boosting,buhlmann2006sparse} in the form of \eqref{eq: general trans matrix}. The boosting methods aggregate a sequence of base procedures  $\{\widehat{g}^{[m]}(\cdot)\}_{m\geq 1}.$ 
For $m\geq 1,$ we construct the base procedure $\widehat{g}^{[m]}$ using the data in $\mathcal{A}_2$ and compute the estimated values given by the $m$-th base procedure $\widehat{g}^{[m]}_{\mathcal{A}_1}=\left(\widehat{g}^{[m]}(X_1,Z_1),\cdots, \widehat{g}^{[m]}(X_{n_1},Z_{n_1})\right)^{\intercal}$. With $\widehat{f}^{[0]}_{\mathcal{A}_1}=0$ and $0<\nu\leq 1$ as the step-length factor {(the default being $\nu = 0.1$)}, we conduct the sequential updates, 
\begin{equation*}
\widehat{f}^{[m]}_{\mathcal{A}_1}=\widehat{f}^{[m-1]}_{\mathcal{A}_1}+\nu \widehat{g}^{[m]}_{\mathcal{A}_1}\quad \text{with}\quad \widehat{g}^{[m]}_{\mathcal{A}_1}=\mathcal{H}^{[m]} (D_{\mathcal{A}_1}-\widehat{f}^{[m-1]}_{\mathcal{A}_1}) \quad \text{for}\quad m\geq 1,
\end{equation*}
where $\mathcal{H}^{[m]}\in \R^{n_1\times n_1}$ is a hat matrix determined by the base procedures. 

%In Section \ref{sec: Omega boosting} in the supplement, we give the exact expression of $\mathcal{H}^{[m]}$ for three different base procedures, including the pairwise regression, the pairwise thin plate, and the decision tree. 

With the stopping time $m_{\rm stop}$, the $L_2$ boosting estimator is $\widehat{f}^{\rm boo}=\widehat{f}^{[m_{\rm stop}]}.$ We now compute the transformation matrix $\Omega$. Set $\Omega^{[0]}=0$ and for $m\geq 1,$ we define 
\begin{equation}
\widehat{f}^{[m]}_{\mathcal{A}_1}=\Omega^{[m]} D_{\mathcal{A}_1} \quad \text{with}\quad \Omega^{[m]}=\nu \mathcal{H}^{[m]}+ ({\rm I}-\nu\mathcal{H}^{[m]})\Omega^{[m-1]}.
\label{eq: Omega def boosting} \end{equation}
Define $\Omega^{\rm boo}=\Omega^{[m_{\rm stop}]}$ and write $\widehat{f}^{\rm boo}=\Omega^{\rm boo} D_{\mathcal{A}_1},$ which is the desired form in \eqref{eq: general trans matrix}. 

In the following, we give two examples of the construction of the base procedures and provide the detailed expression for $\mathcal{H}^{[m]}$ used in \eqref{eq: Omega def boosting}.  We write $C_i=(X_i^{\intercal},Z_i^{\intercal})^{\intercal} \in \R^{p}$ and define the matrix $C\in \R^{n\times p}$ with its $i$-th row as $C_i^{\intercal}.$ An important step of building the base procedures $\{\widehat{g}^{[m]}(Z_i, X_i)\}_{m\geq 1}$ is to conduct the variable selection. 
 That is, each base procedure is only constructed based on a subset of covariates $C_i=(X_i^{\intercal},Z_i^{\intercal}) \in \R^{p}.$ 

%In Algorithm \ref{algo: Boosting 1}, we describe the construction of $\Omega^{\rm boo}$ with the componentwise linear regression as the base procedure. 
\paragraph{Pairwise regression.} In Algorithm \ref{algo: Boosting 2}, we describe the construction of $\Omega^{\rm boo}$ with the pairwise regression and the pairwise thin plate as the base procedure. For both base procedures, we need to specify how to construct the basis functions in steps 3 and 8. For the pairwise regression, we set the first element of $C_i$ as $1$ and define $S^{j,l}_{i}=C_{ij} C_{il}$ for $1\leq i\leq n.$ Then for step 3, we define $\mathcal{P}^{j,l}$ as the projection matrix to the vector  $S^{j,l}_{\mathcal{A}_2};$ for step 8, we define $\mathcal{H}^{[m]}=(S^{\widehat{j}_m,\widehat{l}_m}_{\mathcal{A}_1})^{\intercal}/{\|S^{\widehat{j}_m,\widehat{l}_m}_{\mathcal{A}_1}\|_2^2}.$

For the pairwise thin plate, we follow chapter 7 of \citetsupp{green1993nonparametric} to construct the projection matrix. For $1\leq j<l\leq p,$ we define the matrix $T^{j,l}=\begin{pmatrix} 1&1&\cdots &1\\ X_{1,j}&X_{2,j}&\cdots& X_{n,j}\\ X_{1,l}&X_{2,l}&\cdots& X_{n,l}\end{pmatrix}.$
For $1\leq s\leq n,$ define $E^{j,l}_{s,s}=0$; for $1\leq s\neq t\leq n,$ define $$E^{j,l}_{s,t}=\frac{1}{16\pi} \left\|(X_{s,(j,l)}-X_{t,(j,l)}\right\|_2^2 \log (\left\|(X_{s,(j,l)}-X_{t,(j,l)}\right\|_2^2).$$
Then we define 
$\bar{\mathcal{P}}^{j,l}=\begin{pmatrix}E^{j,l}_{\mathcal{A}_2,\mathcal{A}_2} &(T^{j,l}_{\cdot,\mathcal{A}_2})^{\intercal} \end{pmatrix}
\begin{pmatrix} E^{j,l}_{\mathcal{A}_2,\mathcal{A}_2} &(T^{j,l}_{\cdot,\mathcal{A}_2})^{\intercal} \\T^{j,l}_{\cdot,\mathcal{A}_2}&0 \end{pmatrix}^{-1}\in \R^{n_2\times (n_2+3)}$. In step 3, compute $\mathcal{P}^{j,l}\in \R^{n_2\times n_2}$ as the first $n_2$ columns of $\bar{\mathcal{P}}^{j,l}.$
For step 8, we compute
$$\bar{\mathcal{H}}^{j,l}=\begin{pmatrix}E^{j,l}_{\mathcal{A}_1,\mathcal{A}_1} &(T^{j,l}_{\cdot,\mathcal{A}_1})^{\intercal} \end{pmatrix}
\begin{pmatrix} E^{j,l}_{\mathcal{A}_1,\mathcal{A}_1} &(T^{j,l}_{\cdot,\mathcal{A}_1})^{\intercal} \\T^{j,l}_{\cdot,\mathcal{A}_1}&0 \end{pmatrix}^{-1}\in \R^{n_1\times (n_1+3)},$$
and set  $\mathcal{H}^{[m]}\in \R^{n_1\times n_1}$ as the first $n_1$ columns of $\bar{\mathcal{H}}^{\widehat{j}_m,\widehat{l}_m}.$

\begin{algorithm}[ht!]
\caption{Construction of $\Omega$ in Boosting with non-parametric pairwise regression}
\begin{flushleft}
%\hspace*{\algorithmicindent} 
\textbf{Input:} Data $C\in \R^{n\times p}, D\in \R^{n}$; the step-length factor $0<\nu\leq 1$; the stoping time $m_{\rm stop}.$ \\
%\hspace*{\algorithmicindent} 
\textbf{Output:} $\Omega^{\rm boo}$ 
\end{flushleft}
\begin{algorithmic}[1]
\State Randomly split the data into disjoint subsets $\mathcal{A}_1,\mathcal{A}_2$ with $n_1=\lfloor \frac{2n}{3}\rfloor$ and $|\mathcal{A}_2|=n-n_1$;
    \State Set $m=0,$ $\widehat{f}^{[0]}_{\mathcal{A}_2}=0,$ and $\Omega^{[0]}=0.$
    \State For $1\leq j,l\leq p$, compute $\mathcal{P}^{j,l}$ as the projection matrix to a set of basis functions generated by $C_{\mathcal{A}_2,j}$ and $C_{\mathcal{A}_2,l}.$
  \For{$1\leq m\leq m_{\rm stop}$}
  \State Compute the adjusted outcome $U^{[m]}_{\mathcal{A}_2}=D_{\mathcal{A}_2}-\widehat{f}^{[m-1]}_{\mathcal{A}_2}$
  \State Implement the following base procedure on $\{C_i, U^{[m]}_i\}_{i\in \mathcal{A}_2}$
\begin{equation*}
(\widehat{j}_{m},\widehat{l}_{m})=\argmin_{1\leq j,l\leq p} \left\|U^{[m]}_{\mathcal{A}_2}- \mathcal{P}^{j,l} U^{[m]}_{\mathcal{A}_2}\right\|^2.
\end{equation*}  
  \State Update $\widehat{f}^{[m]}_{\mathcal{A}_2}=\widehat{f}^{[m-1]}_{\mathcal{A}_2}+\nu \mathcal{P}^{\widehat{j}_{m},\widehat{l}_{m}}\left(D_{\mathcal{A}_2}-\widehat{f}^{[m-1]}_{\mathcal{A}_2}\right)$
  \State Construct $\mathcal{H}^{[m]}$ in the same way as $\mathcal{P}^{\widehat{j}_{m},\widehat{l}_{m}}$ but with the data in $\mathcal{A}_1.$
  \State Compute $\Omega^{[m]}=\nu \mathcal{H}^{[m]}+ ({\rm I}-\nu\mathcal{H}^{[m]})\Omega^{[m-1]}$
\EndFor
\State Return $\Omega^{\rm boo}$
 \end{algorithmic}
\label{algo: Boosting 2}
\end{algorithm}

\paragraph{Decision tree.} In Algorithm \ref{algo: Boosting 3}, we describe the construction of $\Omega^{\rm boo}$ with the decision tree as the base procedure. 
\begin{algorithm}[ht!]
\caption{Construction of $\Omega$ in Boosting with Decision tree}
\begin{flushleft}
%\hspace*{\algorithmicindent} 
\textbf{Input:} Data $C\in \R^{n\times p}, D\in \R^{n}$; the step-length factor $0<\nu\leq 1$; the stoping time $m_{\rm stop}.$ \\
%\hspace*{\algorithmicindent} 
\textbf{Output:} $\Omega^{\rm boo}$
\end{flushleft}
\begin{algorithmic}[1]
\State Randomly split the data into disjoint subsets $\mathcal{A}_1,\mathcal{A}_2$ with $n_1=\lfloor \frac{2n}{3}\rfloor$ and $|\mathcal{A}_2|=n-n_1$;
    \State Set $m=0,$ $\widehat{f}^{[0]}_{\mathcal{A}_2}=0,$ and $\Omega^{[0]}=0.$
 %   \State For $1\leq j,l\leq p$, compute $\mathcal{P}^{j,l}$ as the projection matrix to a set of basis functions generated by $C_{\mathcal{A}_2,j}$ and $C_{\mathcal{A}_2,l}.$
  \For{$1\leq m\leq m_{\rm stop}$}
  \State Compute the adjusted outcome $U^{[m]}_{\mathcal{A}_2}=D_{\mathcal{A}_2}-\widehat{f}^{[m-1]}_{\mathcal{A}_2}$
  \State Run the decision tree on $\{C_i, U^{[m]}_i\}_{i\in \mathcal{A}_2}$ and partition $\R^{p}$ as leaves $\{\mathcal{R}^{[m]}_1,\cdots,\mathcal{R}^{[m]}_L\};$ 
%  \State For $C_{\rm new}\in \R^p$ belonging to $\mathcal{R}^{[m]}_{l(C_{\rm new})},$ we obtain the predicted value as
%\begin{equation}
%\widehat{g}_i(C_{\rm new})=\sum_{j\in \mathcal{A}_2} \omega_{j}(C_{\rm new}) U^{[m]}_j \quad \text{with}\quad \omega_{j}(C_{\rm new})=\frac{{\bf{1}}\left[C_i\in \mathcal{R}_{l(C_{\rm new})}\right]}{\sum_{k\in \mathcal{A}_2}{\bf{1}}\left[C_k\in \mathcal{R}_{l(C_{\rm new})}\right]}.
%%(\widehat{j}_{m},\widehat{l}_{m})=\argmin_{1\leq j,l\leq p} \left\|U^{[m]}_{\mathcal{A}_2}- \mathcal{P}^{j,l} U^{[m]}_{\mathcal{A}_2}\right\|^2.
%\end{equation}  
\State For any $j\in \mathcal{A}_2,$ we identify the leaf $\mathcal{R}^{[m]}_{l(C_{j})}$ containing $C_j$ and compute $$\mathcal{P}^{[m]}_{j,t}=\frac{{\bf{1}}\left[C_t\in \mathcal{R}^{[m]}_{l(C_{j})}\right]}{\sum_{k\in \mathcal{A}_2}{\bf{1}}\left[C_k\in \mathcal{R}^{[m]}_{l(C_{j})}\right]} \quad \text{for} \quad t\in \mathcal{A}_2$$
\State Compute the matrix $(\mathcal{P}^{[m]}_{j,t})_{j,t\in \mathcal{A}_2}$ and  update $$\widehat{f}^{[m]}_{\mathcal{A}_2}=\widehat{f}^{[m-1]}_{\mathcal{A}_2}+\nu \mathcal{P}^{[m]}
\left(D_{\mathcal{A}_2}-\widehat{f}^{[m-1]}_{\mathcal{A}_2}\right)$$
  \State Construct the matrix $(\mathcal{H}^{[m]}_{j,t})_{j,t\in { \mathcal{A}_1}}$ as 
 $\mathcal{H}^{[m]}_{j,t}=\frac{{\bf{1}}\left[C_t\in \mathcal{R}^{[m]}_{l(C_{j})}\right]}{{\sum_{k\in \mathcal{A}_1}}{\bf{1}}\left[C_k\in \mathcal{R}^{[m]}_{l(C_{j})}\right]}.$
 \State Compute $\Omega^{[m]}=\nu \mathcal{H}^{[m]}+ ({\rm I}-\nu\mathcal{H}^{[m]})\Omega^{[m-1]}$
\EndFor
\State Return $\Omega^{\rm boo}$
 \end{algorithmic}
\label{algo: Boosting 3}
\end{algorithm}

%%%%%%%%%%%%%%%%%%%%%%%%%%%%
%%%%%%%%%%%%%%%%%%%%%%%%%%%%
\subsection{\texttt{TSCI} with deep neural network}
\label{sec: TS-DNN}
%%%%%%%%%%%%%%%%%%%%%%%%%%%%
%%%%%%%%%%%%%%%%%%%%%%%%%%%%
In the following, we demonstrate how to calculate $\Omega$ for {a}
%the
deep neural network \citepsupp{james2013introduction}. 
We define the first hidden layer as $H^{(1)}_{i,k}=\sigma\left(\omega^{(1)}_{k,0}+\sum_{l=1}^{\pz}\omega^{(1)}_{k,l} Z_{il}+\sum_{l=1}^{\px}\omega^{(1)}_{k,l+\pz}X_{i,l}\right)$ for $1\leq k\leq K_1,$
where $\sigma(\cdot)$ is the activation function and $\{\omega^{(1)}_{k,l}\}_{1\leq k\leq K_1, 1\leq l\leq p_x+p_z}$ are parameters.
For $m\geq 2$, we define the $m$-th hidden layer as 
$H^{(m)}_{i,k}=\sigma\left(\omega^{(m)}_{k,0}+\sum_{l=1}^{K_{m-1}}\omega^{(m)}_{k,l} H^{(m-1)}_{i,l}\right)$ for $1\leq k\leq K_m,$ where $\{\omega^{(m)}_{k,l}\}_{1\leq k\leq K_m, 1\leq l\leq K_{m-1}}$ are unknown parameters. For given $M\geq 1,$ we estimate the unknown parameters based on the data $\{X_i, Z_i, D_i\}_{i\in \mathcal{A}_2}$,
\begin{equation*}
\left\{\widehat{\beta}, \{\widehat{\omega}^{(m)}\}_{1\leq m\leq M}\right\}\coloneqq \argmin_{\{\beta_k\}_{0\leq k\leq K_M},\{\omega^{(m)}\}_{1\leq m\leq M}}\sum_{i\in \mathcal{A}_2}\left(Y_i-\beta_0-\sum_{k=1}^{K_M}\beta_k H^{(M)}_{i,k}\right)^2.
\end{equation*}
With $\{\widehat{\omega}^{(m)}\}_{1\leq m\leq M},$ for $1\leq m \leq M$ and $1\leq i\leq n_1$, we define 
\begin{equation*}
\widehat{H}^{(m)}_{i,k}=\sigma\left(\widehat{\omega}^{(m)}_{k0}+\sum_{l=1}^{K_{m-1}}\widehat{\omega}^{(m)}_{kl} H^{(m-1)}_{i,l}\right) \quad \text{for} \quad 1\leq k\leq K_m
\end{equation*}
with $\widehat{H}^{(1)}_{i,k}=\sigma\left(\widehat{\omega}^{(1)}_{k0}+\sum_{l=1}^{\pz}\widehat{\omega}^{(1)}_{kl} Z_{il}+\sum_{l=1}^{\px}\widehat{\omega}^{(1)}_{k,l+\pz}X_{il}\right)$ for $1\leq k\leq K_1.$ We use $\Omega^{\rm DNN}=\widehat{H}^{(M)}\left([\widehat{H}^{(M)}]^{\intercal}\widehat{H}^{(M)}\right)^{-1}[\widehat{H}^{(M)}]^{\intercal}$ to denote the projection to the column space of the matrix $\widehat{H}^{(M)}$ and express the deep neural network estimator as $\Omega^{\rm DNN}D_{\mathcal{A}_1}.$   With $ \widehat{V}_{\mathcal{A}_1}=\Omega^{\rm DNN} V_{\mathcal{A}_1},$ we define  $\T_{\rm DNN}(V)=\left[\Omega^{\rm DNN}\right]^{\intercal}P^{\perp}_{\widehat{V}_{\mathcal{A}_1}} \Omega^{\rm DNN},$ which is shown to be an orthogonal projection matrix in Lemma \ref{lem: general transform} in the supplement.

%%%%%%%%%%%%%%%%%%%%%%%%%%%%
%%%%%%%%%%%%%%%%%%%%%%%%%%%%
\subsection{\texttt{TSCI} with B-spline}
\label{sec: TS-Bspline}
%%%%%%%%%%%%%%%%%%%%%%%%%%%%
%%%%%%%%%%%%%%%%%%%%%%%%%%%%
As a simplification, we consider the additive model $
\E (D_{i}\mid Z_i, X_{i})=\gamma_1(Z_{i})+\gamma_2(X_{i}),
$ and assume that $\gamma_1(\cdot)$ can be well approximated by a set of B-spline functions $\{b_1(\cdot),\cdots, b_{M}(\cdot)\}.$ In practice, the number $M$ can be chosen by cross-validation.  
We define the matrix $B\in \R^{n\times M}$ with its $i$-th row $B_{i}=\left(b_1(Z_{i}),\cdots, b_{M}(Z_{i})\right)^{\intercal}.$ Without loss of generality, we approximate $\gamma_2(X_i)$ by $W_{i}\in \R^{p_w}$, which is generated by the same set of basis functions for $\phi(X_i)$.  Define $\Omega^{\rm ba}=\Pm\in \R^{n\times n}$ as the projection matrix to the space spanned by the columns of $B\in \R^{n\times k}$ and $W \in \R^{n\times p_{w}}$. We write the first-stage estimator as $\Omega^{\rm ba} D$ and compute $\T_{\rm ba}(V)=
\left[\Omega^{\rm ba}\right]^{\intercal}P^{\perp}_{\widehat{V},{W}} \Omega^{\rm ba}$ with $\widehat{V}=\Omega^{\rm ba} V.$ The transformation matrix $\T_{\rm ba}(V)$ is a projection matrix with 
$M-{\rm rank}(V)\leq {\rm rank}\left[\T_{\rm ba}(V)\right]\leq M.$ When the basis number $M$ is small and the degree of freedom $M+p_w$ is much smaller than $n$, sample splitting is not even needed for if the B-spline is used for fitting the treatment model, which is different from the general machine learning algorithms.  

%%%%%%%%%%%%%%%%%%%%%%%%%%%%%%
\subsection{Properties of $\T(V)$}
%%%%%%%%%%%%%%%%%%%%%%%%%%%%%%
 The following lemma is about the property of 
the transformation matrix $\T(V)$, whose proof can be found in Section \ref{sec: lemma 4 5 proof}. Recall that 
$$\T_{\rm RF}({V})=\Omega^{\intercal}P^{\perp}_{\widehat{V}_{\mathcal{A}_1}} \Omega \quad \text{with}\quad \Omega_{ij}=\omega_j(Z_i, X_i),$$
where $\omega_j(z, x)$ is defined in \eqref{eq: weight function}.

\begin{Lemma}
The transformation matrix $\T_{\rm RF}({V})$  satisfies 
\begin{equation}
\lambda_{\max}(\T_{\rm RF}({V}))\leq 1\quad \text{and}\quad
b^{\intercal} [\T_{\rm RF}({V})]^2 b\leq b^{\intercal} \T_{\rm RF}({V}) b \quad \text{for any}\quad b\in \R^{n_1}.
\label{eq: key inequality for RF}
\end{equation}
As a consequence, we establish 
${\rm Tr}\left([\T_{\rm RF}({V})]^2\right)\leq {\rm Tr}[\T(V)].
$
The transformation matrixs $\T_{\rm ba}(V)$ and $\T_{\rm DNN}({V})$ are orthogonal projection matrices with
\begin{equation}
[\T_{\rm ba}(V)]^2=\T_{\rm ba}(V) \quad \text{and}\quad [\T_{\rm DNN}({V})]^2=\T_{\rm DNN}({V}).
\label{eq: basis equi}
\end{equation}
The transformation matrix $\T_{\rm boo}({V})$  satisfies 
\begin{equation}
\lambda_{\max}(\T_{\rm boo}({V}))\leq \|\Omega^{\rm boo}\|_2^2,\quad \text{and}\quad b^{\intercal} [\T_{\rm boo}({V})]^2 b\leq \|\Omega^{\rm boo}\|_2^2\cdot b^{\intercal} \T_{\rm boo}({V}) b,
\label{eq: key inequality for general}
\end{equation}
for any $b\in \R^{n_1}.$ 
As a consequence, we establish 
${\rm Tr}\left([\T_{\rm boo}({V})]^2\right)\leq \|\Omega^{\rm boo}\|_2^2 \cdot {\rm Tr}[\T(V)].
$
\label{lem: general transform}
\end{Lemma}

\subsection{Homoscadastic correlation}
\label{sec: homo extra}
In the following, we discuss the simplified method and theory in the homoscadastic correlation setting, that is, ${\rm Cov}(\epsilon_i,\delta_i\mid Z_i, X_i)={\rm Cov}(\epsilon_i,\delta_i)$. With this extra assumption, we present the following bias-corrected estimator as an alternative to the estimator defined in \eqref{eq: ML corrected hetero},  
\begin{equation*}
\betaRF(V)=\widehat{\beta}_{\rm init}(V)-\frac{\widehat{\rm Cov}(\delta_i,\epsilon_i)\cdot {\rm Tr}[\T(V)]}{{D}_{\mathcal{A}_1}^{\intercal}\T(V) {D}_{\mathcal{A}_1}},
\end{equation*}
where $\widehat{\beta}_{\rm init}(V)$ is defined in \eqref{eq: RF init} and the estimator of ${\rm Cov}(\delta_i,\epsilon_i)$ is defined as,
\begin{equation*}
\widehat{\rm Cov}(\delta_i,\epsilon_i)=\frac{1}{n_1-r} (D_{\mathcal{A}_1}-\widehat{f}_{\mathcal{A}_1})^{\intercal}P^{\perp}_{V_{\mathcal{A}_1}}[Y_{\mathcal{A}_1}-D_{\mathcal{A}_1}\widehat{\beta}_{\rm init}(V)],
\label{eq: cov est}
\end{equation*}
with $r$ denoting the rank of  the matrix $(V, W)$. The correction in constructing $\betaRF(V)$ 
implicitly requires ${\rm Cov}(\epsilon_i,\delta_i\mid Z_i, X_i)={\rm Cov}(\epsilon_i,\delta_i),$ which might limit practical applications.

We present a simplified version of Theorem
\ref{thm: bias correction effect} by assuming homoscadastic correlation. We shall only present the results for the \texttt{TSCI} with random forests but the extension to the general machine learning methods is straightforward.
\begin{Theorem} Consider the model \eqref{eq: outcome model} and \eqref{eq: treatment model} with ${\rm Cov}(\epsilon_i,\delta_i\mid Z_i, X_i)={\rm Cov}(\epsilon_i,\delta_i)$. Suppose that Conditions {\rm (R1)} and {\rm (R2)} hold, then 
\begin{equation}
\left|\widehat{\rm Cov}(\delta_i,\epsilon_i)-{\rm Cov}(\delta_i,\epsilon_i)\right|\leq \eta^0_n(V)\leq \eta_n(V),
\label{eq: correlation accuracy}
\end{equation}
where $\eta_n(V)$ is defined in \eqref{eq: approximation rate hetero} and
\begin{equation}
\eta^0_n(V)=\frac{\|f_{\mathcal{A}_1}-\widehat{f}_{\mathcal{A}_1}\|_{2}}{\sqrt{n}}+\sqrt{\frac{\log n}{n}}+\left(|\beta-\widehat{\beta}_{\rm init}(V)|+\frac{\|R(V)\|_2}{\sqrt{n}}\right)\left(1+\frac{\|f_{\mathcal{A}_1}-\widehat{f}_{\mathcal{A}_1}\|_{2}}{\sqrt{n}}\right).
\label{eq: approximation rate}
\end{equation}
Furthermore, if we assume \eqref{eq: linder condition} holds, then $\betaRF(V)$ 
 satisfies 
\begin{equation}
\frac{1}{{\rm SE}(V)}\left(\betaRF(V)-\beta\right)={\mathcal{G}}(V)+\mathcal{E}(V),
\label{eq: limiting RF-corrected homo}
\end{equation}
where ${\rm SE}(V)$ is defined in \eqref{eq: limiting RF-corrected}, ${\mathcal{G}}(V)\cid N(0,1)$ and there exist positive constants $C>0$ and $t_0>0$ such that 
{\small
\begin{equation}
\liminf_{n\rightarrow\infty}\PP\left(\left|\mathcal{E}(V)\right|\leq C\frac{\eta^0_n(V)\cdot {\rm Tr}[\T(V)]+\|R(V)\|_2+t_0\sqrt{{\rm Tr}([\T_{\rm RF}({V})]^2)}}{\sqrt{{f}_{\mathcal{A}_1}[\T(V)]^2 {f}_{\mathcal{A}_1}}}\right)\geq 1-\exp(-t_0^2),
\label{eq: error bound homo}
\end{equation}}
with $\eta^0_n(V)$ defined in \eqref{eq: approximation rate}.
\label{thm: homo bias correction effect}
\end{Theorem}

As a remark, if $\|R(V)\|_2/\sqrt{n}\rightarrow 0$, $\widehat{\beta}_{\rm init}(V)\cip \beta$, and ${\|f_{\mathcal{A}_1}-\widehat{f}_{\mathcal{A}_1}\|_{2}}/{\sqrt{n}}\rightarrow 0,$ then we have $\eta^0_n(V)\rightarrow 0.$

\subsection{Consistency of variance estimators}
The following lemma controls the variance consistency, whose proof can be found in Section \ref{sec: lemma 7 proof}.
\begin{Lemma} 
Suppose that Conditions {\rm (R1)} and {\rm (R2)} hold and $\kappa_n(V)^2+\sqrt{\log n} \kappa_n(V)\cip 0,$ with $\kappa_n(V)= \sqrt{\log n}\left(\|R(V)\|_{\infty}+|\beta-\widehat{\beta}_{\rm init}(V)|+{\log n}/{\sqrt{n}}\right).$ If  $$\frac{{\sum_{i=1}^{n_1}\sigma^2_i[\T(V) {D}_{\mathcal{A}_1}]_i^2}
}{{\sum_{i=1}^{n_1}\sigma^2_i[\T(V) {f}_{\mathcal{A}_1}]_i^2}}\cip 1, \quad \text{and}\quad \frac{\max_{1\leq i\leq n_1}[\T(V) {D}_{\mathcal{A}_1}]_i^2}{{{\sum_{i=1}^{n_1}[\T(V) {D}_{\mathcal{A}_1}]_i^2}}}\cip 0,$$
then we have ${\widehat{\rm SE}(V)}/{{\rm SE}(V)}\cip 1.$
\label{lem: variance consistency}
\end{Lemma}

\subsection{Size and Power of $\mathcal{C}(V_q, V_{q'})$}

The limiting distribution in Theorem \ref{thm: comparison RF} implies the size and power of the comparison test $\mathcal{C}(V_q, V_{q'})$ in \eqref{eq: comparison test}, stated in the following theorem.

\begin{Theorem}
Suppose that {the} Conditions of Theorem \ref{thm: comparison RF} hold,  $\widehat{H}(V_{q},V_{q'})/H(V_{q},V_{q'})\cip 1$, 
and $\mathcal{L}_n(V_{q}, V_{q'})$ defined in \eqref{eq: sig difference RF} satisfies $\mathcal{L}_n(V_{q}, V_{q'})\cip \mathcal{L}^*(V_{q}, V_{q'})$ for $\mathcal{L}^*(V_{q}, V_{q'})\in \R \cup \{-\infty,\infty\},$ then the test $\mathcal{C}(V_q, V_{q'})$ in \eqref{eq: comparison test} {with corresponding $\alpha_0$} satisfies, 
\begin{equation}
\lim_{n\rightarrow \infty}\mathbf{P}\left(\mathcal{C}(V_q, V_{q'})=1\right)=1-\Phi\left(z_{\alpha_0}-\mathcal{L}^*(V_{q}, V_{q'})\right)+\Phi\left(-z_{\alpha_0}-\mathcal{L}^*(V_{q}, V_{q'})\right).
\label{eq: asymp power RF}
\end{equation}
%\Zijian{where $\alpha_0$ is used in \eqref{eq: comparison test}.}
With $\left|\mathcal{L}^*(V_{q}, V_{q'})\right|=0$, then we have $
\lim_{n\rightarrow \infty}\mathbf{P}\left(\mathcal{C}(V_q, V_{q'})=1\right)=2\alpha_0.
$
With $\left|\mathcal{L}^*(V_{q}, V_{q'})\right|=\infty$, then we have $\lim_{n\rightarrow \infty}\mathbf{P}\left(\mathcal{C}(V_q, V_{q'})=1\right)=1.$
\label{thm: power RF}
\end{Theorem}

%%%%%%%%%%%%%%%%%%%%%%%%%%%%%
%%%%%%%%%%%%%%%%%%%%%%%%%%%%
%\subsection{Calculation of $\Omega$ for the boosting method}
%\label{sec: Omega boosting}
%%%%%%%%%%%%%%%%%%%%%%%%%%%%%
%%%%%%%%%%%%%%%%%%%%%%%%%%%%

%%%%%%%%%%%%%%%%%%%%%%%%%%%%%%%%%%%%%%%%%%%
%\section{Additional proofs} %of Proposition \ref{prop: initial consistency} and Theorems \ref{thm: bias correction effect}, \ref{thm: limiting TSRF hetero}, and \ref{thm: homo bias correction effect}}
%%%%%%%%%%%%%%%%%%%%%%%%%%%%%%%%%%%%%%%%%%%

%%%%%%%%%%%%%%%%%%%%%%%%%%%%%%%%%%%%%%%%%%%%%%
\section{Proofs}
%%%%%%%%%%%%%%%%%%%%%%%%%%%%%%%%%%%%%%%%%%%%%%
We establish Proposition \ref{prop: special identification}
in Section \ref{sec: identification}.
In Section \ref{sec: limiting init proof}, we establish Proposition \ref{prop: initial consistency}. In Section \ref{sec: limiting TSRF proof hetero}, we establish Theorems  \ref{thm: bias correction effect} and \ref{thm: limiting TSRF hetero}. We prove Theorems \ref{thm: comparison RF} and \ref{thm: power RF} in Section \ref{sec: comparison two}. We establish Theorem \ref{thm: post-selection RF} in Section \ref{sec: multiple selection} and prove Theorem \ref{thm: homo bias correction effect} in Section \ref{sec: correlation consistency}.

%We first introduce some notations and lemmas used throughout the proof. 
We define the conditional covariance matrices $\Lambda, \Sigma^{\delta},\Sigma^{\epsilon} \in \R^{n_1\times n_1}$ as 
$
\Lambda=\E({\delta}_{\mathcal{A}_1} {\epsilon}_{\mathcal{A}_1}^{\intercal}\mid X_{\mathcal{A}_1}, Z_{\mathcal{A}_1}),$ $\Sigma^{\delta}=\E({\delta}_{\mathcal{A}_1} {\delta}_{\mathcal{A}_1}^{\intercal}\mid X_{\mathcal{A}_1}, Z_{\mathcal{A}_1}),$ and $\Sigma^{\epsilon}=\E({\epsilon}_{\mathcal{A}_1} {\epsilon}_{\mathcal{A}_1}^{\intercal}\mid X_{\mathcal{A}_1}, Z_{\mathcal{A}_1})$.
For any $i,j\in \mathcal{A}_1$ and $i\neq j,$ we have 
$\Lambda_{i,j}=\E\left[\delta_j\E(\epsilon_i \mid X_{\mathcal{A}_1}, Z_{\mathcal{A}_1},\delta_j) \mid X_{\mathcal{A}_1}, Z_{\mathcal{A}_1}\right].$
Since $\E(\epsilon_i \mid X_{\mathcal{A}_1}, Z_{\mathcal{A}_1},\delta_j)=\E(\epsilon_i\mid X_{i}, Z_{i})=0$ for $i\neq j$, we have $\Lambda_{i,j}=0$ and $\Lambda$ is a diagonal matrix. Similarly, we can show $\Sigma^{\delta}$ and $\Sigma^{\epsilon}$ are diagonal matrices. The conditional sub-gaussian condition in (R1) implies that 
$
\max_{1\leq j\leq n_1}\max\left\{|\Lambda_{j,j}|,|\Sigma^{\delta}_{j,j}|,|\Sigma^{\epsilon}_{j,j}|\right\}\leq K^2.
$ We shall use $\mathcal{O}$ to denote the set of random variables $\{Z_i,X_i\}_{1\leq i\leq n}$ and $\{D_i\}_{i\in \mathcal{A}_2}$.  

We introduce the following lemma about the concentration of quadratic forms, which is Theorem 1.1 in \citetsupp{rudelson2013hanson}. 
\begin{Lemma} {\rm (Hanson-Wright inequality)} Let $\epsilon\in \R^{n}$ be a random vector with independent sub-gaussian components $\epsilon_i$ with zero mean and sub-gaussian norm $K$. Let $A$ be an $n\times n$ matrix. For every $t\geq 0$, 
$
\mathbf{P}\left(|\epsilon^{\intercal}A \epsilon- \E \epsilon^{\intercal} A\epsilon|>t\right)\leq 2\exp\left[-c\min\left(\frac{t^2}{K^{4}\|A\|_{F}^2}, \frac{t}{K^2\|A\|_2}\right)\right].
%\label{eq: HW-bound}
$
\label{lem: HW-bound}
\end{Lemma}
Note that $2\epsilon^{\intercal} A\delta=(\epsilon+\delta)^{\intercal} A(\epsilon+\delta)-\epsilon^{\intercal} A\epsilon-\delta^{\intercal} A\delta.$ If both $\epsilon_i$ and $\delta_i$ are sub-gaussian, we apply the union bound, Lemma \ref{lem: HW-bound} for both $\epsilon$ and $\delta$ and then establish,
\begin{equation}
\mathbf{P}\left(|\epsilon^{\intercal}A \delta- \E \epsilon^{\intercal} A\delta|>t\right)\leq 6\exp\left[-c\min\left(\frac{t^2}{K^{4}\|A\|_{F}^2}, \frac{t}{K^2\|A\|_2}\right)\right].
\label{eq: HW-bound imp}
\end{equation}

%
 
%%%%%%%%%%%%%%%%%%%%%%%%%%%%%%%%%%%%%%%%%%%%%%%%%%%%%
\subsection{Proof of Proposition \ref{prop: special identification}}
\label{sec: identification}
%%%%%%%%%%%%%%%%%%%%%%%%%%%%%%%%%%%%%%%%%%%%%%%%%%%%%
Note that 
\begin{equation}
\begin{aligned}
&\E(Y_i\mid Z_i=z, X_i=1)\\
&=\E\left[D^{(z)}_i(x=1) Y^{(1,z)}_i(x=1)+(1-D^{(z)}_i(x=1) )Y^{(0,z)}_i(x=1)\mid Z_i=z, X_i=1\right]\\
&=\E\left[D^{(z)}_i(x=1) Y^{(1,z)}_i(x=1)+(1-D^{(z)}_i(x=1) )Y^{(0,z)}_i(x=1)\mid X_i=1\right]
\end{aligned}
\end{equation}
We assume that $Y^{(1,z)}_i(x=1)-Y^{(1,w)}_i(x=1)=Y^{(1,z)}_i(x=0)-Y^{(1,w)}_i(x=0)=(z-w)\pi$ and obtain 
\begin{equation}
\begin{aligned}
&{\rm ITT}_{Y}(x=1)\coloneqq \E(Y_i\mid Z_i=z, X_i=1)-\E(Y_i\mid Z_i=w, X_i=1)\\
&= \E\left[(D^{(z)}_i(x=1)-D^{(w)}_i(x=1))(Y^{(1,w)}_i(x=1)-Y^{(0,w)}_i(x=1))\mid X_i=1\right]+(z-w)\pi\\
&= \beta \E\left[D^{(z)}_i(x=1)-D^{(w)}_i(x=1)\mid X_i=1\right]+(z-w)\pi\\
&= \beta\left(\E(D_i\mid Z_i=z, X_i=1)-\E(D_i\mid Z_i=w, X_i=1)\right)+(z-w)\pi\\
&= \beta \cdot {\rm ITT}_{D}(x=1)+(z-w)\pi.
\end{aligned}
\label{eq: x=1}
\end{equation}
Similarly, we have 
\begin{equation}
{\rm ITT}_{Y}(x=0)=\beta \cdot {\rm ITT}_{D}(x=0)+(z-w)\pi.
\label{eq: x=0}
\end{equation}
We combine \eqref{eq: x=1} and \eqref{eq: x=0} to establish the results.

%%%%%%%%%%%%%%%%%%%%%%%%%%%%%%%%%%%%%%%%%%%%%%%%%%%%%%%%%%%%%%%%%%%%%%%%%%%%%%%%%%%%%
\subsection{Proof of Proposition \ref{prop: initial consistency}}
\label{sec: limiting init proof}
%%%%%%%%%%%%%%%%%%%%%%%%%%%%%%%%%%%%%%%%%%%%%%%%%%%%%%%%%%%%%%%%%%%%%%%%%%%%%%%%%%%%
%By the transformed outcome model \eqref{eq: transformed subset}, 
Recall that we are analyzing $\widehat{\beta}_{\rm init}(V)$ defined in \eqref{eq: RF init} with replacing $\T(V)$ by $\T(V).$
We have decomposed the estimation error $\widehat{\beta}_{\rm init}(V)-\beta$ as 
\begin{equation}
\widehat{\beta}_{\rm init}(V)-\beta=\frac{{{\epsilon}_{\mathcal{A}_1}^{\intercal}\T(V) {\delta}_{\mathcal{A}_1}+{\epsilon}_{\mathcal{A}_1}^{\intercal}\T(V) {f}_{\mathcal{A}_1}+R_{\mathcal{A}_1}(V)^{\intercal}\T(V) {D}_{\mathcal{A}_1}}}{{D}_{\mathcal{A}_1}^{\intercal}\T(V) {D}_{\mathcal{A}_1}}.
\label{eq: init decomposition}
\end{equation}

The following Lemma controls the key components of the above decomposition, whose proof can be found in Section \ref{sec: lemma 5 proof} in the supplement.
\begin{Lemma}
Under Condition {\rm (R1)}, with probability larger than $1-\exp\left(-c \min\left\{t_0^2,t_0\right\}\right)$ for some positive constants $c>0$ and $t_0>0,$
{\small
\begin{equation}
\left|{{\epsilon}_{\mathcal{A}_1}^{\intercal}\T(V) {\delta}_{\mathcal{A}_1}}-{\rm Tr}[\T({V})\Lambda]\right|\leq t_0 K^2\sqrt{{\rm Tr}([\T({V})]^2)}, \; \left|{\epsilon}_{\mathcal{A}_1}^{\intercal}\T(V) {f}_{\mathcal{A}_1}\right|\leq t_0 K\sqrt{{f}_{\mathcal{A}_1}^{\intercal}[\T(V)]^2 {f}_{\mathcal{A}_1}}
\label{eq: inter bound 1 general}
\end{equation}
}
{\small
\begin{equation}
\begin{aligned}
\left|\frac{D^{\intercal}_{\mathcal{A}_1}\T(V) D_{\mathcal{A}_1}}{f^{\intercal}_{\mathcal{A}_1}\T(V) f_{\mathcal{A}_1}}-1\right|\lesssim  \frac{K^2 {\rm Tr}[\T({V})]+t_0 K^2\sqrt{{\rm Tr}([\T({V})]^2)}+t_0K\sqrt{f_{\mathcal{A}_1}^{\intercal}[\T(V)]^2 f_{\mathcal{A}_1}}}{{f^{\intercal}_{\mathcal{A}_1}\T(V) f_{\mathcal{A}_1}}}.%+\frac{t_0 }{\sqrt{f_{\mathcal{A}_1}^{\intercal}\T(V) f_{\mathcal{A}_1}}}.
\end{aligned}
\label{eq: bottom consistency RF}
\end{equation}}
where $\Lambda=\E({\delta}_{\mathcal{A}_1} {\epsilon}_{\mathcal{A}_1}^{\intercal}\mid X_{\mathcal{A}_1}, Z_{\mathcal{A}_1}).$ %the matrix $\Lambda$ is defined in \eqref{eq: cov def}. 
\label{lem: IV strength 1}
\end{Lemma}

Define $\tau_n={f}_{\mathcal{A}_1}^{\intercal}\T(V) {f}_{\mathcal{A}_1}.$ Together with  \eqref{eq: key inequality for RF} and the {generalized} IV strength condition {\rm (R2)}, we apply \eqref{eq: bottom consistency RF} with $t_0=\tau_n^{1/2-c_0}$ for some $0<c_0<1/2$. Then there exists positive constants $c>0$ and $C>0$ such that with probability larger than $1-\exp\left(-c\tau_n^{1/2-c_0}\right),$
\begin{equation}
\left|\frac{D^{\intercal}_{\mathcal{A}_1}\T(V) D_{\mathcal{A}_1}}{f^{\intercal}_{\mathcal{A}_1}\T(V) f_{\mathcal{A}_1}}-1\right|\leq C \frac{K^2{\rm Tr}[\T({V})]}{{{{f}_{\mathcal{A}_1}^{\intercal}\T(V) {f}_{\mathcal{A}_1}}}} +C\frac{K+K^2}{(f_{\mathcal{A}_1}^{\intercal}\T(V) f_{\mathcal{A}_1})^{c_0}}\leq 0.1.
\label{eq: finite-sample strength}
\end{equation}
Together with \eqref{eq: init decomposition}, we establish 
\begin{equation}
\left|\widehat{\beta}_{\rm init}(V)-\beta\right|\lesssim  \frac{\left|{\epsilon}_{\mathcal{A}_1}^{\intercal}\T(V) {f}_{\mathcal{A}_1}\right|}{{{{f}_{\mathcal{A}_1}^{\intercal}\T(V) {f}_{\mathcal{A}_1}}}}+ \frac{\left|{\epsilon}_{\mathcal{A}_1}^{\intercal}\T(V) {\delta}_{\mathcal{A}_1}\right|}{{{{f}_{\mathcal{A}_1}^{\intercal}\T(V) {f}_{\mathcal{A}_1}}}}+\sqrt{\frac{{R_{\mathcal{A}_1}(V)^{\intercal}\T(V)R_{\mathcal{A}_1}(V)}}{{{f}_{\mathcal{A}_1}^{\intercal}\T(V) {f}_{\mathcal{A}_1}}}}.
\label{eq: RF triangle}
\end{equation}
By applying the decomposition \eqref{eq: RF triangle} and the upper bounds \eqref{eq: inter bound 1 general}, and \eqref{eq: inter bound 1 general} with $t_0=(f_{\mathcal{A}_1}^{\intercal}\T(V) f_{\mathcal{A}_1})^{1/2-c_0},$
 we establish that, conditioning on $\mathcal{O}$, with probability lager than $1-\exp\left(-c\tau_n^{1/2-c_0}\right)$ for some positive constants $c>0$ and $c_0\in(0,1/2),$
\begin{equation}
|\widehat{\beta}_{\rm init}(V)-\beta|\lesssim \frac{K^2{\rm Tr}[\T({V})]}{{{f}_{\mathcal{A}_1}^{\intercal}\T(V) {f}_{\mathcal{A}_1}}} +\frac{K+K^2}{(f_{\mathcal{A}_1}^{\intercal}\T(V) f_{\mathcal{A}_1})^{c_0}}+\frac{\|R_{\mathcal{A}_1}(V)\|_2}{\sqrt{{{f}_{\mathcal{A}_1}^{\intercal}\T(V) {f}_{\mathcal{A}_1}}}}.
\label{eq: init rate}
\end{equation} 
%where $C>0$ is some positive constant independent of $n$ and $p$. 
The above concentration bound, the assumption {\rm (R2)}, and the assumption ${f}_{\mathcal{A}_1}^{\intercal}\T(V) {f}_{\mathcal{A}_1}\gg \|R_{\mathcal{A}_1}(V)\|_2^2$ imply $\PP(|\widehat{\beta}_{\rm init}(V)-\beta|\geq c\mid \mathcal{O})\rightarrow 0$. Then for any constant $c>0$, we have 
$\PP(|\widehat{\beta}_{\rm init}(V)-\beta|\geq c)=\E\left(\PP(|\widehat{\beta}_{\rm init}(V)-\beta|\geq c\mid \mathcal{O})\right).$
 We apply the bounded convergence theorem and establish $\PP(|\widehat{\beta}_{\rm init}(V)-\beta|\geq c)\rightarrow 0$ and hence $\widehat{\beta}_{\rm init}(V)\cip\beta.$

%The bounds \eqref{eq: inter bound 1 general} and %\eqref{eq: cov matrix bound} imply 
%$
%\left|\frac{{\epsilon}_{\mathcal{A}_1}^{\intercal}\T(V) {\delta}_{\mathcal{A}_1}}{{{{f}_{\mathcal{A}_1}^{\intercal}\T(V) {f}_{\mathcal{A}_1}}}}\right|\leq \frac{K^2 {\rm Tr}[\T({V})]+t_0 K^2 \sqrt{{\rm Tr}([\T({V})]^2)}}{{{{f}_{\mathcal{A}_1}^{\intercal}\T(V) {f}_{\mathcal{A}_1}}}}.
%$
%The error bound \eqref{eq: inter bound 1 general} implies
%$
%\left|\frac{{\epsilon}_{\mathcal{A}_1}^{\intercal}\T(V) {f}_{\mathcal{A}_1}}{{{{f}_{\mathcal{A}_1}^{\intercal}\T(V) {f}_{\mathcal{A}_1}}}}\right|\leq \frac{t_0 K\sqrt{{f}_{\mathcal{A}_1}^{\intercal}[\T(V)]^2 {f}_{\mathcal{A}_1}}}{{{{f}_{\mathcal{A}_1}^{\intercal}\T(V) {f}_{\mathcal{A}_1}}}}.
%$
%, we establish \eqref{eq: init rate}.

%%%%%%%%%%%%%%%%%%%%%%%%%%%%%%%%%%%%%%%%%%%%%%%%%%%%%%%%%%%%%%%%%%%%%%%%%%%%%%%%%%%%
\subsection{Proof of Theorem  \ref{thm: limiting TSRF hetero}}%\eqref{eq: limiting final RF} and \eqref{eq: error bound}}
\label{sec: limiting TSRF proof hetero}
%%%%%%%%%%%%%%%%%%%%%%%%%%%%%%%%%%%%%%%%%%%%%%%%%%%%%%%%%%%%%%%%%%%%%%%%%%%%%%%%%%%%
In the following, we shall prove Theorem \ref{thm: bias correction effect}, which implies Theorem \ref{thm: limiting TSRF hetero} together with Condition (R2-I).

\begin{Theorem} Suppose that the same conditions of Proposition \ref{prop: initial consistency} and 
\eqref{eq: linder condition} hold.
Then $\widehat{\beta}(V)$ defined in \eqref{eq: ML corrected hetero}
 satisfies 
\begin{equation}
\frac{1}{{\rm SE}(V)}\left(\widehat{\beta}(V)-\beta\right)={\mathcal{G}}(V)+\mathcal{E}(V) \quad \text{with}\quad {\rm SE}(V)=\frac{\sqrt{\sum_{i=1}^{n_1}\sigma^2_i[\T(V) {f}_{\mathcal{A}_1}]_i^2}
}{{f}_{\mathcal{A}_1}^{\intercal}\T(V) {f}_{\mathcal{A}_1}},
\label{eq: limiting RF-corrected}
\end{equation}
where ${\mathcal{G}}(V)\cid N(0,1)$ and there exist positive constants $t_0>0$ and $C>0$ such that 
%$
%\liminf_{n\rightarrow\infty}\PP\left(\left|\mathcal{E}(V)\right|\leq C U_n(V)\right)\geq 1-\exp(-t_0^2),
%$
%with 
{\small
\begin{equation}
\liminf_{n\rightarrow\infty}\PP\left(\left|\mathcal{E}(V)\right|\leq\frac{\sqrt{\log n}\cdot\eta_n(V)\cdot {\rm Tr}[\T(V)]+t_0\sqrt{{\rm Tr}([\T({V})]^2)}+\|R(V)\|_2}{\sqrt{{f}_{\mathcal{A}_1}[\T(V)]^2 {f}_{\mathcal{A}_1}}}\right)\geq 1-\exp(-t_0^2),
\label{eq: error bound}
\end{equation}
}
with $\eta_n(V)$ defined in \eqref{eq: approximation rate hetero}.
\label{thm: bias correction effect}
\end{Theorem}

We start with the following error decomposition, 
\begin{equation}
\begin{aligned}
\betaHe(V)-\beta%=\widehat{\beta}_{\rm init}(V)-\frac{\sum_{i=1}^{n_1}[\T(V)]_{ii}  \widehat{\delta}_i[\widehat{\epsilon}(V)]_i}{{{D}_{\mathcal{A}_1}^{\intercal}\T(V) {D}_{\mathcal{A}_1}}}-\beta\\
%&=\widehat{\beta}_{\rm init}(V)-\beta-\frac{}{{{D}_{\mathcal{A}_1}^{\intercal}\T(V) {D}_{\mathcal{A}_1}}}\\
%&=\frac{{\epsilon}_{\mathcal{A}_1}^{\intercal}\T(V) {f}_{\mathcal{A}_1}+R_{\mathcal{A}_1}(V)^{\intercal}\T(V) {D}_{\mathcal{A}_1}}{{D}_{\mathcal{A}_1}^{\intercal}\T(V) {D}_{\mathcal{A}_1}}-\frac{\sum_{i=1}^{n_1}[\T(V)]_{ii}  \widehat{\delta}_i[\widehat{\epsilon}(V)]_i-\delta_{\mathcal{A}_1}^{\intercal}\T(V)\epsilon_{\mathcal{A}_1}}{{D}_{\mathcal{A}_1}^{\intercal}\T(V) {D}_{\mathcal{A}_1}}.
=\frac{{\epsilon}_{\mathcal{A}_1}^{\intercal}\T(V) {f}_{\mathcal{A}_1}+R_{\mathcal{A}_1}(V)^{\intercal}\T(V) {D}_{\mathcal{A}_1}}{{D}_{\mathcal{A}_1}^{\intercal}\T(V) {D}_{\mathcal{A}_1}}+\frac{\Err_1+\Err_2}{{D}_{\mathcal{A}_1}^{\intercal}\T(V) {D}_{\mathcal{A}_1}},
\end{aligned}
\label{eq: key decomp hetero}
\end{equation}
with
$
\Err_1=\sum_{1\leq i\neq j\leq n_1}^{n_1}[\T(V)]_{ij}  \delta_{i}\epsilon_j$ and 
$
\Err_2=\sum_{i=1}^{n_1}[\T(V)]_{ii}  (f_{i}-\widehat{f}_{i})\left(\epsilon_i+[\widehat{\epsilon}(V)]_i-\epsilon_i\right)+\sum_{i=1}^{n_1}[\T(V)]_{ii}  \delta_{i}\left([\widehat{\epsilon}(V)]_i-\epsilon_i\right).
$  Define 
$${\mathcal{G}}(V)=\frac{1}{{\rm SE}(V)}\frac{{\epsilon}_{\mathcal{A}_1}^{\intercal}\T(V) {f}_{\mathcal{A}_1}}{{D}_{\mathcal{A}_1}^{\intercal}\T(V) {D}_{\mathcal{A}_1}},
\quad
\mathcal{E}(V)=\frac{1}{{\rm SE}(V)}\frac{R_{\mathcal{A}_1}(V)^{\intercal}\T(V) {D}_{\mathcal{A}_1}-\Err_1-\Err_2}{{D}_{\mathcal{A}_1}^{\intercal}\T(V) {D}_{\mathcal{A}_1}}.
$$
Then the decomposition \eqref{eq: key decomp hetero} implies \eqref{eq: limiting RF-corrected}. 
We apply \eqref{eq: finite-sample strength} together with the {generalized} IV strength condition {\rm (R2)} and establish 
$
\frac{D^{\intercal}_{\mathcal{A}_1}\T(V) D_{\mathcal{A}_1}}{f^{\intercal}_{\mathcal{A}_1}\T(V) f_{\mathcal{A}_1}}\cip 1.
$
%Under the homoscedastic error assumption, we have 
%$${{\rm Var}({\epsilon}_{\mathcal{A}_1}^{\intercal}\T(V) {f}_{\mathcal{A}_1}\mid \mathcal{O})}=\sigma^2 {f}_{\mathcal{A}_1}^{\intercal}[\T(V)]^2 {f}_{\mathcal{A}_1}.$$
Condition \eqref{eq: linder condition} implies the Linderberg condition. Hence, we establish 
$\mathcal{G}(V)\mid \mathcal{O}\cid N(0,1),$
and $\mathcal{G}(V)\cid N(0,1).$

%We now provide the proof of the upper bound for $\left|\mathcal{E}(V)\right|$ in \eqref{eq: error bound}. 
Since $\E(\Err_1\mid \mathcal{O})=0$, we apply the similar argument as in \eqref{eq: inter bound 1 general} and obtain that, conditioning on $\mathcal{O}$, with probability larger than $1-\exp\left(-c t_0^2\right)$ for some constants $c>0$ and $t_0>0,$
$
\left|\Err_1\right|\leq t_0 K^2 \sqrt{\sum_{1\leq i\neq j\leq n_1} [\T(V)]_{ij}^2} \leq t_0 K^2\sqrt{{\rm Tr}([\T({V})]^2)}.
$
Then we establish 
{\small \begin{equation}
\begin{aligned}
\PP\left(\left|\Err_1\right|\geq t_0 K^2\sqrt{{\rm Tr}([\T({V})]^2)}\right)&=\E\left[\PP\left(\left|\Err_1\right|\geq t_0 K^2\sqrt{{\rm Tr}([\T({V})]^2)}\mid \mathcal{O}\right)\right]\leq \exp\left(-c t_0^2\right).
\end{aligned}
\label{eq: inter bound 1 imp}
\end{equation}}
We introduce the following Lemma to control $\Err_2,$ whose proof is presented in Section \ref{sec: lemma 6 proof} in the supplement.

\begin{Lemma}
If Condition {\rm (R2)} holds, then with probability larger than $1-n_1^{-c},$ \begin{equation*}
\begin{aligned}
\max_{1\leq i\leq n_1}\left|[\widehat{\epsilon}(V)]_i-\epsilon_i\right|%&\leq \max_{1\leq i\leq n_1}\{|D_i|,\|W_i\|_2, \|V_i\|_2\}\cdot \sqrt{k}\left(\|R(V)\|_{\infty}+|\beta-\widehat{\beta}_{\rm init}(V)|+\frac{1}{\sqrt{n}}\right)\\
\leq C \sqrt{\log n}\left(\|R(V)\|_{\infty}+|\beta-\widehat{\beta}_{\rm init}(V)|+\frac{\log n}{\sqrt{n}}\right).
\end{aligned}
\end{equation*}
%where $C>0$ and $c>0$ are positive constants independent of $n$ and $p.$
\label{lem: standard OLS lemma}
\end{Lemma}

\vspace{-7mm}

The sub-gaussianity of $\epsilon_i$ implies that, with probability larger than $1-n_1^{-c}$ for some positive constant $c>0,$  $\max_{1\leq i\leq n_1} |\epsilon_i|+\max_{1\leq i\leq n_1} |\delta_i|\leq C\sqrt{\log n}$ for some positive constant $C>0.$ {Since $\T(V)$ is positive definite, we have $[\T(V)]_{ii}\geq 0$ for any $1\leq i\leq n_1.$} We have 
{\small
\begin{equation}
\left|\Err_2\right|\lesssim \sum_{i=1}^{n_1}[\T(V)]_{ii} \left[ |f_{i}-\widehat{f}_{i}|\left(\sqrt{\log n}+\left|[\widehat{\epsilon}(V)]_i-\epsilon_i\right|\right)+ \sqrt{\log n}\left|[\widehat{\epsilon}(V)]_i-\epsilon_i\right|\right].
\label{eq: second approximation error}
\end{equation}}
By Lemma \ref{lem: standard OLS lemma}, we have $
|\Err_2|\lesssim \sqrt{\log n} \cdot \eta_n(V) \cdot {\rm Tr}[\T(V)]
$
with $\eta_n(V)$ defined in \eqref{eq: approximation rate hetero}.
Together with \eqref{eq: inter bound 1 imp}, we establish $\liminf_{n\rightarrow\infty}\PP\left(\left|\mathcal{E}(V)\right|\leq C U_n(V)\right)\geq 1-\exp(-t_0^2).$ 

%%%%%%%%%%%%%%%%%%%%%%%%%%%%%%%%%%%%%%%%%%%%%%%%%%%%%%%%%%%%
\subsection{Proofs of Theorems \ref{thm: comparison RF} and \ref{thm: power RF}}
\label{sec: comparison two}
%%%%%%%%%%%%%%%%%%%%%%%%%%%%%%%%%%%%%%%%%%%%%%%%%%%%%%%%%%%%
By conditional sub-gaussianity and ${\rm Var}(\delta_i\mid Z_i,X_i)\geq c,$ we have 
$c\leq {\rm Var}(\delta_i\mid Z_i,X_i)\leq C.$ Hence, we have 
$
\conpara\asymp {{f}_{\mathcal{A}_1}^{\intercal}\T(V) {f}_{\mathcal{A}_1}}.
$
With $H(V_{q},V_{q'})$ defined in \eqref{eq: difference variance}, we have 
\begin{equation}
\frac{\widehat{\beta}(V_q)-\widehat{\beta}(V_{q'})}{\sqrt{H(V_{q},V_{q'})
}}=\mathcal{G}_n(V_{q}, V_{q'})+\mathcal{L}_n(V_{q}, V_{q'})
+\frac{1}{\sqrt{H(V_q,V_{q'})}}\left(\widetilde{\mathcal{E}}(V_{q})-\widetilde{\mathcal{E}}(V_{q'})\right),
\label{eq: single diff decomp}
\end{equation}
where $\mathcal{L}_n(V_{q}, V_{q'})$ is defined in \eqref{eq: sig difference RF},
\begin{equation}
\mathcal{G}_n(V_{q}, V_{q'})=\frac{1}{\sqrt{H(V_q,V_{q'})}}\left(\frac{{f}^{\intercal}_{\mathcal{A}_1} \T(V_q)\epsilon_{\mathcal{A}_1}}{{D}^{\intercal}_{\mathcal{A}_1} \T(V_q)D_{\mathcal{A}_1}}-\frac{{f}^{\intercal}_{\mathcal{A}_1} \T(V_{q'})\epsilon_{\mathcal{A}_1}}{{D}^{\intercal}_{\mathcal{A}_1} \T(V_{q'})D_{\mathcal{A}_1}}\right),
\label{eq: main term RF}
\end{equation}
$$\widetilde{\mathcal{E}}(V_{q})=\frac{\sum_{i=1}^{n_1}[\T(V_q)]_{ii}  \widehat{\delta}_i[\widehat{\epsilon}(V_{Q_{\max}})]_i-\delta_{\mathcal{A}_1}^{\intercal}\T(V_q)\epsilon_{\mathcal{A}_1}}{{D}_{\mathcal{A}_1}^{\intercal}\T(V_q) {D}_{\mathcal{A}_1}}.$$

The remaining proof relies on the following lemma, whose proof can be found at Section \ref{sec: lemma 4 proof} in the supplement.
\begin{Lemma} Suppose that Conditions {\rm (R1)}, {\rm (R2)}, and {\rm (R3)} hold, then 
\begin{equation*}
\frac{1}{\sqrt{H(V_q,V_{q'})}}\left|\widetilde{\mathcal{E}}(V_{q})-\widetilde{\mathcal{E}}(V_{q'})\right|\cip 0, \quad \text{and} \quad \mathcal{G}_n(V_{q}, V_{q'}) \cid N(0,1).
\label{eq: small approximation error}
\end{equation*}
\label{lem: diff key terms}
\end{Lemma}
\vspace{-6mm}
%Note that 
%$
%\left|\frac{D_{\mathcal{A}_1}^{\intercal} \T(V_{q}) [R(V_{q})]_{\mathcal{A}_1} }{{D}^{\intercal}_{\mathcal{A}_1} \T_{\rm RF}(V_{q})D_{\mathcal{A}_1}}\right|\leq \frac{\|[R(V_{q})]_{\mathcal{A}_1}\|_2}{\sqrt{{D}^{\intercal}_{\mathcal{A}_1} \T_{\rm RF}(V_{q})D_{\mathcal{A}_1}}}.
%$
By applying \eqref{eq: finite-sample strength}, we establish that, conditioning on $\mathcal{O}$, with probability larger than $1-\exp\left(-c\tau_n^{1/2-c_0}\right),$
$
\left|\mathcal{L}_n(V_{q}, V_{q'})\right|\lesssim \frac{1}{\sqrt{H(V_{q},V_{q'})}}\left(\frac{\|[R(V_{q})]_{\mathcal{A}_1}\|_2}{\sqrt{\mu(V_{q})}}+\frac{\|[R(V_{q'})]_{\mathcal{A}_1}\|_2}{\sqrt{\mu(V_{q'})}}\right).
$
%where $\mu(V_{q})$ is defined in \eqref{eq: IV strength hetero}. 
Then we apply the above inequality and the condition $\sqrt{H(V_q, V_{q'})} \gg \max_{V\in\{V_{q}, V_{q'}\}}{\|R(V)\|_2}/{\sqrt{\mu(V)}}$ and establish that 
$
\left|\mathcal{L}_n(V_{q}, V_{q'})\right|\cip 0.$
Together with the decomposition \eqref{eq: single diff decomp}, and Lemma \ref{lem: diff key terms}, we establish the asymptotic limiting distribution of $\widehat{\beta}(V_q)-\widehat{\beta}(V_{q'})$
in Theorem \ref{thm: comparison RF}.

 With the decomposition \eqref{eq: single diff decomp}, we establish \eqref{eq: asymp power RF} in Theorem \ref{thm: power RF} by the following Lemma \ref{lem: diff key terms} and applying the condition $\mathcal{L}_n(V_{q}, V_{q'})\cip \mathcal{L}^*(V_{q}, V_{q'})$ and $\widehat{H}(V_q,V_{q'})\cip {H}(V_q,V_{q'}).$

%%%%%%%%%%%%%%%%%%%%%%%%%%%%%%%%%%%%%%%%%%%%%%%%%%%%%%%%%%%%%%%%%%%%%%%%%%
\subsection{Proof of Theorem \ref{thm: post-selection RF}}
\label{sec: multiple selection}
In the following, we shall prove 
\begin{equation}
\liminf_{n\rightarrow \infty} \PP\left[\beta \in {\rm CI}(V_{\widehat{q}})\right]\geq 1-\alpha-2\alpha_0 \quad \text{with} \quad \widehat{q}=\widehat{q}_c \;\; \text{or} \;\; \widehat{q}_r,
\label{eq: post-selection RF}
\end{equation}
Recall that the test $\mathcal{C}(V_q)$ is defined in \eqref{eq: lawyer test RF}
and note that 
\begin{equation}
\left\{\widehat{q}_c=q^{*}\right\}=\left\{Q_{\max}\geq q^{*}\right\}\cap \left(\cap_{q=0}^{q^{*}-1}\left\{\mathcal{C}(V_q)=1\right\}\right)\cap \left\{\mathcal{C}(V_{q^{*}})=0\right\}.
\label{eq: prop decomposition RF}
\end{equation}
%where $\mathcal{C}(V_q)$ is defined in \eqref{eq: lawyer test RF}. 
Define the events 
$$
\mathcal{B}_1=\left\{\max_{0\leq q< q'\leq Q_{\max}}\left|\mathcal{G}_n(V_{q},V_{q'})\right|\leq \rho(\alpha_0)\right\} \quad \text{and}\quad \mathcal{B}_2=\left\{\left|\widehat{\rho}/\rho(\alpha_0)-1\right|\leq \tau_0\right\},
$$
$$\mathcal{B}_3=\left\{\max_{0\leq q<q'\leq Q_{\max}}\frac{1}{\sqrt{H(V_q,V_{q'})}}\left|\widetilde{\mathcal{E}}(V_{q})-\widetilde{\mathcal{E}}(V_{q'})\right|\leq \tau_1\rho(\alpha_0)\right\}.$$
where $\rho(\alpha_0)$ is defined in \eqref{eq: quantile}, $\widehat{\rho}$ is defined in \eqref{eq: sample quantile} with $\T_{\rm RF}$ replaced by $\T$, and $\mathcal{G}_n(V_{q},V_{q'})$ is defined in \eqref{eq: main term RF}.
By the definition of $\rho(\alpha_0)$ in \eqref{eq: quantile} and the following \eqref{eq: diff normal approximation}, we control the probability of the event 
$
\lim_{n\rightarrow\infty}\mathbf{P}\left(\mathcal{B}_1\right)=1-2\alpha_0.
$
By the following \eqref{eq: small approximation error} and $\widehat{\rho}/\rho(\alpha_0)\cip 1$, we establish that, for any positive constants $\tau_0>0$ and $\tau_1>0,$
$\liminf_{n\rightarrow \infty}\mathbf{P}\left(\mathcal{B}_2\cap \mathcal{B}_3\right)=1.$
Combing the above two equalities, we have 
\begin{equation}
\liminf_{n\rightarrow \infty}\mathbf{P}\left(\mathcal{B}_1\cap \mathcal{B}_2\cap \mathcal{B}_3\right)\geq 1-2\alpha_0.
\label{eq: high probability}
\end{equation}
% defined in 
%\eqref{eq: sample quantile} satisfies $\widehat{\rho}\cip \rho(\alpha_0)$ with $\rho(\alpha_0)$ defined 

For any $0\leq q \leq q^{*}-1,$ the condition (R4) implies that there exists some $q+1\leq q'\leq q^{*}$ such that 
$
\left|\mathcal{L}_n(V_{q}, V_{q'})\right|\geq A \rho(\alpha_0).
$
On the event $\mathcal{B}_1\cap\mathcal{B}_2\cap\mathcal{B}_3$,  we apply the expression \eqref{eq: single diff decomp} and obtain 
\begin{equation}
\left|[\widehat{\beta}(V_q)-\widehat{\beta}(V_{q'})]/{\sqrt{H(V_{q},V_{q'})
}}\right|\geq A \rho(\alpha_0)-\rho(\alpha_0) 
-\tau_1\rho(\alpha_0)\geq (A-1-\tau_1)(1-\tau_0) \widehat{\rho}.
\label{eq: implication 1}
\end{equation}

For any $q^{*}\leq q'\leq Q_{\max},$ we have $R(V_{q'})=0.$ 
Then on the event $\mathcal{B}_1\cap\mathcal{B}_2\cap\mathcal{B}_3$, we apply the expression \eqref{eq: single diff decomp} and obtain that, 
\begin{equation*}
\left|[\widehat{\beta}(V_{Q*})-\widehat{\beta}(V_{q'})]/{\sqrt{H(V_{Q*},V_{q'})
}}\right|\leq \rho(\alpha_0) 
+\tau_1\rho(\alpha_0)\leq (1+\tau_1)(1+\tau_0)\widehat{\rho}\leq 1.01 \widehat{\rho}.
\end{equation*}
Together with \eqref{eq: implication 1}, the event $\mathcal{B}_1\cap\mathcal{B}_2\cap\mathcal{B}_3$ implies  $\cap_{q=0}^{q^{*}-1}\left\{\mathcal{C}(V_q)=1\right\}\cap \left\{\mathcal{C}(V_{q^{*}})=0\right\}.$
By \eqref{eq: high probability}, we establish 
$\liminf_{n\rightarrow \infty}\mathbf{P}\left[\left(\cap_{q=0}^{q^{*}-1}\left\{\mathcal{C}(V_q)=1\right\}\right)\cap \left\{\mathcal{C}(V_{q^{*}})=0\right\}\right]\geq 1-2\alpha_0.
$
Together with \eqref{eq: prop decomposition RF},  we have
$
\liminf_{n\rightarrow \infty} \mathbf{P}\left(\widehat{q}_c\neq q^{*}\right)\leq 2\alpha_0.
$
To control the coverage probability, we decompose $\mathbf{P}\left(\frac{1}{{\rm SE}(V_{\widehat{q}_c})}\left|\widehat{\beta}(V_{\widehat{q}_c})-\beta\right|\geq z_{\alpha/2}\right)$ as
\begin{equation}
\begin{aligned}
&\mathbf{P}\left(\left\{\frac{1}{{\rm SE}(V_{\widehat{q}_c})}\left|\widehat{\beta}(V_{\widehat{q}_c})-\beta\right|\geq z_{\alpha/2}\right\}\cap \left\{\widehat{q}_c= q^{*}\right\}\right)\\
&+\mathbf{P}\left(\left\{\frac{1}{{\rm SE}(V_{\widehat{q}_c})}\left|\widehat{\beta}(V_{\widehat{q}_c})-\beta\right|\geq z_{\alpha/2}\right\}\cap \left\{\widehat{q}_c\neq q^{*}\right\}\right).
\end{aligned}
\label{eq: type I error decomp}
\end{equation}
Note that 
{\small
\begin{equation*}
\begin{aligned}
\mathbf{P}\left(\left\{\frac{1}{{\rm SE}(V_{\widehat{q}_c})}\left|\widehat{\beta}(V_{\widehat{q}_c})-\beta\right|\geq z_{\alpha/2}\right\}\cap \left\{\widehat{q}_c= q^{*}\right\}\right)
\leq \mathbf{P}\left(\left\{\frac{1}{{\rm SE}(V_{q^{*}})}\left|\widehat{\beta}(V_{q^{*}})-\beta\right|\geq z_{\alpha/2}\right\}\right)\leq \alpha,
\end{aligned}
\label{eq: type I error a}
\end{equation*}
}
and 
$
\mathbf{P}\left(\left\{\frac{1}{{\rm SE}(V_{\widehat{q}_c})}\left|\widehat{\beta}(V_{\widehat{q}_c})-\beta\right|\geq z_{\alpha/2}\right\}\cap \left\{\widehat{q}_c\neq q^{*}\right\}\right)
\leq \mathbf{P}\left(\widehat{q}_c\neq q^{*}\right)\leq 2\alpha_0.
$
By the decomposition \eqref{eq: type I error decomp}, we combine the above two inequalities and establish 
$$\mathbf{P}\left(\frac{1}{{\rm SE}(V_{\widehat{q}_c})}\left|\widehat{\beta}(V_{\widehat{q}_c})-\beta\right|\geq z_{\alpha/2}\right)\leq \alpha+2\alpha_0,$$
which implies \eqref{eq: post-selection RF} with $\widehat{q}=\widehat{q}_c.$ By the definition $\widehat{q}_r=\min\{\widehat{q}_c+1,Q_{\max}\},$ we apply a similar argument and establish \eqref{eq: post-selection RF} with $\widehat{q}=\widehat{q}_r.$ 

%%%%%%%%%%%%%%%%%%%%%%%%%%%%%%%%%%%%%%%%%%%%%%%%%%%%%%%%%%%%%%%%%%%%%%%%%%%%%%%%%%%%%
\subsection{Proof of Theorem \ref{thm: homo bias correction effect}}
\label{sec: correlation consistency}
%%%%%%%%%%%%%%%%%%%%%%%%%%%%%%%%%%%%%%%%%%%%%%%%%%%%%%%%%%%%%%%%%%%%%%%%%%%%%%%%%%%%
If ${\rm Cov}(\epsilon_i,\delta_i\mid Z_i, X_i)={\rm Cov}(\epsilon_i,\delta_i)$, then \eqref{eq: inter bound 1 general} further implies
\begin{equation}
\left|{{\epsilon}_{\mathcal{A}_1}^{\intercal}\T(V) {\delta}_{\mathcal{A}_1}}-{\rm Cov}(\epsilon_i,\delta_i)\cdot{\rm Tr}[\T_{\rm RF}({V})]\right|\leq t_0 K^2\sqrt{{\rm Tr}([\T_{\rm RF}({V})]^2)}.
\label{eq: inter bound 1}
\end{equation}
\paragraph*{Proof of \eqref{eq: correlation accuracy}.} We decompose the error $\widehat{\rm Cov}(\delta_i,\epsilon_i)-{\rm Cov}(\delta_i,\epsilon_i)$ as,
\begin{equation}
\begin{aligned}
&\frac{1}{n_1-r} (f_{\mathcal{A}_1}-\widehat{f}_{\mathcal{A}_1}+\delta_{\mathcal{A}_1})^{\intercal}P^{\perp}_{V_{\mathcal{A}_1}}[\epsilon_{\mathcal{A}_1}+D_{\mathcal{A}_1}(\beta-\widehat{\beta}_{\rm init}(V))+R_{\mathcal{A}_1}(V)]-{\rm Cov}(\delta_i,\epsilon_i)\\
&=T_1+T_2+T_3,
\end{aligned}
\label{eq: decomp est C}
\end{equation}
where $T_1=\frac{1}{n_1-r} \left[(\delta_{\mathcal{A}_1})^{\intercal}P^{\perp}_{V_{\mathcal{A}_1}^{\intercal}}\epsilon_{\mathcal{A}_1}-(n_1-r){\rm Cov}(\delta_i,\epsilon_i)\right],$ and
\begin{equation*}
\begin{aligned}
T_2&=\frac{1}{n_1-r} (\delta_{\mathcal{A}_1})^{\intercal}P^{\perp}_{V_{\mathcal{A}_1}^{\intercal}}\left[D_{\mathcal{A}_1}(\beta-\widehat{\beta}_{\rm init}(V))+R_{\mathcal{A}_1}(V)\right],\\
T_3&=\frac{1}{n_1-r} (f_{\mathcal{A}_1}-\widehat{f}_{\mathcal{A}_1})^{\intercal}P^{\perp}_{V_{\mathcal{A}_1}^{\intercal}}[\epsilon_{\mathcal{A}_1}+D_{\mathcal{A}_1}(\beta-\widehat{\beta}_{\rm init}(V))+R_{\mathcal{A}_1}(V)].
\end{aligned}
\end{equation*}
We apply \eqref{eq: HW-bound imp} with $A=P^{\perp}_{V_{\mathcal{A}_1}^{\intercal}}$ and $t=t_0 K^2 \sqrt{n_1-r}$ for some $0<t_0\leq  \sqrt{n_1-r},$ and establish  %we establish \eqref{eq: inter bound 1} and
%2\exp\left(-{c}\min\{\frac{\log n}{K^4},\frac{\sqrt{\log n}}{K^2}\}\right)$ for some $c>0,$
\begin{equation*}
\begin{aligned}
\PP\left(\left|\epsilon_{\mathcal{A}_1}^{\intercal}P^{\perp}_{V_{\mathcal{A}_1}^{\intercal}} \delta_{\mathcal{A}_1}- {\rm Cov}(\epsilon_i,\delta_i)\cdot {\rm Tr}\left(P^{\perp}_{V_{\mathcal{A}_1}^{\intercal}}\right)\right| \geq t_0 K^2 \sqrt{n_1-r}\mid \mathcal{O}\right)\leq 6\exp\left(-c t_0^2\right).
\end{aligned}
\end{equation*}
By choosing $t_0=\log (n_1-r),$ we establish that, with probability larger than $1-(n_1-r)^{-c}$ for some positive constant $c>0,$ then 
\begin{equation}
\left|T_1\right|\lesssim \sqrt{{\log (n_1-r)}/{(n_1-r)}}.
\label{eq: T1 bound}
\end{equation}
We apply \eqref{eq: HW-bound imp} with $A=P^{\perp}_{V_{\mathcal{A}_1}^{\intercal}}$ and $t=t_0 K^2 \sqrt{n_1-r}$ for some $0<t_0\leq  \sqrt{n_1-r},$ 
\begin{equation*}
\begin{aligned}
\PP\left(\left|\epsilon_{\mathcal{A}_1}^{\intercal}P^{\perp}_{V_{\mathcal{A}_1}^{\intercal}} D_{\mathcal{A}_1}- {\rm Cov}(\epsilon_i,D_i)\cdot {\rm Tr}\left(P^{\perp}_{V_{\mathcal{A}_1}^{\intercal}}\right)\right| \geq t_0 K^2 \sqrt{n_1-r}\mid \mathcal{O}\right)\leq 6\exp\left(-c t_0^2\right).
\end{aligned}
\end{equation*}
The above concentration bound implies 
\begin{equation}
\PP\left(\frac{1}{n_1-r}\left|\epsilon_{\mathcal{A}_1}^{\intercal}P^{\perp}_{V_{\mathcal{A}_1}^{\intercal}} D_{\mathcal{A}_1}\right| \geq C\left(1+\sqrt{\frac{\log (n_1-r)}{n_1-r}}\right) \mid \mathcal{O}\right)\leq 6(n_1-r)^{-c}.
\label{eq: inter concentration}
\end{equation}
Hence, we establish that, with probability larger than $1-(n_1-r)^{-c},$%-\exp\left(-c[\tau_n]^{1/2-c_0}\right),$
\begin{equation}
\left|T_2\right|\lesssim \left|\beta-\widehat{\beta}_{\rm init}(V)\right|+{\|R(V)\|_2}/{\sqrt{n_1-r}},%\lesssim \frac{{\rm Tr}[\T_{\rm RF}({V})]}{\mu(V)} +C\frac{1}{(\mu(V))^{c_0}}+\frac{\|R(V)\|_2}{\sqrt{n_1-r}},
\label{eq: T2 bound}
\end{equation}
where the last inequality follows from \eqref{eq: init rate}. Regarding $T_3,$ we have 
{\small
\begin{equation}
\begin{aligned}
&|T_3|=\left|\frac{1}{n_1-r}(f_{\mathcal{A}_1}-\widehat{f}_{\mathcal{A}_1})^{\intercal}P^{\perp}_{V_{\mathcal{A}_1}^{\intercal}}[\epsilon_{\mathcal{A}_1}+D_{\mathcal{A}_1}(\beta-\widehat{\beta}_{\rm init}(V))+R_{\mathcal{A}_1}(V)]\right|\\
&\lesssim \frac{\|P^{\perp}_{V_{\mathcal{A}_1}^{\intercal}}(f_{\mathcal{A}_1}-\widehat{f}_{\mathcal{A}_1})\|_{2}}{\sqrt{n_1-r}} \frac{\|P^{\perp}_{V_{\mathcal{A}_1}^{\intercal}}\epsilon_{\mathcal{A}_1}\|_2+\|P^{\perp}_{V_{\mathcal{A}_1}^{\intercal}}D_{\mathcal{A}_1}\|_2\cdot|\beta-\widehat{\beta}_{\rm init}(V)|+\|R_{\mathcal{A}_1}(V)\|_2}{\sqrt{n_1-r}}.
\end{aligned}
\label{eq: CS inter bound}
\end{equation}}
By a similar argument as in \eqref{eq: inter concentration}, we establish that, with probability larger than $1-(n_1-r)^{-c},$
$\frac{1}{\sqrt{n_1}}\|P^{\perp}_{V_{\mathcal{A}_1}^{\intercal}}\epsilon_{\mathcal{A}_1}\|_2+\frac{1}{\sqrt{n_1}}\|P^{\perp}_{V_{\mathcal{A}_1}^{\intercal}}D_{\mathcal{A}_1}\|_2\leq C,$
for some positive constant $C>0.$ %Consequently, $$\frac{\|P^{\perp}_{V_{\mathcal{A}_1}^{\intercal}}\epsilon_{\mathcal{A}_1}\|_2+\|P^{\perp}_{V_{\mathcal{A}_1}^{\intercal}}D_{\mathcal{A}_1}\|_2\cdot|\beta-\widehat{\beta}_{\rm init}(V)|+\|R(V)\|_2}{\sqrt{n_1-r}}\lesssim 1+|\beta-\widehat{\beta}_{\rm init}(V)|+\frac{\|R(V)\|_2}{\sqrt{n_1-r}}.$$
Together with \eqref{eq: CS inter bound}, we establish that, with probability larger than $1-(n_1-r)^{-c},$
\begin{equation*}
\left|T_3\right|\lesssim \frac{\|P^{\perp}_{V_{\mathcal{A}_1}^{\intercal}}(f_{\mathcal{A}_1}-\widehat{f}_{\mathcal{A}_1})\|_{2}}{\sqrt{n_1-r}}\cdot  \left(1+|\beta-\widehat{\beta}_{\rm init}(V)|+\frac{\|R(V)\|_2}{\sqrt{n_1-r}}\right).
\end{equation*}
Together with \eqref{eq: decomp est C} and the upper bounds \eqref{eq: T1 bound} and \eqref{eq: T2 bound}, we establish \eqref{eq: correlation accuracy}.
%{eq: approximation rate}
 
 %Proposition \ref{thm: bias correction effect}. 
%We can further upper bound $\eta_n$ defined in \eqref{eq: approximation rate} by 

%Together with \eqref{eq: init rate}, we establish that, with probability lager than $1-\exp\left(-c\tau_n^{1/2-c_0}\right)$ for some positive constants $c>0$ and $0<c_0<1/2,$
%\begin{equation}
%\begin{aligned}
%\eta_n&\leq \left(\frac{{\rm Tr}[\T_{\rm RF}({V})]}{\mu(V)} +C\frac{1}{(\mu(V))^{c_0}}+\frac{\|R(V)\|_2}{\sqrt{\mu(V)}}\right)\left(1+\frac{\|f_{\mathcal{A}_1}-\widehat{f}_{\mathcal{A}_1}\|_{2}}{\sqrt{n}}\right)\\
%&+\frac{\|f_{\mathcal{A}_1}-\widehat{f}_{\mathcal{A}_1}\|_{2}}{\sqrt{n}}\left(1+\frac{\|R(V)\|_2}{\sqrt{n}}\right)+\frac{\|R(V)\|_2+\sqrt{\log n}}{\sqrt{n}}.
%\end{aligned}
%\label{eq: approximation rate upper}
%\end{equation}

\paragraph*{Proof of \eqref{eq: error bound homo}.} By \eqref{eq: init decomposition}, we obtain the following decomposition for $\betaRF(V)$ in \eqref{eq: ML corrected hetero},
\begin{equation}
\begin{aligned}
&\betaHe(V)-\beta=\frac{{\epsilon}_{\mathcal{A}_1}^{\intercal}\T(V) {f}_{\mathcal{A}_1}-[\widehat{\rm Cov}(\delta_i,\epsilon_i)-{\rm Cov}(\delta_i,\epsilon_i)]{\rm Tr}[\T(V)]}{{D}_{\mathcal{A}_1}^{\intercal}\T(V) {D}_{\mathcal{A}_1}}\\
&+\frac{{\epsilon}_{\mathcal{A}_1}^{\intercal}\T(V) {\delta}_{\mathcal{A}_1}-{\rm Cov}(\delta_i,\epsilon_i){\rm Tr}[\T(V)]}{{D}_{\mathcal{A}_1}^{\intercal}\T(V) {D}_{\mathcal{A}_1}}+\frac{R_{\mathcal{A}_1}(V)^{\intercal}\T(V) {D}_{\mathcal{A}_1}}{{D}_{\mathcal{A}_1}^{\intercal}\T(V) {D}_{\mathcal{A}_1}}.
\end{aligned}
\label{eq: decomp betaRF}
\end{equation}
The above decomposition implies \eqref{eq: limiting RF-corrected} with 
$
\mathcal{G}(V)=\frac{1}{{\rm SE}(V)}\frac{{\epsilon}_{\mathcal{A}_1}^{\intercal}\T(V) {f}_{\mathcal{A}_1}}{{{D}_{\mathcal{A}_1}^{\intercal}\T(V) {D}_{\mathcal{A}_1}}},$
and
\begin{equation*}
\begin{aligned}
&{\rm SE}(V)\cdot\mathcal{E}(V)= \frac{R_{\mathcal{A}_1}(V)^{\intercal}\T(V) {D}_{\mathcal{A}_1}}{{D}_{\mathcal{A}_1}^{\intercal}\T(V) {D}_{\mathcal{A}_1}}\\
&+\frac{[{\rm Cov}(\delta_i,\epsilon_i)-\widehat{\rm Cov}(\delta_i,\epsilon_i)]{\rm Tr}[\T(V)]+{\epsilon}_{\mathcal{A}_1}^{\intercal}\T(V) {\delta}_{\mathcal{A}_1}-{\rm Cov}(\delta_i,\epsilon_i){\rm Tr}[\T(V)]}{{D}_{\mathcal{A}_1}^{\intercal}\T(V) {D}_{\mathcal{A}_1}}.
\end{aligned}
\label{eq: approximation error}
\end{equation*}

We establish $\mathcal{G}(V)\cid N(0,1)$ by applying the same arguments as in Section \ref{sec: limiting TSRF proof hetero} in the main paper. %in \eqref{eq: bottom consistency}. 
%In the following, we analyze the term $\mathcal{E}_1(V)$ and $\mathcal{E}_2(V)$ defined in \eqref{eq: approximation error}.
We establish \eqref{eq: error bound homo} by combining \eqref{eq: correlation accuracy}, \eqref{eq: finite-sample strength}, and \eqref{eq: inter bound 1}.

%%%%%%%%%%%%%%%%%%%%%%%%%%%%%%%%%%%%%%%%%%%%%%%%%%%%%%%%%%%%%%%%%%
%\section{Proofs of Theorem \ref{thm: comparison RF}, Corollary \ref{thm: power RF}, and Theorem \ref{thm: post-selection RF}}
%%%%%%%%%%%%%%%%%%%%%%%%%%%%%%%%%%%%%%%%%%%%%%%%%%%%%%%%%%%%%%%%%%

% $$\frac{1}{\sqrt{H(V_q,V_{q'})}}\left(\frac{D_{\mathcal{A}_1}^{\intercal} \T_{\rm RF}(V_{q}) [R(V_{q})]_{\mathcal{A}_1} }{{D}^{\intercal}_{\mathcal{A}_1} \T_{\rm RF}(V_{q})D_{\mathcal{A}_1}}+\widetilde{\mathcal{E}}(V_{q})
%-\frac{D_{\mathcal{A}_1}^{\intercal} \T_{\rm RF}(V_{q'}) R_{\mathcal{A}_1}(V) }{{D}^{\intercal}_{\mathcal{A}_1} \T_{\rm RF}(V_{q'})D_{\mathcal{A}_1}}-\widetilde{\mathcal{E}}(V_{q'})\right)\cip 0$$

%
%\begin{equation}
%\quad \sqrt{H(V_q, V_{q'})} \gg 
%\frac{{\rm Tr}[\T_{\rm RF}(V_q)-\T_{\rm RF}(V_{q'})]+\T_{\rm RF}(V_{q'})(\mu(V_q)/\mu(V_{q'})-1)}{\mu(V_q)}
%\label{eq: approx error assump 1}
%\end{equation}
%and 

%\begin{equation}
%\sqrt{H(V_q, V_{q'})} \gg \max_{V\in\{V_{q}, V_{q'}\}}\left\{\frac{\|R(V)\|_2}{\sqrt{\mu(V)}}+\frac{1}{\mu(V)}\left(1+\frac{\T(V)}{\sqrt{\mu(V)}}\right)\right\}
%\label{eq: approx error assump 2}
%\end{equation}

%As long as \eqref{eq: signal diff condition} holds,
%which also implies $g^{\intercal} \Ptn g-g^{\intercal} \Pt g\gg M-q,$
%then we have 
%\begin{equation}
%\end{equation}

%%%%%%%%%%%%%%%%%%%%%%%%%%%%%%%%%%%%%%%%%%%%%%%%%
%%%%%%%%%%%%%%%%%%%%%%%%%%%%%%%%%%%%%%%%%%%%%%%%
\section{Proof of Extra Lemmas}
%%%%%%%%%%%%%%%%%%%%%%%%%%%%%%%%%%%%%%%%%%%%%%%%%%%%%%%%%%%%%%%%%%%%%%%%%%%%%%%%%%%%%%%%%%%%%%%%%%%%%%

%%%%%%%%%%%%%%%%%%%%%%%%%%%%%%%%%%%%%%%%%%%%%%%%%%%%%%%%%%%%%%%%%%%%%%%%%%%%%%%%%%%%%%%%%%%%%%%%%%%%%%
\subsection{Proof of Lemma \ref{lem: IV strength 1}}
\label{sec: lemma 5 proof}
%%%%%%%%%%%%%%%%%%%%%%%%%%%%%%%%%%%%%%%%%%%%%%%%%%%%%%%%%%%%%%%%%%%%%%%%%%%%%%%%%%%%%%%%%%%%%%%%%%%%%%
Note that 
$
\E\left[{\epsilon}_{\mathcal{A}_1}^{\intercal}\T(V) {\delta}_{\mathcal{A}_1}
\mid \mathcal{O}\right]={\rm Tr}\left(\T(V)\Lambda\right).
$
We apply \eqref{eq: HW-bound imp} with $A=\T(V)$ and $t=t_0 K^2 \|\T(V)\|_F$ for some $t_0>0$ and establish  
\begin{equation}
\begin{aligned}
&\PP\left(\left|\epsilon_{\mathcal{A}_1}^{\intercal}\T(V) \delta_{\mathcal{A}_1}- {\rm Tr}\left(\T(V)\Lambda\right)\right| \geq t_0 K^2 \|\T(V)\|_F\mid \mathcal{O}\right)\\
& \leq 6\exp\left(-c \min\left\{t_0^2,t_0\frac{\|\T(V)\|_{F}}{\|\T(V)\|_2}\right\}\right)\leq 6\exp\left(-c \min\left\{t_0^2,t_0\right\}\right),
\end{aligned}
\label{eq: correction concentration}
\end{equation}
where the last inequality follows from $\|\T(V)\|_{F}\geq \|\T(V)\|_2$.  The above concentration bound implies \eqref{eq: inter bound 1 general} by taking an expectation with respect to $\mathcal{O}.$

Since
$
\E\left[{\delta}_{\mathcal{A}_1}^{\intercal}\T(V) {\delta}_{\mathcal{A}_1}
\mid \mathcal{O}\right]={\rm Tr} \left(\T(V)\Sigma^{\delta}\right),
$
we apply a similar argument to \eqref{eq: correction concentration} and establish 
\begin{equation}
\begin{aligned}
\PP\left(\left|\delta_{\mathcal{A}_1}^{\intercal}\T(V) \delta_{\mathcal{A}_1}- {\rm Tr}\left(\T(V)\Sigma^{\delta}\right)\right| \geq t_0 K^2 \|\T(V)\|_F\mid \mathcal{O}\right)\leq 2\exp\left(-c \min\left\{t_0^2,t_0\right\}\right).
\end{aligned}
\label{eq: correction concentration 2}
\end{equation}

Note that 
$
\E\left[{\delta}_{\mathcal{A}_1}^{\intercal}\T(V) {f}_{\mathcal{A}_1}
\mid \mathcal{O}\right]=0,
$
and conditioning on $\mathcal{O}$, $\{{\delta}_i\}_{i\in \mathcal{A}_1}$ are independent sub-gaussian random variables. We apply Proposition 5.16 of \citetsupp{vershynin2010introduction} and establish that, with probability larger than $1-\exp(-t_0^2)$ for some positive constant $c>0,$
\begin{equation}
\PP\left(
{\delta}_{\mathcal{A}_1}^{\intercal}\T(V) {f}_{\mathcal{A}_1}\geq C t_0 K \sqrt{{f}_{\mathcal{A}_1}^{\intercal}[\T(V)]^2 {f}_{\mathcal{A}_1}}\mid \mathcal{O}\right)\leq \exp(-ct_0^2).
\label{eq: RF upper 2}
\end{equation}
The above concentration bound implies \eqref{eq: inter bound 1 general} by taking an expectation with respect to $\mathcal{O}.$

By the decomposition
$
{D}_{\mathcal{A}_1}^{\intercal}\T(V) {D}_{\mathcal{A}_1}-{f}_{\mathcal{A}_1}^{\intercal}\T(V) {f}_{\mathcal{A}_1}={\delta}_{\mathcal{A}_1}^{\intercal}\T(V) {f}_{\mathcal{A}_1}+{\delta}_{\mathcal{A}_1}^{\intercal}\T(V) {\delta}_{\mathcal{A}_1},
$
we establish \eqref{eq: bottom consistency RF} by applying the concentration bounds \eqref{eq: correction concentration 2} and \eqref{eq: RF upper 2}.

%%%%%%%%%%%%%%%%%%%%%%%%%%%%%%%%%%%%%%%%%%%%%%%%%%%%%%%%%%%%%%%%%%%%%%%%%%%%%%%%%%%%%%%%%%%%%%%%%%%%%%
\subsection{Proof of Lemma \ref{lem: standard OLS lemma} }
\label{sec: lemma 6 proof}
%%%%%%%%%%%%%%%%%%%%%%%%%%%%%%%%%%%%%%%%%%%%%%%%%%%%%%%%%%%%%%%%%%%%%%%%%%%%%%%%%%%%%%%%%%%%%%%%%%%%%%

%\paragraph{Proof of Lemma \ref{lem: standard OLS lemma}.} 
Under the model $Y_i=D_i\beta+V_i^{\intercal} \pi+R_i+\epsilon_i$, we have 
\begin{equation}
Y_{\mathcal{A}_1}-\widehat{\beta}_{\rm init}(V)D_{\mathcal{A}_1}=V_{\mathcal{A}_1} \pi+R_{{\mathcal{A}_1}}+D_{\mathcal{A}_1} (\beta-\widehat{\beta}_{\rm init}(V))+\epsilon_{\mathcal{A}_1}.
\label{eq: minus beta D}
\end{equation}
The least square estimator of $\pi$ is expressed as,  
$
\widehat{\pi}=(V^{\intercal} V)^{-1}V^{\intercal}\left(Y_{\mathcal{A}_1}-\widehat{\beta}_{\rm init}(V)D_{\mathcal{A}_1}\right).
$
Note that 
\begin{equation*}
\begin{aligned}
&(V^{\intercal} V)^{-1}V^{\intercal}\left(Y_{\mathcal{A}_1}-\widehat{\beta}_{\rm init}(V)D_{\mathcal{A}_1}\right)-\pi\\
&=(V^{\intercal} V)^{-1}V^{\intercal}\left([R(V)]_{{\mathcal{A}_1}}+D_{\mathcal{A}_1} (\beta-\widehat{\beta}_{\rm init}(V))+\epsilon_{\mathcal{A}_1}\right)\\
&=(\sum_{i=1}^{n_1}V_i V_i^{\intercal})^{-1}\sum_{i=1}^{n_1}V_i\left([R(V)]_{i}+D_{i} (\beta-\widehat{\beta}_{\rm init}(V))+\epsilon_{i}\right).
\end{aligned}
\end{equation*}
Note that $\E V_i \epsilon_i=0$ and conditioning on $\mathcal{O}$, $\{{\epsilon}_i\}_{i\in \mathcal{A}_1}$ are independent sub-gaussian random variables. By Proposition 5.16 of \citetsupp{vershynin2010introduction}, there exist positive constants  $C>0$ and $c>0$ such that
$
\PP\left(\frac{1}{n_1}\left|\sum_{i=1}^{n_1}V_{ij}\epsilon_i\right|\geq C t_0 {\sqrt{\sum_{i=1}^{n_1}V_{ij}^2}}/{n_1}\mid \mathcal{O}\right)\leq \exp(-ct_0^2).
$
By Condition {\rm (R1)}, we have $\max_{i,j}V_{ij}^2\leq C\log n$ and apply the union bound and establish 
\begin{equation}
\PP\left(\left\|\sum_{i=1}^{n_1}V_{i}\epsilon_i/n_1\right\|_{\infty}\geq C {\log n_1}/{\sqrt{n_1}}\right)\leq n_1^{-c}.
\label{eq: concentration zero}
\end{equation}
Similarly to \eqref{eq: concentration zero}, we establish 
$
\PP\left(\left\|\frac{1}{n_1}\sum_{i=1}^{n_1}V_{i}\delta_i\right\|_{\infty}\geq C \frac{\log n_1}{\sqrt{n_1}}\right)\leq n_1^{-c}.$
Together with the expression
$
\frac{1}{n_1}\sum_{i=1}^{n_1}V_i D_i=\frac{1}{n_1}\sum_{i=1}^{n_1}V_i f_i+\frac{1}{n_1}\sum_{i=1}^{n_1}V_i \delta_i,
$
we apply Condition {\rm (R1)} and establish 
$
\PP\left(\left|\frac{1}{n_1}\sum_{i=1}^{n_1}V_i D_i\right|\geq C\right)\leq n_1^{-c}.
$
Together with \eqref{eq: concentration zero}, and Condition {\rm (R1)}, we establish that, with probability larger than $1-n_1^{-c},$ 
\begin{equation}
\|\widehat{\pi}-\pi\|_2\lesssim \|R(V)\|_{\infty}+\left|\beta-\widehat{\beta}_{\rm init}(V)\right|+{\log n}/{\sqrt{n}}.
\label{eq: parameter rate}
\end{equation}
Our proposed estimator $\widehat{\epsilon}(V)$ defined in \eqref{eq: TSRF SE hetero} has the following equivalent expression,
\begin{equation*}
\begin{aligned}
\widehat{\epsilon}(V)&=P^{\perp}_{V_{\mathcal{A}_1}^{\intercal}}[Y_{\mathcal{A}_1}-D_{\mathcal{A}_1}\widehat{\beta}_{\rm init}(V)]\\
&=Y_{\mathcal{A}_1}-D_{\mathcal{A}_1}\widehat{\beta}_{\rm init}(V)-V(V^{\intercal} V)^{-1}V^{\intercal}\left(Y_{\mathcal{A}_1}-\widehat{\beta}_{\rm init}(V)D_{\mathcal{A}_1}\right)\\
&=Y_{\mathcal{A}_1}-D_{\mathcal{A}_1}\widehat{\beta}_{\rm init}(V)-V_{\mathcal{A}_1} \widehat{\pi}.
\end{aligned}
\end{equation*}
Then we apply \eqref{eq: minus beta D} and obtain
$
[\widehat{\epsilon}(V)]_i-\epsilon_i=D_i(\beta-\widehat{\beta}_{\rm init}(V))+V_i^{\intercal}(\pi-\widehat{\pi})+R_i(V).
$
Together with Condition {\rm (R1)} and \eqref{eq: parameter rate}, we establish Lemma \ref{lem: standard OLS lemma}.

\subsection{Proof of Lemma \ref{lem: diff key terms}} \label{sec: lemma 4 proof}
%In the following, we shall prove
%\begin{equation}
%\frac{1}{\sqrt{H(V_q,V_{q'})}}\left|\widetilde{\mathcal{E}}(V_{q})-\widetilde{\mathcal{E}}(V_{q'})\right|\cip 0,
%\label{eq: small approximation error}
%\end{equation}
%and 

%\paragraph*{Proof of \eqref{eq: small approximation error}.} 
Similarly to \eqref{eq: key decomp hetero}, we decompose $\widetilde{\mathcal{E}}(V_{q})$ as
$\widetilde{\mathcal{E}}(V_{q})=\frac{\Err_1+\Err_2}{{D}_{\mathcal{A}_1}^{\intercal}\T(V_q) {D}_{\mathcal{A}_1}},$
where 
$
\Err_1=\sum_{1\leq i\neq j\leq n_1}^{n_1}[\T(V_q)]_{ij}  \delta_{i}\epsilon_j$, and 
\begin{equation*}
\Err_2=\sum_{i=1}^{n_1}[\T(V_q)]_{ii}  (f_{i}-\widehat{f}_{i})\left(\epsilon_i+[\widehat{\epsilon}(V_{Q_{\max}})]_i-\epsilon_i\right)+\sum_{i=1}^{n_1}[\T(V_q)]_{ii}  \delta_{i}\left([\widehat{\epsilon}(V_{Q_{\max}})]_i-\epsilon_i\right).
\label{eq: error 2}
\end{equation*}
 We apply the same analysis as that of \eqref{eq: second approximation error} and establish  
\begin{equation*}
\begin{aligned}
\left|\Err_2\right|&\lesssim \sum_{i=1}^{n_1}[\T(V)]_{ii}  \left[|f_{i}-\widehat{f}_{i}|\left(\sqrt{\log n}+\left|[\widehat{\epsilon}(V_{Q_{\max}})]_i-\epsilon_i\right|\right)+\sqrt{\log n}\left|[\widehat{\epsilon}(V_{Q_{\max}})]_i-\epsilon_i\right|\right].
\end{aligned}
\end{equation*}
By Lemma \ref{lem: standard OLS lemma} with $V=V_{Q_{\max}},$ we apply the similar argument as that of \eqref{eq: error bound} and establish that 
\begin{equation*}
\PP\left(\left|\widetilde{\mathcal{E}}(V_{q})\right|\geq C \frac{\sqrt{\log n}\cdot\eta_n(V_{Q_{\max}})\cdot {\rm Tr}[\T(V_q)]+t_0\sqrt{{\rm Tr}([\T({V_q})]^2)}}{{{f}_{\mathcal{A}_1}\T(V_q) {f}_{\mathcal{A}_1}}}\right)\geq 1-\exp(-t_0^2)
\end{equation*}
where $\eta_n(V)$ is defined in \eqref{eq: approximation rate hetero}. We establish $\left|\widetilde{\mathcal{E}}(V_{q})\right|/{\sqrt{H(V_q,V_{q'})}}\cip 0$ under Condition (R3). Similarly, we establish $\left|\widetilde{\mathcal{E}}(V_{q'})\right|/{\sqrt{H(V_q,V_{q'})}}\cip 0$ under Condition (R3). That is, we establish $\frac{1}{\sqrt{H(V_q,V_{q'})}}\left|\widetilde{\mathcal{E}}(V_{q})-\widetilde{\mathcal{E}}(V_{q'})\right|\cip 0.$

Now we prove $\mathcal{G}_n(V_{q}, V_{q'}) \cid N(0,1)
$ and start with the decomposition,
\begin{equation}
\begin{aligned}
&\frac{f_{\mathcal{A}_1}^{\intercal} \T(V_{q'}) \epsilon_{\mathcal{A}_1} }{D_{\mathcal{A}_1}^{\intercal} \T(V_{q'}) D_{\mathcal{A}_1}}-\frac{{f}^{\intercal}_{\mathcal{A}_1} \T(V_q)\epsilon_{\mathcal{A}_1}}{{D}^{\intercal}_{\mathcal{A}_1} \T(V_q)D_{\mathcal{A}_1}}
=\frac{f_{\mathcal{A}_1}^{\intercal} \T(V_{q'}) \epsilon_{\mathcal{A}_1} }{f_{\mathcal{A}_1}^{\intercal} \T(V_{q'}) f_{\mathcal{A}_1}}-\frac{{f}^{\intercal}_{\mathcal{A}_1} \T(V_q)\epsilon_{\mathcal{A}_1}}{{f}^{\intercal}_{\mathcal{A}_1} \T(V_q)f_{\mathcal{A}_1}}
\\
&+\frac{f_{\mathcal{A}_1}^{\intercal} \T(V_{q'}) \epsilon_{\mathcal{A}_1} }{f_{\mathcal{A}_1}^{\intercal} \T(V_{q'}) f_{\mathcal{A}_1}}\left(\frac{{f_{\mathcal{A}_1}^{\intercal} \T(V_{q'}) f_{\mathcal{A}_1}}}{{D_{\mathcal{A}_1}^{\intercal} \T(V_{q'}) D_{\mathcal{A}_1}}}-1\right)-\frac{{f}^{\intercal}_{\mathcal{A}_1} \T(V_q)\epsilon_{\mathcal{A}_1}}{{f}^{\intercal}_{\mathcal{A}_1} \T(V_q)f_{\mathcal{A}_1}}\left(\frac{{f_{\mathcal{A}_1}^{\intercal} \T(V_{q}) f_{\mathcal{A}_1}}}{{D_{\mathcal{A}_1}^{\intercal} \T(V_{q}) D_{\mathcal{A}_1}}}-1\right).
\end{aligned}
\label{eq: var decomp RF}
\end{equation}

Since the vector $S$ defined in \eqref{eq: vector diff} satisfies $\max_{i\in \mathcal{A}_1}{S_i^2}/{\sum_{i\in \mathcal{A}_1}S_i^2}\rightarrow 0,$ we verify the Linderberg condition and  establish 
\begin{equation}
\frac{1}{\sqrt{H(V_q,V_{q'})}}\left(\frac{f_{\mathcal{A}_1}^{\intercal} \T(V_{q'}) \epsilon_{\mathcal{A}_1} }{f_{\mathcal{A}_1}^{\intercal} \T(V_{q'}) f_{\mathcal{A}_1}}-\frac{{f}^{\intercal}_{\mathcal{A}_1} \T(V_q)\epsilon_{\mathcal{A}_1}}{{f}^{\intercal}_{\mathcal{A}_1} \T(V_q)f_{\mathcal{A}_1}}\right)\cid N(0,1).
\label{eq: normal lind}
\end{equation}

We apply \eqref{eq: inter bound 1 general} and \eqref{eq: bottom consistency RF} and establish that with probability larger than $1-\exp\left(-c \min\left\{t_0^2,t_0\right\}\right)$ for some positive constants $c>0$ and $t_0>0,$
\begin{equation*}
\left|\frac{f_{\mathcal{A}_1}^{\intercal} \T(V_{q}) \epsilon_{\mathcal{A}_1} }{f_{\mathcal{A}_1}^{\intercal} \T(V_{q}) f_{\mathcal{A}_1}}\right|\lesssim \frac{t_0}{\sqrt{\mu(V_q)}}, \quad\text{and}\quad \left|\frac{{f_{\mathcal{A}_1}^{\intercal} \T(V_{q}) f_{\mathcal{A}_1}}}{{D_{\mathcal{A}_1}^{\intercal} \T(V_{q}) D_{\mathcal{A}_1}}}-1\right|\lesssim \frac{{\rm Tr}[\T({V_q})]}{\mu(V_q)}+ \frac{t_0}{\sqrt{\mu(V_q)}}.
\end{equation*}
We apply the above inequalities and establish that with probability larger than $1-\exp\left(-c \min\left\{t_0^2,t_0\right\}\right)$ for some positive constants $c>0$ and $t_0>0,$
\begin{equation*}
\begin{aligned}
&\frac{1}{\sqrt{H(V_q,V_{q'})}}\left|\frac{f_{\mathcal{A}_1}^{\intercal} \T(V_{q'}) \epsilon_{\mathcal{A}_1} }{f_{\mathcal{A}_1}^{\intercal} \T(V_{q'}) f_{\mathcal{A}_1}}\left(\frac{{f_{\mathcal{A}_1}^{\intercal} \T(V_{q'}) f_{\mathcal{A}_1}}}{{D_{\mathcal{A}_1}^{\intercal} \T(V_{q'}) D_{\mathcal{A}_1}}}-1\right)\right|\\
&\lesssim  \frac{1}{\sqrt{H(V_q,V_{q'})}}\frac{t_0}{\sqrt{\mu(V_q)}}\left(\frac{{\rm Tr}[\T({V_q})]}{\mu(V_q)}+ \frac{t_0}{\sqrt{\mu(V_q)}}\right).
\end{aligned}
\end{equation*}
Under the condition {\rm (R3)}, we establish 
$$\frac{1}{\sqrt{H(V_q,V_{q'})}}\left|\frac{f_{\mathcal{A}_1}^{\intercal} \T(V_{q}) \epsilon_{\mathcal{A}_1} }{f_{\mathcal{A}_1}^{\intercal} \T(V_{q}) f_{\mathcal{A}_1}}\left(\frac{{f_{\mathcal{A}_1}^{\intercal} \T(V_{q}) f_{\mathcal{A}_1}}}{{D_{\mathcal{A}_1}^{\intercal} \T(V_{q}) D_{\mathcal{A}_1}}}-1\right)\right|\cip 0.$$
Similarly, we have 
$\frac{1}{\sqrt{H(V_q,V_{q'})}}\left|\frac{f_{\mathcal{A}_1}^{\intercal} \T(V_{q'}) \epsilon_{\mathcal{A}_1} }{f_{\mathcal{A}_1}^{\intercal} \T(V_{q'}) f_{\mathcal{A}_1}}\left(\frac{{f_{\mathcal{A}_1}^{\intercal} \T(V_{q'}) f_{\mathcal{A}_1}}}{{D_{\mathcal{A}_1}^{\intercal} \T(V_{q'}) D_{\mathcal{A}_1}}}-1\right)\right|\cip 0.$
The above inequalities and the decomposition \eqref{eq: var decomp RF} imply 
\begin{equation}
\left|\mathcal{G}_n(V_{q},V_{q'})-\frac{1}{\sqrt{H(V_q,V_{q'})}}\left(\frac{f_{\mathcal{A}_1}^{\intercal} \T(V_{q'}) \epsilon_{\mathcal{A}_1} }{f_{\mathcal{A}_1}^{\intercal} \T(V_{q'}) f_{\mathcal{A}_1}}-\frac{{f}^{\intercal}_{\mathcal{A}_1} \T(V_q)\epsilon_{\mathcal{A}_1}}{{f}^{\intercal}_{\mathcal{A}_1} \T(V_q)f_{\mathcal{A}_1}}\right)\right|\cip 0.
\label{eq: diff normal approximation}
\end{equation}
Together with \eqref{eq: normal lind}, we establish $\mathcal{G}_n(V_{q}, V_{q'}) \cid N(0,1)$.

\subsection{Proof of Lemma \ref{lem: general transform}}
\label{sec: lemma 4 5 proof}
Note that 
$[\T_{\rm RF}({V})]^2=(\Omega)^{\intercal}P^{\perp}_{\widehat{V}_{\mathcal{A}_1}} \Omega
(\Omega)^{\intercal}P^{\perp}_{\widehat{V}_{\mathcal{A}_1}} \Omega,
$ and $\T(V)$ and $[\T_{\rm RF}({V})]^2$ are positive definite. For a vector $b\in \R^{n}$, we have 
\begin{equation}
\begin{aligned}
b^{\intercal} [\T_{\rm RF}({V})]^2 b=(P^{\perp}_{\widehat{V}_{\mathcal{A}_1}} \Omega b)^{\intercal} \Omega
(\Omega)^{\intercal}
P^{\perp}_{\widehat{V}_{\mathcal{A}_1}} \Omega b\leq \|P^{\perp}_{\widehat{V}_{\mathcal{A}_1}} \Omega b\|_2^2 \|\Omega\|_2^2. %\leq b^{\intercal} \T_{\rm RF}({V}) b.
%&\leq b^{\intercal} \T_{\rm RF}({V}) b\cdot \|\Omega\|_{1}\cdot \|\Omega\|_{\infty},
\end{aligned}
\label{eq: relation upper}
\end{equation}
Let $\|\Omega\|_{1}$ and $\|\Omega\|_{\infty}$ denote the matrix $1$ and $\infty$ norm, respectively. 
Since $\|\Omega\|_{1}=1$ and $\|\Omega\|_{\infty}=1$ for the random forests setting, we have the upper bound for the spectral norm $\|\Omega\|_2^2\leq \|\Omega\|_{1}\cdot \|\Omega\|_{\infty} \leq 1.$
Together with \eqref{eq: relation upper}, we establish \eqref{eq: key inequality for RF}.
Note that $b^{\intercal}\T_{\rm RF}({V})b=b^{\intercal}(\Omega)^{\intercal}P^{\perp}_{\widehat{V}_{\mathcal{A}_1}} \Omega b\leq \|\Omega\|_2^2\|b\|_2^2,
$ we establish that $\lambda_{\max}(\T_{\rm RF}({V}))\leq 1.$

We apply the minimax expression of eigenvalues and obtain that %$k$-th largest eigenvalue of $[\T_{\rm RF}({V})]^2$ admits the expression
\begin{equation*}
\begin{aligned}
\lambda_{k}([\T_{\rm RF}({V})]^2)&=\max_{U: {\rm dim} (U)=k}\min_{u \in U} \frac{u^{\intercal} [\T_{\rm RF}({V})]^2 u}{\|u\|_2^2}\\
&\leq \max_{U: {\rm dim} (U)=k}\min_{u \in U} \frac{u^{\intercal} \T_{\rm RF}({V}) u}{\|u\|_2^2}=\lambda_{k}(\T_{\rm RF}({V})).
\end{aligned}
\end{equation*}
where the inequality follows from \eqref{eq: key inequality for RF}. The above inequality leads to ${\rm Tr}\left([\T_{\rm RF}({V})]^2\right)\leq {\rm Tr}[\T(V)].
$ Since $\Pm^{\perp} \Pb=\Pm^{\perp}$, we have 
\begin{equation*}
\begin{aligned}
[\T_{\rm ba}({V})]^2&=
%\Pm P_{V_{BW}, W}^{\perp}\Pm \Pm P_{V_{BW}, W}^{\perp}\Pm\\
\Pm P_{V_{BW}, W}^{\perp}\Pm P_{V_{BW}, W}^{\perp}\Pm\\
&=\Pm P_{V_{BW}, W}^{\perp}\left({\rm I}-\Pm^{\perp}\right)P_{V_{BW}, W}^{\perp}\Pm=\Pm P_{V_{BW}, W}^{\perp} \Pm=\T_{\rm ba}({V})
\end{aligned}
%\label{eq: key equation}
\end{equation*}
Similar to the above proof, we establish $[\T_{\rm DNN}({V})]^2=\T_{\rm DNN}({V}).$ The proof of \eqref{eq: key inequality for general} is the same as that of \eqref{eq: key inequality for RF} by replacing \eqref{eq: relation upper} with 
$
b^{\intercal} [\T_{\rm RF}({V})]^2 b\leq b^{\intercal} \T_{\rm RF}({V}) b\cdot \|\Omega\|_2^2.
$
%The other arguments are the same as those in the proof of Lemma \ref{lem: general transform}.

\subsection{Proof of Lemma \ref{lem: variance consistency}}
\label{sec: lemma 7 proof}
By rewriting Lemma \ref{lem: standard OLS lemma}, we have that, with probability larger than $1-n^{-c},$
\begin{equation}
\max_{1\leq i\leq n_1}\left|[\widehat{\epsilon}(V)]_i-\epsilon_i\right|%&\leq \max_{1\leq i\leq n_1}\{|D_i|,\|W_i\|_2, \|V_i\|_2\}\cdot \sqrt{k}\left(\|R(V)\|_{\infty}+|\beta-\widehat{\beta}_{\rm init}(V)|+\frac{1}{\sqrt{n}}\right)\\
\leq C \kappa_n(V), \quad  \kappa_n(V)= \sqrt{\log n}\left(\|R(V)\|_{\infty}+|\beta-\widehat{\beta}_{\rm init}(V)|+\frac{\log n}{\sqrt{n}}\right).
\label{eq: key bound variance}
\end{equation}
It is sufficient to show 
\begin{equation}
\frac{({f}_{\mathcal{A}_1}^{\intercal}\T(V) {f}_{\mathcal{A}_1})^2}{{\sum_{i=1}^{n_1}\sigma^2_i[\T(V) {f}_{\mathcal{A}_1}]_i^2}
}\frac{{\sum_{i=1}^{n_1}[\widehat{\epsilon}(V)]^2_i[\T(V) {D}_{\mathcal{A}_1}]_i^2}
}{({D}_{\mathcal{A}_1}^{\intercal}\T(V) {D}_{\mathcal{A}_1})^2}\cip 1.
\label{eq: target 1}
\end{equation}
Note that 
\begin{equation}
\frac{{\sum_{i=1}^{n_1}[\widehat{\epsilon}(V)]^2_i[\T(V) {D}_{\mathcal{A}_1}]_i^2}
}{{\sum_{i=1}^{n_1}\sigma^2_i[\T(V) {f}_{\mathcal{A}_1}]_i^2}}=\frac{{\sum_{i=1}^{n_1}\sigma^2_i[\T(V) {D}_{\mathcal{A}_1}]_i^2}
}{{\sum_{i=1}^{n_1}\sigma^2_i[\T(V) {f}_{\mathcal{A}_1}]_i^2}}\cdot \frac{{\sum_{i=1}^{n_1}[\widehat{\epsilon}(V)]^2_i[\T(V) {D}_{\mathcal{A}_1}]_i^2}
}{{\sum_{i=1}^{n_1}\sigma^2_i[\T(V) {D}_{\mathcal{A}_1}]_i^2}}.
\label{eq: decomp variance}
\end{equation}
We further decompose 
\begin{equation}
\begin{aligned}
&\frac{{\sum_{i=1}^{n_1}[\widehat{\epsilon}(V)]^2_i[\T(V) {D}_{\mathcal{A}_1}]_i^2}
}{{\sum_{i=1}^{n_1}\sigma^2_i[\T(V) {D}_{\mathcal{A}_1}]_i^2}}-1=\frac{{\sum_{i=1}^{n_1}[\widehat{\epsilon}(V)-\epsilon_i+\epsilon_i]^2[\T(V) {D}_{\mathcal{A}_1}]_i^2}
}{{\sum_{i=1}^{n_1}\sigma^2_i[\T(V) {D}_{\mathcal{A}_1}]_i^2}}-1\\
&=\frac{{\sum_{i=1}^{n_1}[\widehat{\epsilon}(V)-\epsilon_i]^2[\T(V) {D}_{\mathcal{A}_1}]_i^2}
}{{\sum_{i=1}^{n_1}\sigma^2_i[\T(V) {D}_{\mathcal{A}_1}]_i^2}}+\frac{{\sum_{i=1}^{n_1}\epsilon_i[\widehat{\epsilon}(V)-\epsilon_i][\T(V) {D}_{\mathcal{A}_1}]_i^2}
}{{\sum_{i=1}^{n_1}\sigma^2_i[\T(V) {D}_{\mathcal{A}_1}]_i^2}}+\frac{{\sum_{i=1}^{n_1}(\epsilon_i^2-\sigma^2_i)[\T(V) {D}_{\mathcal{A}_1}]_i^2}
}{{\sum_{i=1}^{n_1}\sigma^2_i[\T(V) {D}_{\mathcal{A}_1}]_i^2}}
\end{aligned}
\end{equation}
Note that 
$\left|\frac{{\sum_{i=1}^{n_1}(\epsilon_i^2-\sigma^2_i)[\T(V) {D}_{\mathcal{A}_1}]_i^2}
}{{\sum_{i=1}^{n_1}\sigma^2_i[\T(V) {D}_{\mathcal{A}_1}]_i^2}}\right|\lesssim \left|\frac{{\sum_{i=1}^{n_1}(\epsilon_i^2-\sigma^2_i)[\T(V) {D}_{\mathcal{A}_1}]_i^2}
}{{\sum_{i=1}^{n_1}[\T(V) {D}_{\mathcal{A}_1}]_i^2}}\right|.$
Define the vector $a\in \R^{n_1}$ with $a_i=\frac{[\T(V) {D}_{\mathcal{A}_1}]_i^2}{{{\sum_{i=1}^{n_1}[\T(V) {D}_{\mathcal{A}_1}]_i^2}}}$ and we have $\|a\|_1=1$ and $\|a\|_2^2\leq \|a\|_{\infty}.$
By applying Proposition 5.16 of \citetsupp{vershynin2010introduction}, we establish that, \begin{equation}
\PP\left(\left|\sum_{i=1}^{n_1}a_i(\epsilon_i^2-\sigma^2_i)\right|\geq t_0 K\|a\|_{\infty}\mid \mathcal{O}\right)\leq \exp(-c\min\{t_0^2,t_0\}).
\end{equation}
By the condition $\|a\|_{\infty}\cip 0$ and $\kappa_n(V)^2+\sqrt{\log n} \kappa_n(V)\cip 0,$ we establish 
$${{\sum_{i=1}^{n_1}[\widehat{\epsilon}(V)]^2_i[\T(V) {D}_{\mathcal{A}_1}]_i^2}
}/{{\sum_{i=1}^{n_1}\sigma^2_i[\T(V) {D}_{\mathcal{A}_1}]_i^2}}\cip 1.$$
By \eqref{eq: decomp variance} and ${{\sum_{i=1}^{n_1}\sigma^2_i[\T(V) {D}_{\mathcal{A}_1}]_i^2}
}/{{\sum_{i=1}^{n_1}\sigma^2_i[\T(V) {f}_{\mathcal{A}_1}]_i^2}}\cip 1$, and $\frac{({f}_{\mathcal{A}_1}^{\intercal}\T(V) {f}_{\mathcal{A}_1})^2}{({D}_{\mathcal{A}_1}^{\intercal}\T(V) {D}_{\mathcal{A}_1})^2}\cip 1,$ we establish \eqref{eq: target 1}.

%%%%%%%%%%%%%%%%%%%%%%%%%%%%%%%%%%%%%%%%%%%%%%%%%%%%%%%%%%%%%%%%%%%%%%%%%%%%%%%%%%%%%%%%%%%%%%%%%
\section{Additional Simulation Results}
%%%%%%%%%%%%%%%%%%%%%%%%%%%%%%%%%%%%%%%%%%%%%%%%%%%%%%%%%%%%%%%%%%%%%%%%%%%%%%%%%%%%%%%%%%%%%%%%%

%%%%%%%%%%%%%%%%%%%%
\subsection{The appropriate threshold of IV strength for reliable inference}
\label{sec: thol 40}
%%%%%%%%%%%%%%%%%%%%%%
In this section, we show the performance of \texttt{TSCI} in terms of RMSE, bias, and CI coverage with different IV strengths. We generate $Z_i$, $X_i$ and errors $\{(\delta_i, \epsilon_i)^\intercal\}_{1\leq i\leq n}$ following the same procedure as in Settings B1 and B2 in Section \ref{sec: invalid IV sim} but with $Z_i=\Psi(X^*_{i,p_x+1})\in(0,1)$. We generate the treatment and outcome $\{D_i,Y_i\}_{1\leq i\leq n}$ following $D_i=1/2\cdot Z_i + a\cdot\left(\sin(2\pi Z_i)+3/2\cdot\cos(2\pi Z_i)\right) - 3/10\cdot \sum_{j=1}^{10} X_{i,j}+\delta_i$ and $Y_i=1/2\cdot D_i+1/5\cdot\sum_{j=1}^{10} X_{i,j}+ \epsilon_i$. With fixing the sample size $n=3000$, we control the IV strength by varying $a$ across $\{0.15,0.17,\cdots,0.35\}$. Since we consider the valid IV setting, we implement \texttt{TSCI} with random forests for 500 times and specify $\mathcal{V}_0=\{\aw(x)\}$ with $\aw(x)=\{1,x_1,\cdots,x_{p_x}\}$. In Figure \ref{fig: thol 40}, we show that when IV strength is above 40, \texttt{TSCI} has a smaller RMSE and bias, and its confidence interval achieves the desired coverage.
\begin{figure}[ht!]
    \centering
    \includegraphics[width=\linewidth]{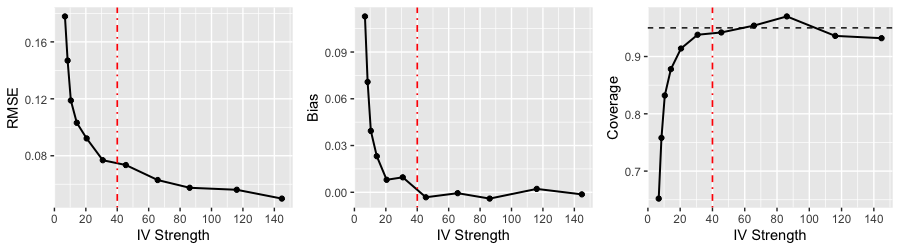}
    \caption{\small RMSE, bias and CI coverage of \texttt{TSCI} with different IV strengths. The red dashed line indicates the IV strength as 40. The black dashed line indicates the 95\% coverage level. }
    \label{fig: thol 40}
\end{figure}

We also use the setting with $a\in\{0.25, 0.3, 0.33\}$ to illustrate the bias of $\widehat{\beta}_{\rm init}(V)$ and the effect of the proposed bias correction in Figure \ref{fig:bias correction} in the main paper.

%%%%%%%%%%%%%%%%%%%%%%%%%%%%%%%%%%%%%%%%%%%%%%%%%%%%%%%%%%%%%%%%%%%%%%%%%%%%%%%%%%%%%%%%%%%%%%%%%%%%%%%
\subsection{Comparison with machine-learning IV}
\label{sec: ML comparison}
%%%%%%%%%%%%%%%%%%%%%%%%%%%%%%%%%%%%%%%%%%%%%%%%%%%%
%%%%%%%%%%%%%%%%%%%%%%%%%%%%%%%%%%%%%%%%%%%%%%%%%%%%

%In this section, we focus on the valid IV setting since most machine-learning IV methods are proposed in such a context. 
In the following, we compare our \texttt{TSCI} estimator with the \texttt{MLIV} estimator described in \eqref{eq: MLIV} {while} setting the full set of observed covariates as adjusted forms {in our method.} 
%The comparison results are expected to hold for the general invalid IV setting by correctly specifying the matrix $V.$ {We demonstrate the performance of TSCI in Section \ref{sec: invalid IV sim} when IVs are not valid.}
%We set $\beta=0.5$ as the treatment effect, $p_z=1$ and $p_x=20$, and  
% We generate $X^{*}_{i}\in \R^{p_x+1}$ following a multivariate normal distribution with zero mean and covariance matrix $\Sigma$ where $\Sigma_{i,j}=0.5^{|i-j|}$ for $1\leq i,j\leq p_x+1$. With $\Phi$ {denoting} 
% the standard normal cumulative distribution function, we define $X_{i,j}=\Phi(X^{*}_{i,j})$ for $1\leq j\leq p_x$ and $Z_i=\Phi(X^{*}_{i,p_x+1}).$
% We generate the data $Z_i$ and $X_i$ following the same procedure in Section \ref{sec: thol 40}.
% We generate the treatment and outcome $\{D_i,Y_i\}_{1\leq i\leq n}$ following $D_i=1/2\cdot Z_i + a\cdot\left(\sin(2\pi Z_i)+3/2\cdot\cos(2\pi Z_i)\right) - 3/10\cdot \sum_{j=1}^{10} X_{i,j}+\delta_i$ and $Y_i=1/2\cdot D_i+1/5\cdot\sum_{j=1}^{10} X_{i,j}+ \epsilon_i$. We set the errors $\{(\delta_i, \epsilon_i)^\intercal\}_{1\leq i\leq n}$ as heteroscedastic following \citet{bekker2015jackknife}: for $1\leq i\leq n,$ we generate $\delta_i \sim N(0,Z_i^2+0.25)$ and  
% $
% \epsilon_i=0.6 \delta_i+\sqrt{{[1-0.6^2]}/[0.86^4+1.38072^2]}(1.38072\cdot \tau_{1,i}+0.86^2\cdot \tau_{2,i}),
% $
% where conditioning on $Z_i$,  and $\tau_{1,i}$ and $\tau_{2,i}$ are generated to be independent of $\delta_i$, with $\tau_{1,i}\sim N(0, Z_i^2+0.25)$ and $\tau_{2,i} \sim N(0,1).$ 
We use the exactly same setting in Section \ref{sec: thol 40}. In Table \ref{tab: ML Pitfall}, we compare \texttt{TSCI} with \texttt{MLIV} for different $n\in\{1000, 3000, 5000\}$ and different $a\in\{0.2,0.25,0.3,0.35\}$ with a larger value of $a$ corresponding to a stronger IV. We implement 500 rounds of simulations and report the average measure.

\begin{table}[ht!]
\centering
\caption{\small Comparison of \texttt{TSCI} and \texttt{MLIV} with the treatment model fitted by RF. The columns indexed with ``without self-prediction" stands for the split RF being implemented without self-prediction, while ``with self-prediction" stands for the split RF being implemented with self-prediction. The columns indexed by ``\texttt{TSCI}", ``Init", ``\texttt{MLIV}" correspond to our proposed \texttt{TSCI} estimator in \eqref{eq: ML corrected hetero}, the initial estimator in \eqref{eq: RF init}, and the \texttt{MLIV} estimator in \eqref{eq: MLIV}. The column indexed with ``RMSE Ratio" represents the MSE ratio of the MLLV estimator to \texttt{TSCI} estimator; the column indexed with ``IV Str" stands for our proposed IV strength in \eqref{eq: IV strength homo}.} %\begin{adjustbox}{center}
\resizebox{\columnwidth}{!}{
\begin{tabular}{|c|c|ccc|c|c|ccc|c|c|}
\hline
& & \multicolumn{5}{c|}{without self-prediction} & \multicolumn{5}{c|}{with self-prediction} \\
  \hline
& & \multicolumn{3}{c|}{Bias} & RMSE &  & \multicolumn{3}{c|}{Bias} & RMSE & \\
  \hline
a & n & \texttt{TSCI} & Init & \texttt{MLIV} & Ratio & IV Str & \texttt{TSCI} & Init & \texttt{MLIV} & Ratio  & IV Str \\ 
 \hline
% 0.10 & 1000 & 0.35 & 0.45 & 2.77 & 95.80 & 1.31 & 0.44 & 0.50 & 0.56 & 1.21 & 8.97 \\ 
%   0.10 & 3000 & 0.20 & 0.34 & -0.55 & 41.80 & 4.76 & 0.36 & 0.45 & 0.54 & 1.43 & 26.52 \\ 
%   0.10 & 5000 & 0.14 & 0.29 & -0.89 & 49.41 & 8.82 & 0.33 & 0.43 & 0.54 & 1.55 & 44.10 \\ 
%   0.15 & 1000 & 0.28 & 0.40 & 0.64 & 38.92 & 1.47 & 0.37 & 0.45 & 0.52 & 1.32 & 10.19 \\ 
%   0.15 & 3000 & 0.12 & 0.26 & 0.15 & 39.29 & 6.65 & 0.28 & 0.39 & 0.51 & 1.65 & 31.18 \\ 
%   0.15 & 5000 & 0.06 & 0.20 & -0.18 & 3.43 & 13.30 & 0.23 & 0.35 & 0.50 & 1.89 & 54.96 \\ 
  \multirow{3}{*}{0.20} & 1000 & 0.17 & 0.30 & -4.31 & 374.06 & 2.18 & 0.28 & 0.38 & 0.48 & 1.51 & 11.90 \\ 
  & 3000 & 0.03 & 0.14 & -0.15 & 2.94 & 12.11 & 0.17 & 0.29 & 0.45 & 2.17 & 42.06 \\ 
  & 5000 & 0.01 & 0.10 & -0.07 & 1.67 & 27.97 & 0.12 & 0.23 & 0.42 & 2.69 & 76.41 \\ 
 \hline
  \multirow{3}{*}{0.25} & 1000 & 0.06 & 0.18 & -0.38 & 23.14 & 3.84 & 0.17 & 0.28 & 0.41 & 1.88 & 15.79 \\ 
  & 3000 & 0.00 & 0.06 & -0.05 & 1.37 & 30.35 & 0.08 & 0.17 & 0.35 & 2.97 & 74.95 \\ 
  & 5000 & 0.00 & 0.04 & -0.02 & 1.11 & 77.97 & 0.04 & 0.12 & 0.32 & 3.84 & 156.57 \\ 
 \hline
  \multirow{3}{*}{0.30} & 1000 & 0.02 & 0.09 & -0.11 & 2.01 & 7.49 & 0.09 & 0.18 & 0.32 & 2.22 & 23.92 \\ 
  & 3000 & -0.00 & 0.02 & -0.02 & 1.09 & 74.82 & 0.03 & 0.09 & 0.25 & 3.74 & 151.09 \\ 
  & 5000 & 0.00 & 0.02 & -0.01 & 1.06 & 194.74 & 0.01 & 0.06 & 0.21 & 4.51 & 316.40 \\ 
 \hline
  \multirow{3}{*}{0.35} & 1000 & 0.00 & 0.05 & -0.05 & 1.29 & 14.32 & 0.04 & 0.11 & 0.24 & 2.57 & 41.62 \\ 
  & 3000 & -0.01 & 0.01 & -0.01 & 0.99 & 144.72 & 0.01 & 0.04 & 0.17 & 3.55 & 255.65 \\ 
  & 5000 & -0.00 & 0.01 & -0.01 & 1.00 & 337.34 & 0.01 & 0.04 & 0.16 & 4.44 & 526.98 \\ 
 %  0.40 & 1000 & -0.00 & 0.03 & -0.03 & 1.03 & 23.65 & 0.02 & 0.07 & 0.18 & 2.57 & 65.49 \\ 
 %  0.40 & 3000 & -0.00 & 0.01 & -0.01 & 0.96 & 218.28 & 0.00 & 0.03 & 0.13 & 3.39 & 386.66 \\ 
 %  0.40 & 5000 & -0.00 & 0.01 & -0.00 & 0.96 & 484.72 & 0.00 & 0.02 & 0.12 & 3.85 & 759.93 \\ 
   \hline
\end{tabular}}
\label{tab: ML Pitfall}
\end{table}

In Table \ref{tab: ML Pitfall}, we observe that the \texttt{MLIV} has a larger bias and standard error than \texttt{TSCI}, leading to an inflated MSE of \texttt{MLIV}. We have explained in Section \ref{sec: dml comp} that this happens when the IV strength captured by RF is relatively weak. For a stronger IV (with $a=0.35$), when the sample size is 3000 or 5000 and the self-prediction is excluded, the \texttt{MLIV} and \texttt{TSCI} have comparable MSE, but our proposed \texttt{TSCI} estimator generally exhibits a smaller bias. Moreover, by comparing \texttt{TSCI} and initial estimators, our proposed bias correction step effectively removes the {bias}
 %overfitting bias 
 due to the {high complexity} of RF algorithm. Another interesting observation is that the removal of self-prediction helps reducing the bias of both \texttt{TSCI} and \texttt{MLIV} estimators for the reason of reducing the correlation between the ML predicted value and the unmeasured confounders. 
 % In Section \ref{sec: additional res for MLIV}, we consider other settings with different non-linear terms in the treatment model. 

We then show how the coefficient $c_f$ in \eqref{eq: MLIV} in the main paper inflates the estimation error of \texttt{MLIV} when IV is weak. In Figure \ref{fig: freq of cf}, we display the distribution of \texttt{TSCI} and \texttt{MLIV} estimators in box plots and the frequency of $c_f$ in the histogram in the setting with $n=1000$ and $a=0.2$. We can see that there are certain proportion of the coefficient $c_f$ close to 0, which would inflate the estimator to extremely large values and thus increase the \texttt{MLIV} estimator variance when IV is weak.
\begin{figure}[ht!]
    \centering
    \includegraphics[width=0.9\linewidth]{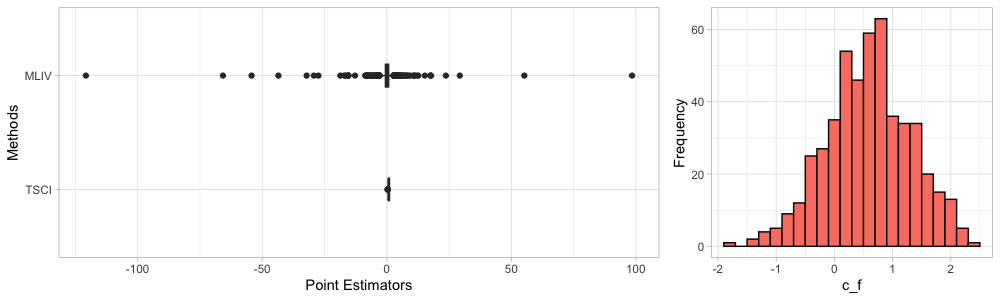}
    \caption{\small %Distributions of \texttt{MLIV} and \texttt{TSCI} estimators, and the coefficient $c_f$ of the ML fitted IV $\widehat{f}$. 
    The left panel shows the boxplot of \texttt{TSCI} and \texttt{MLIV} estimators and the right panel shows the histogram of the coefficient $c_f$ of $\widehat{f}$ in the first stage regression $D\sim \widehat{f}+X$ in the \texttt{MLIV} method.}
    \label{fig: freq of cf}
\end{figure}

%%%%%%%%%%%%%%%%%%%%%%%%%%%%%%%%%%%%%%%%%%%%%%%%%%%%%%%%%%%%
\subsection{Multiple IV (with non-linearity)}
\label{sec: multiple IV}
%%%%%%%%%%%%%%%%%%%%%%%%%%%%%%%%%%%%%%%%%%%%%%%%%%%%%%%%%%%%
In this section, we consider multiple IVs and compare \texttt{TSCI} with existing methods dealing with invalid IVs, including \texttt{TSHT} \citepsupp{guo2018confidence} and \texttt{CIIV} \citepsupp{windmeijer2019confidence}. We consider the setting with 10 IVs. With fixing the sample size $n=3000$, we generate $X^*_i\in\R^{p_x+p_z}$ following the multivariate normal distribution with zero mean and covariance $\Sigma$ where $\Sigma_{i,j}=0.5^{|i-j|}$ for $1\leq i,j\leq p_x+p_z$. We define the first $p_z$ columns of $X^*$ as IVs and denote it by $Z_i=X^*_{i,1:p_z}$. The remaining columns are defined as observed covariates, that is $X_i = X^*_{i,(p_z+1):(p_z+p_x)}$. We generate errors $\{(\delta_i, \epsilon_i)^\intercal\}_{1\leq i\leq n}$ following the bivariate normal distribution with zero mean, unit variance and covariance as 0.5. We generate $\{D_i, Y_i\}_{1\leq i\leq n}$ following $D_i=\sum_{j=1}^{p_z}|Z_{i,j}| + \sum_{j=1}^{p_z}\tanh(Z_{i,j}) + 1/2\cdot\sum_{j=1}^{p_x}X_{i,j}+\delta_i$ and $Y_i=D_i+Z_i\pi+1/2\cdot\sum_{j=1}^{p_x}X_{i,j}+\epsilon_i$ where the vector $\pi\in\R^{p_z}$ indicates the invalidity level of each IV. The $j$-th IV is valid if $\pi_j=0$; otherwise, it is invalid. A larger $|\pi_j|$ value indicates that the $j$-th IV is more severely invalid.  We consider the following two settings,  
%types of $\pi$ to accommodate the majority rule/plurality rule assumption in \texttt{TSHT} and \texttt{CIIV} as the following.
\begin{itemize}
    \item Setting D1: $\pi=a\cdot(0,0,0,0,{0.5},0.5,0.5,0.5,0.5,0.5)^\intercal$,
    \item Setting D2: $\pi=a\cdot(0,0,0,0,0.5,0.5,0.5, 1,1,1)^\intercal$.
\end{itemize}
For Setting D1, neither the majority nor the plurality rule is satisfied. For Setting D2, the plurality rule is satisfied. We vary the value of $a\in\{0.1,0.2,\cdots,1\}$  to simulate the different levels of invalidity. We specify the sets of basis as $\mathcal{V}_0=\{\aw(x)\}$ with $\aw(x)=\{1,x_1,\cdots,x_{p_x}\}$, $\mathcal{V}_1=\{\mathbf{z},\aw(x)\}$ and $\mathcal{V}_2=\{\mathbf{z},\mathbf{z}^2,\aw(x)\}$ with $\mathbf{z}^2=\{z_1^2,\cdots, z_{10}^2\}$. We run 500 rounds of simulation and report the results in Figure \ref{fig: multiple IVs sim}. 

\begin{figure}[hbt!]
    \centering
    \includegraphics[width=\linewidth]{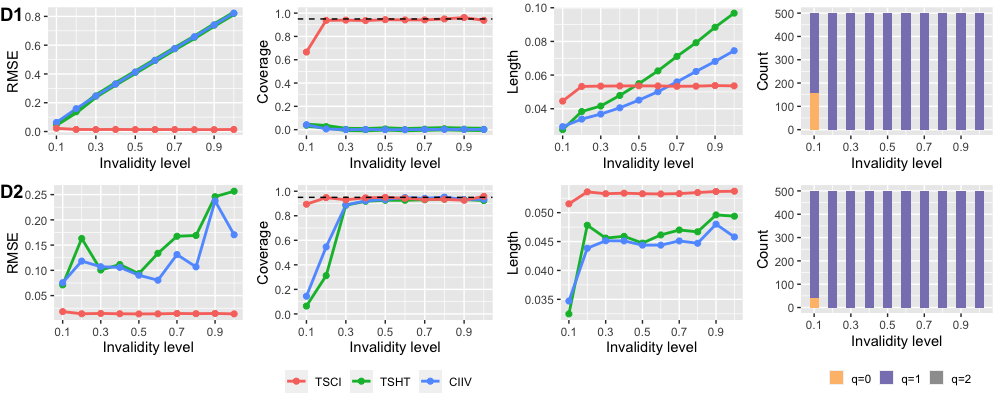}
    \caption{\small Comparison of \texttt{TSCI} with \texttt{TSHT} and \texttt{CIIV} in terms of RMSE, CI coverage and length under Settings D1 and D2. Setting D1 corresponds to a setting that the plurality rule fails while Setting D2 corresponds to a setting that the plurality holds. The stacked bar charts in the last column show the basis selection of \texttt{TSCI} where $\mathcal{V}_{q=1}$ is the oracle basis set. ``\texttt{TSCI}'' is our proposed method detailed in Algorithm \ref{algo: TSCI selection}. ``\texttt{TSHT}'' and ``\texttt{CIIV}'' are the methods proposed in \protect\citetsupp{guo2018confidence} and \protect\citetsupp{windmeijer2019confidence}.}
    \label{fig: multiple IVs sim}
\end{figure}

In Setting D1, neither the majority rule nor the plurality rule is satisfied, so \texttt{TSHT} and \texttt{CIIV} cannot select valid IVs and do not achieve a desired coverage level of 95\%. In contrast, \texttt{TSCI} is able to detect the existing invalidity and obtain valid inference. When the invalidity level is low (with $a=0.1$), \texttt{TSCI} may not be able to identify the invalidity and thus lead to coverage below 95\%.

In Setting D2, \texttt{TSCI} achieves the desired coverage level while \texttt{TSHT} and \texttt{CIIV} achieve the desired coverage for a relatively large invalidity level. When the invalidity level is low, say $a=0.1$ and $a=0.2$, \texttt{TSHT} and \texttt{CIIV} are significantly impacted by locally invalid IVs and generate poor coverage. In comparison to \texttt{TSHT} and \texttt{CIIV}, \texttt{TSCI} provides much more robust performance to the low invalidity levels because it evaluates the total invalidity levels accumulated by all ten invalid IVs. %by using the specified basis set. %Both \texttt{TSHT} and \texttt{CIIV} detect the invalidity on the single IV basis. 

%%%%%%%%%%%%%%%%%%%%%%%%%%%%%%%%
\subsection{Additional Results in Section \ref{sec: invalid IV sim}}
\label{sec: additional invalid IV setting}
%%%%%%%%%%%%%%%%%%%%%%%%%%%%%%%%
In this section, we further report the additional results for Setting B1 and results for Setting B2, and introduce a setting with binary IV denoted as Setting B3.

For Setting B1, we report the coverage for Vio=2 in Table \ref{tab: Coverage Model 1 vio2} and we further report the mean absolute bias and the confidence interval length in for both violation forms in Tables \ref{tab: bias and V selctn invalid IV model1} and \ref{tab: length and V selctn invalid IV model1}, respectively. In Table \ref{tab: bias and V selctn invalid IV model1}, \texttt{TSLS} has a large bias due to the existence of invalid IVs; even in the oracle setting with the prior knowledge of the best $\mathcal{V}_q$ approximating $g$, {RF-Full}, and {RF-Plug} have a large bias. Compared with RF-Init, \texttt{TSCI} corrects the bias effectively and thus have the desired 95\% coverage. In Table \ref{tab: length and V selctn invalid IV model1}, the CI of \texttt{TSCI} is shorter than the oracle method in settings with small sample size or weak interaction strength, because it fails to select the correct violation forms.

\begin{table}[h]
\centering
\resizebox{\linewidth}{!}{
\begin{tabular}{|c|c|c|cccc|cccc|c|ccc|}
  \hline
\multicolumn{3}{|c|}{} & \multicolumn{4}{c|}{\texttt{TSCI-RF}} & \multicolumn{4}{c|}{Proportions of selection} & \texttt{TSLS} & \multicolumn{3}{c|}{Other RF(oracle)} \\
  \hline
vio & a & n & Oracle & Comp & Robust & Invalidity & $\widehat{q}_c=0$ & $\widehat{q}_c=1$ & $\widehat{q}_c=2$ & $\widehat{q}_c=3$ &  & Init & Plug & Full \\ 
  \hline
  % \multirow{9}{*}{1} & \multirow{3}{*}{0.0} & 1000 & 0.91 & 0.01 & 0.01 & 0.01 & 0.99 & 0.01 & 0.00 & 0.00 & 0.00 & 0.80 & 0.38 & 0.01 \\ 
  % &  & 3000 & 0.92 & 0.92 & 0.92 & 1.00 & 0.00 & 0.84 & 0.16 & 0.00 & 0.00 & 0.91 & 0.64 & 0.00 \\ 
  % &  & 5000 & 0.91 & 0.92 & 0.93 & 1.00 & 0.00 & 0.85 & 0.15 & 0.00 & 0.00 & 0.89 & 0.74 & 0.00 \\ 
  % \cline{2-15}
  % & \multirow{3}{*}{0.5} & 1000 & 0.91 & 0.23 & 0.25 & 0.24 & 0.76 & 0.24 & 0.01 & 0.00 & 0.00 & 0.84 & 0.56 & 0.02 \\ 
  % &  & 3000 & 0.95 & 0.94 & 0.94 & 1.00 & 0.00 & 0.97 & 0.02 & 0.01 & 0.00 & 0.91 & 0.43 & 0.00 \\ 
  % &  & 5000 & 0.92 & 0.92 & 0.91 & 1.00 & 0.00 & 0.98 & 0.01 & 0.01 & 0.00 & 0.88 & 0.09 & 0.00 \\ 
  % \cline{2-15}
  % & \multirow{3}{*}{1.0} & 1000 & 0.96 & 0.92 & 0.92 & 0.95 & 0.05 & 0.93 & 0.01 & 0.00 & 0.00 & 0.91 & 0.52 & 0.08 \\ 
  % &  & 3000 & 0.94 & 0.94 & 0.95 & 1.00 & 0.00 & 0.99 & 0.01 & 0.00 & 0.00 & 0.92 & 0.00 & 0.00 \\ 
  % &  & 5000 & 0.94 & 0.94 & 0.94 & 1.00 & 0.00 & 0.98 & 0.02 & 0.01 & 0.00 & 0.92 & 0.00 & 0.00 \\ 
  % \hline
  \multirow{9}{*}{2} & \multirow{3}{*}{0.0} & 1000 & 0.92 & 0.01 & 0.01 & 0.00 & 1.00 & 0.00 & 0.00 & 0.00 & 0.00 & 0.82 & 0.30 & 0.01 \\ 
  &  & 3000 & 0.94 & 0.94 & 0.94 & 1.00 & 0.00 & 0.00 & 1.00 & 0.00 & 0.00 & 0.91 & 0.63 & 0.00 \\ 
  &  & 5000 & 0.94 & 0.94 & 0.94 & 1.00 & 0.00 & 0.00 & 1.00 & 0.00 & 0.00 & 0.88 & 0.77 & 0.00 \\ 
  \cline{2-15}
  & \multirow{3}{*}{0.5} & 1000 & 0.92 & 0.11 & 0.12 & 0.21 & 0.79 & 0.08 & 0.13 & 0.00 & 0.00 & 0.86 & 0.49 & 0.01 \\ 
  &  & 3000 & 0.94 & 0.93 & 0.92 & 1.00 & 0.00 & 0.00 & 0.97 & 0.03 & 0.00 & 0.90 & 0.37 & 0.00 \\ 
  &  & 5000 & 0.94 & 0.94 & 0.94 & 1.00 & 0.00 & 0.00 & 0.99 & 0.01 & 0.00 & 0.88 & 0.03 & 0.00 \\ 
  \cline{2-15}
  & \multirow{3}{*}{1.0} & 1000 & 0.95 & 0.89 & 0.89 & 0.96 & 0.04 & 0.02 & 0.93 & 0.01 & 0.00 & 0.89 & 0.40 & 0.01 \\ 
  &  & 3000 & 0.93 & 0.93 & 0.93 & 1.00 & 0.00 & 0.00 & 0.99 & 0.01 & 0.00 & 0.92 & 0.00 & 0.00 \\ 
  &  & 5000 & 0.93 & 0.93 & 0.93 & 1.00 & 0.00 & 0.00 & 1.00 & 0.00 & 0.00 & 0.90 & 0.00 & 0.00 \\ 
   \hline
\end{tabular}}
\caption{\small Coverage {(at nominal level 0.95)} for Setting B1 with Vio=2. The columns indexed with ``\texttt{TSCI-RF}" correspond to our proposed \texttt{TSCI} with random forests, where the columns indexed with ``Oracle", ``Comp" and ``Robust" correspond to {the estimators with $\mathcal{V}_q$ selected by the oracle knowledge, the comparison method,} and the robust method. The column indexed with ``Invalidity" reports the proportion of detecting the proposed IV as invalid. The columns indexed with ``Proportions of selection'' reports the proportions of basis sets $\mathcal{V}_q$ for $0\leq q\leq 3$ selected by \texttt{TSCI-RF}. The column indexed with ``\texttt{TSLS}" corresponds to the \texttt{TSLS} estimator. The columns indexed with ``Init", ``Plug", ``Full" correspond to the RF estimators without bias correction, the plug-in RF estimator and the no data-splitting RF estimator, {with the oracle knowledge of the best $\mathcal{V}_q$}.}
\label{tab: Coverage Model 1 vio2}
\end{table}

\begin{table}[H]
\centering
\resizebox{\linewidth}{!}{
\begin{tabular}{|c|c|c|cccc|cccc|c|ccc|}
  \hline
\multicolumn{3}{|c|}{} & \multicolumn{4}{c|}{\texttt{TSCI-RF}} & \multicolumn{4}{c|}{Proportions of selection} & \texttt{TSLS} & \multicolumn{3}{c|}{Other RF(oracle)} \\
  \hline
  vio & a & n & Oracle & Comp & Robust & Invalidity & $\widehat{q}_c=0$ & $\widehat{q}_c=1$ & $\widehat{q}_c=2$ & $\widehat{q}_c=3$ &  & Init & Plug & Full \\ 
  \hline
  \multirow{9}{*}{1} & \multirow{3}{*}{0.0} & 1000 & 0.02 & 0.53 & 0.53 & 0.01 & 0.99 & 0.01 & 0.00 & 0.00 & 0.56 & 0.13 & 0.48 & 0.30 \\ 
  &  & 3000 & 0.01 & 0.01 & 0.01 & 1.00 & 0.00 & 0.84 & 0.16 & 0.00 & 0.56 & 0.04 & 0.14 & 0.25 \\ 
  &  & 5000 & 0.00 & 0.00 & 0.00 & 1.00 & 0.00 & 0.85 & 0.15 & 0.00 & 0.56 & 0.03 & 0.05 & 0.23 \\ 
  \cline{2-15}
  & \multirow{3}{*}{0.5} & 1000 & 0.01 & 0.26 & 0.25 & 0.24 & 0.76 & 0.24 & 0.01 & 0.00 & 0.33 & 0.08 & 0.24 & 0.22 \\ 
  &  & 3000 & 0.00 & 0.00 & 0.00 & 1.00 & 0.00 & 0.97 & 0.02 & 0.01 & 0.33 & 0.03 & 0.16 & 0.19 \\ 
  &  & 5000 & 0.00 & 0.00 & 0.00 & 1.00 & 0.00 & 0.98 & 0.01 & 0.01 & 0.33 & 0.02 & 0.22 & 0.18 \\ 
  \cline{2-15}
  & \multirow{3}{*}{1.0} & 1000 & 0.00 & 0.01 & 0.01 & 0.95 & 0.05 & 0.93 & 0.01 & 0.00 & 0.23 & 0.04 & 0.15 & 0.13 \\ 
  &  & 3000 & 0.00 & 0.00 & 0.00 & 1.00 & 0.00 & 0.99 & 0.01 & 0.00 & 0.23 & 0.01 & 0.38 & 0.11 \\ 
  &  & 5000 & 0.00 & 0.00 & 0.00 & 1.00 & 0.00 & 0.98 & 0.02 & 0.01 & 0.23 & 0.01 & 0.37 & 0.10 \\ 
  \hline
  \multirow{9}{*}{2} & \multirow{3}{*}{0.0} & 1000 & 0.00 & 0.54 & 0.54 & 0.00 & 1.00 & 0.00 & 0.00 & 0.00 & 0.56 & 0.11 & 0.53 & 0.29 \\ 
  &  & 3000 & 0.00 & 0.00 & 0.00 & 1.00 & 0.00 & 0.00 & 1.00 & 0.00 & 0.56 & 0.05 & 0.13 & 0.25 \\ 
  &  & 5000 & 0.00 & 0.00 & 0.00 & 1.00 & 0.00 & 0.00 & 1.00 & 0.00 & 0.56 & 0.03 & 0.05 & 0.23 \\ 
  \cline{2-15}
  & \multirow{3}{*}{0.5} & 1000 & 0.01 & 0.38 & 0.38 & 0.21 & 0.79 & 0.08 & 0.13 & 0.00 & 0.33 & 0.08 & 0.34 & 0.23 \\ 
  &  & 3000 & 0.00 & 0.00 & 0.01 & 1.00 & 0.00 & 0.00 & 0.97 & 0.03 & 0.33 & 0.03 & 0.19 & 0.20 \\ 
  &  & 5000 & 0.00 & 0.00 & 0.01 & 1.00 & 0.00 & 0.00 & 0.99 & 0.01 & 0.33 & 0.03 & 0.29 & 0.19 \\ 
  \cline{2-15}
  & \multirow{3}{*}{1.0} & 1000 & 0.01 & 0.04 & 0.04 & 0.96 & 0.04 & 0.02 & 0.93 & 0.01 & 0.23 & 0.04 & 0.25 & 0.15 \\ 
  &  & 3000 & 0.00 & 0.00 & 0.00 & 1.00 & 0.00 & 0.00 & 0.99 & 0.01 & 0.23 & 0.02 & 0.52 & 0.12 \\ 
  &  & 5000 & 0.00 & 0.00 & 0.00 & 1.00 & 0.00 & 0.00 & 1.00 & 0.00 & 0.23 & 0.01 & 0.48 & 0.11 \\ 
   \hline
\end{tabular}}
\caption{\small Absolute Bias for Setting B1. The columns indexed with ``\texttt{TSCI-RF}" correspond to our proposed \texttt{TSCI} with random forests, where the columns indexed with ``Oracle", ``Comp" and ``Robust" correspond to {estimators with $\mathcal{V}_q$ selected by the oracle knowledge, the comparison method,} and the robust method. The column indexed with ``Invalidity" reports the proportion of detecting the proposed IV as invalid. The columns indexed with ``Proportions of selection'' reports the proportions of basis sets $\mathcal{V}_q$ selected by \texttt{TSCI-RF} using the comparison method. The columns indexed with ``\texttt{TSLS}" corresponds to the \texttt{TSLS} estimator. The columns indexed with ``Init", ``Plug", ``Full" correspond to the RF estimators without bias correction, the plug-in RF estimator and the no data-splitting RF estimator, {with the oracle knowledge of the best $\mathcal{V}_q$}.}
\label{tab: bias and V selctn invalid IV model1}
\end{table}

\begin{table}[H]
\centering
\resizebox{\linewidth}{!}{
\begin{tabular}{|c|c|c|cccc|cccc|c|ccc|}
  \hline
\multicolumn{3}{|c|}{} & \multicolumn{4}{c|}{\texttt{TSCI-RF}} & \multicolumn{4}{c|}{Proportions of selection} & \texttt{TSLS} & \multicolumn{3}{c|}{Other RF(oracle)} \\
  \hline
vio & a & n & Oracle & Comp & Robust & Invalidity & $\widehat{q}_c=0$ & $\widehat{q}_c=1$ & $\widehat{q}_c=2$ & $\widehat{q}_c=3$ &  & Init & Plug & Full \\ 
  \hline
  \multirow{9}{*}{1} & \multirow{3}{*}{0.0} & 1000 & 0.49 & 0.11 & 0.11 & 0.01 & 0.99 & 0.01 & 0.00 & 0.00 & 0.08 & 0.49 & 0.82 & 0.22 \\ 
  &  & 3000 & 0.32 & 0.32 & 0.32 & 1.00 & 0.00 & 0.84 & 0.16 & 0.00 & 0.05 & 0.32 & 0.38 & 0.14 \\ 
  &  & 5000 & 0.23 & 0.23 & 0.23 & 1.00 & 0.00 & 0.85 & 0.15 & 0.00 & 0.04 & 0.23 & 0.27 & 0.11 \\ 
  \cline{2-15}
  & \multirow{3}{*}{0.5} & 1000 & 0.38 & 0.13 & 0.14 & 0.24 & 0.76 & 0.24 & 0.01 & 0.00 & 0.05 & 0.38 & 0.60 & 0.19 \\ 
  &  & 3000 & 0.22 & 0.22 & 0.23 & 1.00 & 0.00 & 0.97 & 0.02 & 0.01 & 0.03 & 0.22 & 0.26 & 0.11 \\ 
  &  & 5000 & 0.17 & 0.17 & 0.17 & 1.00 & 0.00 & 0.98 & 0.01 & 0.01 & 0.02 & 0.17 & 0.19 & 0.09 \\ 
  \cline{2-15}
  & \multirow{3}{*}{1.0} & 1000 & 0.25 & 0.24 & 0.24 & 0.95 & 0.05 & 0.93 & 0.01 & 0.00 & 0.04 & 0.25 & 0.33 & 0.14 \\ 
  &  & 3000 & 0.13 & 0.13 & 0.13 & 1.00 & 0.00 & 0.99 & 0.01 & 0.00 & 0.02 & 0.13 & 0.18 & 0.08 \\ 
  &  & 5000 & 0.10 & 0.10 & 0.10 & 1.00 & 0.00 & 0.98 & 0.02 & 0.01 & 0.02 & 0.10 & 0.13 & 0.06 \\ 
  \hline
  \multirow{9}{*}{2} & \multirow{3}{*}{0.0} & 1000 & 0.49 & 0.17 & 0.17 & 0.00 & 1.00 & 0.00 & 0.00 & 0.00 & 0.12 & 0.49 & 0.85 & 0.21 \\ 
  &  & 3000 & 0.31 & 0.31 & 0.31 & 1.00 & 0.00 & 0.00 & 1.00 & 0.00 & 0.07 & 0.31 & 0.38 & 0.14 \\ 
  &  & 5000 & 0.23 & 0.23 & 0.23 & 1.00 & 0.00 & 0.00 & 1.00 & 0.00 & 0.06 & 0.23 & 0.27 & 0.11 \\ 
  \cline{2-15}
  & \multirow{3}{*}{0.5} & 1000 & 0.38 & 0.16 & 0.16 & 0.21 & 0.79 & 0.08 & 0.13 & 0.00 & 0.07 & 0.38 & 0.70 & 0.18 \\ 
  &  & 3000 & 0.23 & 0.23 & 0.26 & 1.00 & 0.00 & 0.00 & 0.97 & 0.03 & 0.04 & 0.23 & 0.28 & 0.11 \\ 
  &  & 5000 & 0.17 & 0.17 & 0.21 & 1.00 & 0.00 & 0.00 & 0.99 & 0.01 & 0.03 & 0.17 & 0.20 & 0.09 \\ 
  \cline{2-15}
  & \multirow{3}{*}{1.0} & 1000 & 0.24 & 0.24 & 0.24 & 0.96 & 0.04 & 0.02 & 0.93 & 0.01 & 0.05 & 0.24 & 0.40 & 0.14 \\ 
  &  & 3000 & 0.13 & 0.13 & 0.13 & 1.00 & 0.00 & 0.00 & 0.99 & 0.01 & 0.03 & 0.13 & 0.22 & 0.08 \\ 
  &  & 5000 & 0.10 & 0.10 & 0.11 & 1.00 & 0.00 & 0.00 & 1.00 & 0.00 & 0.02 & 0.10 & 0.15 & 0.06 \\ 
   \hline
\end{tabular}}
\caption{\small Average CI length for Setting B1. The columns indexed with ``\texttt{TSCI-RF}" correspond to \texttt{TSCI} with random forests, where the columns indexed with ``Oracle", ``Comp" and ``Robust" correspond to {estimators with $\mathcal{V}_q$ selected by the oracle knowledge, the comparison method,} and the robust method. The column indexed with ``Invalidity" reports the proportion of detecting the proposed IV as invalid. The columns indexed with ``Proportions of selection'' reports the proportions of basis sets $\mathcal{V}_q$ selected by \texttt{TSCI-RF} using the comparison method. The columns indexed with ``\texttt{TSLS}" corresponds to the \texttt{TSLS} estimator. The columns indexed with ``Init", ``Plug", ``Full" correspond to the RF estimators without bias correction, the plug-in RF estimator and the no data-splitting RF estimator, {with the oracle knowledge of the best $\mathcal{V}_q$}.}
\label{tab: length and V selctn invalid IV model1}
\end{table}

% In Table \ref{tab: length and V selctn invalid IV model1}, the robust selection methods typically lead to longer confidence intervals since more violation forms might be adjusted with the robust selection. For Setting B2, we report the empirical coverage in Table \ref{tab: Coverage Model 2} which are generally similar to those for Setting B1.

\begin{table}[t]
\centering
\resizebox{\linewidth}{!}{
\begin{tabular}{|c|c|c|cccc|cccc|c|ccc|}
  \hline
\multicolumn{3}{|c|}{} & \multicolumn{4}{c|}{\texttt{TSCI-RF}} & \multicolumn{4}{c|}{Proportions of selection} & \texttt{TSLS} & \multicolumn{3}{c|}{Other RF(oracle)} \\
  \hline
vio & a & n & Oracle & Comp & Robust & Invalidity & $\widehat{q}_c=0$ & $\widehat{q}_c=1$ & $\widehat{q}_c=2$ & $\widehat{q}_c=3$ &  & Init & Plug & Full \\ 
  \hline
  \multirow{9}{*}{1} & \multirow{3}{*}{0} & 1000 & 0.84 & 0.84 & 0.84 & 1.00 & 0.00 & 1.00 & 0.00 & 0.00 & 0.58 & 0.81 & 0.16 & 0.00 \\ 
  &  & 3000 & 0.93 & 0.93 & 0.94 & 1.00 & 0.00 & 0.96 & 0.04 & 0.00 & 0.11 & 0.91 & 0.00 & 0.00 \\ 
  &  & 5000 & 0.93 & 0.94 & 0.94 & 1.00 & 0.00 & 0.95 & 0.05 & 0.00 & 0.01 & 0.93 & 0.00 & 0.00 \\ 
  \cline{2-15}
  & \multirow{3}{*}{0.5} & 1000 & 0.94 & 0.94 & 0.95 & 1.00 & 0.00 & 0.95 & 0.04 & 0.01 & 0.00 & 0.88 & 0.00 & 0.01 \\ 
  &  & 3000 & 0.93 & 0.93 & 0.92 & 1.00 & 0.00 & 0.95 & 0.05 & 0.00 & 0.00 & 0.92 & 0.00 & 0.01 \\ 
  &  & 5000 & 0.93 & 0.93 & 0.93 & 1.00 & 0.00 & 0.99 & 0.01 & 0.00 & 0.00 & 0.92 & 0.00 & 0.00 \\ 
  \cline{2-15}
  & \multirow{3}{*}{1} & 1000 & 0.94 & 0.94 & 0.95 & 1.00 & 0.00 & 0.98 & 0.01 & 0.00 & 0.00 & 0.92 & 0.00 & 0.01 \\ 
  &  & 3000 & 0.95 & 0.96 & 0.96 & 1.00 & 0.00 & 0.98 & 0.02 & 0.00 & 0.00 & 0.93 & 0.00 & 0.00 \\ 
  &  & 5000 & 0.93 & 0.93 & 0.93 & 1.00 & 0.00 & 0.99 & 0.01 & 0.00 & 0.00 & 0.91 & 0.00 & 0.00 \\ 
  \hline
  \multirow{9}{*}{2} & \multirow{3}{*}{0} & 1000 & 0.84 & 0.17 & 0.17 & 0.99 & 0.01 & 0.97 & 0.02 & 0.00 & 0.55 & 0.16 & 0.13 & 0.00 \\ 
  &  & 3000 & 0.96 & 0.96 & 0.94 & 1.00 & 0.00 & 0.00 & 0.99 & 0.01 & 0.14 & 0.93 & 0.00 & 0.00 \\ 
  &  & 5000 & 0.95 & 0.95 & 0.94 & 1.00 & 0.00 & 0.00 & 0.99 & 0.01 & 0.02 & 0.92 & 0.00 & 0.01 \\ 
  \cline{2-15}
  & \multirow{3}{*}{0.5} & 1000 & 0.95 & 0.73 & 0.72 & 0.92 & 0.08 & 0.17 & 0.70 & 0.05 & 0.00 & 0.68 & 0.00 & 0.03 \\ 
  &  & 3000 & 0.94 & 0.94 & 0.94 & 1.00 & 0.00 & 0.00 & 0.95 & 0.05 & 0.00 & 0.93 & 0.00 & 0.00 \\ 
  &  & 5000 & 0.95 & 0.95 & 0.95 & 1.00 & 0.00 & 0.00 & 0.98 & 0.02 & 0.00 & 0.92 & 0.00 & 0.00 \\ 
  \cline{2-15}
  & \multirow{3}{*}{1} & 1000 & 0.96 & 0.95 & 0.96 & 1.00 & 0.00 & 0.01 & 0.93 & 0.06 & 0.00 & 0.94 & 0.00 & 0.01 \\ 
  &  & 3000 & 0.96 & 0.95 & 0.94 & 1.00 & 0.00 & 0.00 & 0.95 & 0.05 & 0.00 & 0.93 & 0.00 & 0.00 \\ 
  &  & 5000 & 0.96 & 0.96 & 0.95 & 1.00 & 0.00 & 0.00 & 0.97 & 0.03 & 0.00 & 0.95 & 0.00 & 0.00 \\ 
   \hline
\end{tabular}}
\caption{\small CI coverage for Setting B2. The columns indexed with ``\texttt{TSCI-RF}" correspond to \texttt{TSCI} with random forests, where the columns indexed with ``Oracle", ``Comp" and ``Robust" correspond to {estimators with $\mathcal{V}_q$ selected by the oracle knowledge, the comparison method,} and the robust method. The column indexed with ``Invalidity" reports the proportion of detecting the proposed IV as invalid. The columns indexed with ``Proportions of selection'' reports the proportions of basis sets $\mathcal{V}_q$ selected by \texttt{TSCI-RF} using the comparison method. The columns indexed with ``\texttt{TSLS}" corresponds to the \texttt{TSLS} estimator. The columns indexed with ``Init", ``Plug", ``Full" correspond to the RF estimators without bias correction, the plug-in RF estimator and the no data-splitting RF estimator, {with the oracle knowledge of the best $\mathcal{V}_q$}.
}
\label{tab: Coverage Model 2}
\end{table}
For Setting B2, we report the empirical coverage in Table \ref{tab: Coverage Model 2} which are generally similar to those for Setting B1.

{To approximate the real data analysis in Section \ref{sec: real}, we further generate a binary IV as $Z_{i}={\bf 1}(\Phi(X^{*}_{i,6})>0.6)$ and the covariates $X_{i,j}=X^{*}_{i,j}$ for $1\leq i\leq n$ and $1\leq j\leq 5.$ We consider the following models for $f(Z_{i},X_{i})$ and $g(Z_i,X_i)$,
\begin{itemize}
%\item[3.] Model 4 (binary IV):  $f(Z_{i},X_{i})= Z_i\cdot(1+a\cdot\sum_{j=1}^{5}X_{ij})-\sum_{j=1}^{p}0.3X_{ij},$ and $g(Z_i,X_i) = Z_i+0.2\cdot \sum_{j=1}^{p}X_{ij}.$
\item Setting B3 (binary IV): $f(Z_i,X_i)=Z_i\cdot(1+a\sum_{j=1}^{4}X_{ij}(1+X_{i,j+1}))-3/10\cdot\sum_{i=1}^{5}X_{ij}$ and $g(Z_i,X_i)=Z_i+1/2\cdot Z_i\cdot (\sum_{i=1}^{3}X_{ij})).$
\end{itemize}
Compared to Settings B1 and B2, the outcome model in Setting B3 involves the interaction between $Z_i$ and $X_{i}$ while the treatment model involves a more complicated interaction term, whose strength is controlled by $a$. We specify $\mathcal{V}_0=\{\aw(x)\}$ with $\aw(x)=\{1,x_1,\cdots,x_5\}$ and $\mathcal{V}_1=\mathcal{V}_0\cup\{z, z\cdot x_1,\cdots,z\cdot x_5\}$. With the specified basis sets, we implement \texttt{TSCI} with random forests as detailed in Algorithm \ref{algo: TSCI selection} in the main paper. In Table \ref{tab: binary IV model 3},  we demonstrate our proposed \texttt{TSCI} method for Setting B3. The observations are coherent with those for Settings B1 and B2. The main difference between the binary IV (Setting B3) and the continuous IV (Settings B1 and B2) is that the treatment effect is not  identifiable for $a=0$, which happens only for the binary IV setting. However, with a non-zero interaction and a relatively large sample size, our proposed \texttt{TSCI} methods achieve the desired coverage. We also observe that the bias correction is effective and improves the coverage when the interaction $a$ is relatively small. 

\begin{table}[H]
\centering
\resizebox{\linewidth}{!}{
\begin{tabular}[t]{|c|c|ccc|ccc|ccc|c|cc|}
  \hline
  \multicolumn{1}{|c}{} & \multicolumn{1}{c|}{} & \multicolumn{10}{c|}{\texttt{TSCI-RF}} & \multicolumn{2}{c|}{RF-Init} \\
  \hline
  \multicolumn{1}{|c}{ } & \multicolumn{1}{c|}{ } & \multicolumn{3}{c|}{Bias} & \multicolumn{3}{c|}{Length} & \multicolumn{3}{c|}{Coverage} & \multicolumn{1}{c|}{Invalidity} & \multicolumn{1}{c}{Bias} & \multicolumn{1}{c|}{Coverage} \\
  \hline
  a & n & Orac & Comp & Robust & Orac & Comp & Robust & Orac & Comp & Robust &  & Orac & Orac\\
  \hline
  \multirow{3}{*}{0.25} & 1000 & 0.01 & 0.56 & 0.56 & 0.42 & 0.23 & 0.23 & 0.90 & 0.28 & 0.28 & 0.31 & 0.10 & 0.82 \\ 
  & 3000 & 0.00 & 0.01 & 0.01 & 0.23 & 0.23 & 0.23 & 0.95 & 0.95 & 0.95 & 0.99 & 0.05 & 0.84 \\ 
  & 5000 & 0.00 & 0.00 & 0.00 & 0.18 & 0.18 & 0.18 & 0.94 & 0.94 & 0.94 & 1.00 & 0.03 & 0.86 \\ 
  \hline
  \multirow{3}{*}{0.50} & 1000 & 0.00 & 0.02 & 0.02 & 0.22 & 0.22 & 0.22 & 0.93 & 0.90 & 0.90 & 0.97 & 0.04 & 0.90 \\ 
  & 3000 & -0.00 & -0.00 & -0.00 & 0.12 & 0.12 & 0.12 & 0.93 & 0.93 & 0.93 & 1.00 & 0.02 & 0.89 \\ 
  & 5000 & 0.00 & 0.00 & 0.00 & 0.09 & 0.09 & 0.09 & 0.95 & 0.95 & 0.95 & 1.00 & 0.02 & 0.89 \\ 
  \hline
  \multirow{3}{*}{0.75} & 1000 & 0.00 & 0.00 & 0.00 & 0.15 & 0.15 & 0.15 & 0.92 & 0.90 & 0.90 & 0.99 & 0.02 & 0.91 \\ 
  & 3000 & 0.00 & 0.00 & 0.00 & 0.08 & 0.08 & 0.08 & 0.94 & 0.94 & 0.94 & 1.00 & 0.01 & 0.89 \\ 
  & 5000 & -0.00 & -0.00 & -0.00 & 0.06 & 0.06 & 0.06 & 0.93 & 0.93 & 0.93 & 1.00 & 0.01 & 0.91 \\ 
   \hline
\end{tabular}}
\caption{\small Bias, length, and coverage {(at nominal level 0.95)} for Setting B3. The columns indexed with ``\texttt{TSCI-RF}" corresponds to our proposed \texttt{TSCI} with random forests, where the columns indexed with ``Bias", ``Length", and ``Coverage" correspond to the absolute bias of the point estimator, the length and empirical coverage of the constructed CI respectively. The columns indexed with ``Oracle", ``Comp" and ``Robust" correspond to \texttt{TSCI} estimators with $\mathcal{V}_q$ selected by the oracle knowledge, the comparison method, and the robust method. The column indexed with ``Invalidity" reports the proportion of detecting the proposed IV as invalid. The columns indexed with ``RF-Init" correspond to the RF estimators without bias correction but with the oracle knowledge of the best $\mathcal{V}_q$.
% \vspace{10pt}
}
\label{tab: binary IV model 3}
\end{table}

%%%%%%%%%%%%%%%%%%%%%%%%%%%%%%%%%%%%%%%%%%%%%%%%%%%%%%%%%%%%%%%%%%%%%%%%%%%%%%%%%%%%%%%%%%%%%%%%%
\subsection{Binary Treatment}
\label{sec: binary treatment}
%%%%%%%%%%%%%%%%%%%%%%%%%%%%%%%%%%%%%%%%%%%%%%%%%%%%%%%%%%%%%%%%%%%%%%%%%%%%%%%%%%%%%%%%%%%%%%%%%

%Simulation Settings: $\beta = 1$, $\tau = 1$, $\gamma = 0.2$
We consider the binary treatment setting and explore the finite-sample performance of our proposed \texttt{TSCI} method. 
We consider the outcome model $Y_i = D_i\beta + Z_i + 0.2\cdot \sum_{j=1}^p{X_{ij}} + \epsilon_i$ with $\beta=1$ and $p_x=20.$ 
We generate 
$\epsilon_i$ and $\delta_i$ following bivariate normal distribution with zero mean, unit variance and covariance as 0.5. 
We generate the binary treatment $D_i$ with the conditional mean as
$$
\E(D_i\mid Z_i,X_i)= \frac{\exp(f(Z_i,X_i) + \delta_i)}{1+\exp(f(Z_i,X_i) + \delta_i)},
$$
where $f(Z_i,X_i)$ is specified in the following two ways.
\begin{itemize}
\item Setting 1 (continuous IV): generate $Z_i, X_i$ following Section \ref{sec: invalid IV sim}. $f(Z_i,X_i)=-25/12+Z_i+Z_i^2+1/8\cdot Z_i^4+Z_i\cdot(a\cdot\sum_{j=1}^5 X_{i,j})-3/10\cdot\sum_{j=1}^{p_x} X_{i,j}$.
\item Setting 2 (binary IV): generate $f(Z_{i},X_{i})= Z_i\cdot(1+a\cdot\sum_{j=1}^{5}X_{ij})-\sum_{j=1}^{p}0.3X_{ij}$.
\end{itemize} 

\begin{table}[ht!]
%\caption{}
\centering
\resizebox{\linewidth}{!}{
\begin{tabular}[t]{|c|c|c|ccc|ccc|ccc|c|cc|c|}
\hline
\multicolumn{1}{|c}{ } & \multicolumn{1}{c}{ } & \multicolumn{1}{c|}{ } & \multicolumn{10}{c|}{\texttt{TSCI-RF}} & \multicolumn{2}{c|}{RF-Init} & \multicolumn{1}{c|}{\texttt{TSCI-RF}} \\
\cline{4-13} \cline{14-15} \cline{16-16}
\multicolumn{1}{|c}{ } & \multicolumn{1}{c}{ } & \multicolumn{1}{c|}{ } & \multicolumn{3}{c|}{Bias} & \multicolumn{3}{c|}{Length} & \multicolumn{3}{c|}{Coverage} & \multicolumn{1}{c|}{Invalidity} & \multicolumn{1}{c}{Bias} & \multicolumn{1}{c|}{Coverage} & \multicolumn{1}{c|}{Weak IV} \\
\hline
Setting & a & n & Orac & Comp & Robust & Orac & Comp & Robust & Orac & Comp & Robust & & Orac & Orac & \\
\hline
 &  & 1000 & 0.00 & 0.00 & 0.00 & 1.08 & 1.08 & 1.08 & 0.92 & 0.92 & 0.92 & 1.00 & 0.05 & 0.94 & 0.99\\

 &  & 3000 & 0.01 & 0.01 & 0.00 & 0.59 & 0.59 & 0.61 & 0.94 & 0.93 & 0.93 & 1.00 & 0.00 & 0.95 & 0.00\\

 & \multirow{-3}{*}{\centering\arraybackslash 0.0} & 5000 & 0.00 & 0.00 & 0.01 & 0.44 & 0.45 & 0.89 & 0.94 & 0.93 & 0.95 & 1.00 & 0.01 & 0.94 & 0.00\\
 \cline{2-16}

 &  & 1000 & 0.03 & 0.03 & 0.03 & 1.12 & 1.12 & 1.12 & 0.90 & 0.90 & 0.90 & 1.00 & 0.07 & 0.94 & 0.99\\

 &  & 3000 & 0.00 & 0.00 & 0.00 & 0.62 & 0.62 & 0.62 & 0.93 & 0.93 & 0.93 & 1.00 & 0.01 & 0.94 & 0.00\\

 & \multirow{-3}{*}{\centering\arraybackslash 0.5} & 5000 & 0.00 & 0.00 & 0.01 & 0.45 & 0.45 & 0.68 & 0.94 & 0.93 & 0.93 & 1.00 & 0.01 & 0.94 & 0.00\\
  \cline{2-16}

 &  & 1000 & 0.01 & 0.01 & 0.01 & 1.22 & 1.22 & 1.22 & 0.92 & 0.92 & 0.92 & 1.00 & 0.04 & 0.95 & 0.99\\

 &  & 3000 & 0.01 & 0.01 & 0.00 & 0.66 & 0.66 & 0.67 & 0.95 & 0.95 & 0.95 & 1.00 & 0.01 & 0.96 & 0.00\\

 & \multirow{-3}{*}{\centering\arraybackslash 1.0} & 5000 & 0.00 & 0.00 & 0.00 & 0.49 & 0.49 & 0.62 & 0.93 & 0.93 & 0.93 & 1.00 & 0.01 & 0.94 & 0.00\\
 \cline{2-16}

 &  & 1000 & 0.01 & 0.01 & 0.01 & 1.30 & 1.30 & 1.30 & 0.89 & 0.89 & 0.89 & 1.00 & 0.06 & 0.94 & 1.00\\

 &  & 3000 & 0.00 & 0.00 & 0.00 & 0.73 & 0.73 & 0.74 & 0.95 & 0.95 & 0.95 & 1.00 & 0.01 & 0.96 & 0.01\\

\multirow{-12}{*}{\centering\arraybackslash 1} & \multirow{-3}{*}{\centering\arraybackslash 1.5} & 5000 & 0.00 & 0.00 & 0.01 & 0.53 & 0.54 & 0.82 & 0.93 & 0.92 & 0.93 & 1.00 & 0.01 & 0.94 & 0.00\\

\hline
 &  & 1000 & 0.37 & 0.37 & 0.37 & 1.62 & 1.62 & 1.62 & 0.66 & 0.66 & 0.66 & 0.93 & 0.41 & 0.78 & 1.00\\

 &  & 3000 & 0.41 & 0.41 & 0.41 & 1.25 & 1.25 & 1.25 & 0.60 & 0.60 & 0.60 & 1.00 & 0.42 & 0.72 & 1.00\\

 & \multirow{-3}{*}{\centering\arraybackslash 0.0} & 5000 & 0.35 & 0.35 & 0.35 & 1.05 & 1.05 & 1.05 & 0.61 & 0.61 & 0.61 & 1.00 & 0.40 & 0.65 & 1.00\\
 \cline{2-16}

 &  & 1000 & 0.30 & 0.30 & 0.30 & 1.21 & 1.21 & 1.21 & 0.65 & 0.65 & 0.65 & 1.00 & 0.36 & 0.77 & 1.00\\

 &  & 3000 & 0.21 & 0.21 & 0.21 & 1.18 & 1.18 & 1.18 & 0.73 & 0.73 & 0.73 & 1.00 & 0.29 & 0.83 & 1.00\\

 & \multirow{-3}{*}{\centering\arraybackslash 0.5} & 5000 & 0.10 & 0.10 & 0.10 & 1.09 & 1.09 & 1.09 & 0.82 & 0.82 & 0.82 & 1.00 & 0.22 & 0.89 & 1.00\\
 \cline{2-16}

 &  & 1000 & 0.16 & 0.20 & 0.16 & 1.35 & 1.30 & 1.35 & 0.77 & 0.77 & 0.77 & 0.93 & 0.26 & 0.87 & 1.00\\

 &  & 3000 & 0.03 & 0.03 & 0.03 & 1.11 & 1.10 & 1.11 & 0.88 & 0.88 & 0.88 & 0.99 & 0.11 & 0.94 & 1.00\\

 & \multirow{-3}{*}{\centering\arraybackslash 1.0} & 5000 & 0.00 & 0.00 & 0.00 & 0.92 & 0.92 & 0.92 & 0.89 & 0.89 & 0.89 & 1.00 & 0.07 & 0.93 & 0.57\\
\cline{2-16}

 &  & 1000 & 0.06 & 0.22 & 0.06 & 1.59 & 1.31 & 1.59 & 0.83 & 0.67 & 0.83 & 0.75 & 0.18 & 0.93 & 1.00\\

 &  & 3000 & 0.01 & 0.02 & 0.01 & 1.14 & 1.13 & 1.14 & 0.90 & 0.90 & 0.90 & 0.98 & 0.08 & 0.93 & 0.87\\

\multirow{-12}{*}{\centering\arraybackslash 2} & \multirow{-3}{*}{\centering\arraybackslash 1.5} & 5000 & 0.00 & 0.00 & 0.00 & 0.91 & 0.91 & 0.91 & 0.93 & 0.93 & 0.93 & 1.00 & 0.04 & 0.95 & 0.02\\
\hline
\end{tabular}}
\caption{\small Bias, length, and coverage {(at nominal level 0.95)} for binary treatment model with Model 1 (continuous IV) and Model 4 (binary IV). The columns indexed with ``\texttt{TSCI-RF}" corresponds to our proposed \texttt{TSCI} with the random forests, where the columns indexed with ``Bias", ``Length", and ``Coverage" correspond to the absolute bias of the point estimator, the length and empirical coverage of the constructed confidence interval respectively. The columns indexed with ``Oracle", ``Comp" and ``Robust" correspond to the \texttt{TSCI} estimators with $\mathcal{V}_q$ selected by the oracle knowledge, the comparison method, and the robust method. The column indexed with ``Invalidity" reports the proportion of detecting the proposed IV as invalid. The columns indexed with ``RF-Init" correspond to the RF estimators without bias correction but with the oracle knowledge of the best $\mathcal{V}_q$. The column indexed with ``Weak IV" stands for the proportion out of 500 simulations reporting $Q_{\max}<1.$}
\label{tab: Model 4}
\end{table}

The binary treatment result is summarized in Table \ref{tab: Model 4}. The main observation is similar to those for the continuous treatment reported in the main paper. We shall point out the major differences in the following. The settings with the binary treatment are in general more challenging since the IV strength is relatively weak. To measure this, we have reported the column indexed with ``weak IV" standing for the proportion of simulations with $Q_{\max}=0.$ For settings where our proposed generalized IV strength is strong such that $Q_{\max}\geq 1$, our proposed \texttt{TSCI} method achieves the desired coverage level. Even when the generalized IV strength leads to $Q_{\max}<1,$ our proposed (oracle) \texttt{TSCI} may still achieve the desired coverage level for setting 1. 

\section{Additional Results for Real Data Analysis}
This section contains the additional results for the real data analysis. In Figure \ref{fig: var_importance}, we first show the importance score of all variables in random forests, based on which the six most important covariates are selected to construct the basis set $\mathcal{V}_1$ in Section \ref{sec: real}.

\begin{figure}[htp!]
    \centering
    \includegraphics[scale=0.85]{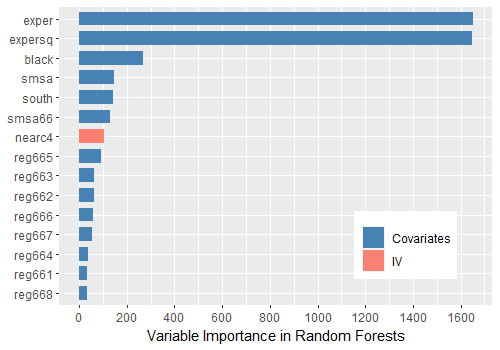}
    \caption{\small The importance score of each variable in random forest fitting.}
    \label{fig: var_importance}
\end{figure}

\subsection{Other specifications of $\mathcal{V}_2$}
\label{sec: add real}

In the following, in addition to the basis set $\mathcal{V}_2$ considered in Section \ref{sec: real}, we consider three more specifications of $\mathcal{V}_2$, as detailed in Table \ref{tab:specifications for v_2}, and test the robustness of \texttt{TSCI}'s selection  process.
\begin{table}[H]
    \renewcommand\arraystretch{1.5}
    \centering
    \resizebox{\linewidth}{!}{
    \begin{tabular}{|c|c|}
        \hline
        $\mathcal{V}_2$ & \makecell{$\mathcal{V}_1\cup$\{\texttt{nearc4}$\cdot$\texttt{reg1}, \texttt{nearc4}$\cdot$\texttt{reg2}, \texttt{nearc4}$\cdot$\texttt{reg3}, \texttt{nearc4}$\cdot$\texttt{reg4},\texttt{nearc4}$\cdot$\texttt{reg5}, \texttt{nearc4}$\cdot$\texttt{reg6}, \texttt{nearc4}$\cdot$\texttt{reg7}, \texttt{nearc4}$\cdot$\texttt{reg8}\}} \\
        \hline
        $\mathcal{V}_2^{(1)}$ & $\mathcal{V}_1\cup\{\texttt{nearc4}\cdot\texttt{exper}^3\}$ \\
        \hline
        $\mathcal{V}_2^{(2)}$ & \makecell{$\mathcal{V}_1\cup\{\texttt{nearc4}\cdot\texttt{exper}\cdot\texttt{black}, \texttt{nearc4}\cdot\texttt{exper}\cdot\texttt{south}, \texttt{nearc4}\cdot\texttt{exper}\cdot\texttt{smsa}, \texttt{nearc4}\cdot\texttt{exper}\cdot\texttt{smsa66}$,\\
        $\texttt{nearc4}\cdot\texttt{expersq}\cdot\texttt{black}, \texttt{nearc4}\cdot\texttt{expersq}\cdot\texttt{south}, \texttt{nearc4}\cdot\texttt{expersq}\cdot\texttt{smsa}, \texttt{nearc4}\cdot\texttt{expersq}\cdot\texttt{smsa66}$, \\
        $\texttt{nearc4}\cdot\texttt{black}\cdot\texttt{south}, \texttt{nearc4}\cdot\texttt{black}\cdot\texttt{smsa}, \texttt{nearc4}\cdot\texttt{black}\cdot\texttt{smsa66}$,\\
        $\texttt{nearc4}\cdot\texttt{south}\cdot\texttt{smsa}, \texttt{nearc4}\cdot\texttt{south}\cdot\texttt{smsa66}, \texttt{nearc4}\cdot\texttt{smsa}\cdot\texttt{smsa66}\}$} \\
        \hline
        $\mathcal{V}_2^{(3)}$ & \makecell{$\mathcal{V}_1\cup\{\texttt{nearc4}\cdot\texttt{exper}^3,\texttt{nearc4}\cdot\texttt{exper}\cdot\texttt{black}, \texttt{nearc4}\cdot\texttt{exper}\cdot\texttt{south}, \texttt{nearc4}\cdot\texttt{exper}\cdot\texttt{smsa}, \texttt{nearc4}\cdot\texttt{exper}\cdot\texttt{smsa66}$,\\
        $\texttt{nearc4}\cdot\texttt{expersq}\cdot\texttt{black}, \texttt{nearc4}\cdot\texttt{expersq}\cdot\texttt{south}, \texttt{nearc4}\cdot\texttt{expersq}\cdot\texttt{smsa}, \texttt{nearc4}\cdot\texttt{expersq}\cdot\texttt{smsa66}$, \\
        $\texttt{nearc4}\cdot\texttt{black}\cdot\texttt{south}, \texttt{nearc4}\cdot\texttt{black}\cdot\texttt{smsa}, \texttt{nearc4}\cdot\texttt{black}\cdot\texttt{smsa66}$,\\
        $\texttt{nearc4}\cdot\texttt{south}\cdot\texttt{smsa}, \texttt{nearc4}\cdot\texttt{south}\cdot\texttt{smsa66}, \texttt{nearc4}\cdot\texttt{smsa}\cdot\texttt{smsa66}\}$} \\
        \hline
    \end{tabular}}
    \caption{\small Different specifications of the second basis set $\mathcal{V}_2$. $\mathcal{V}_2$ is the specification we used in Section \ref{sec: real}. $\mathcal{V}_2^{(1)}$ contains the two-way interaction of the IV with covariates \texttt{exper} and \texttt{expersq} and $\mathcal{V}_1$; $\mathcal{V}_2^{(2)}$ includes all the two-way interactions of the IV with 6 most important covariates excluding the interaction of the IV with \texttt{exper}, \texttt{expersq}; $\mathcal{V}_2^{(3)}$ includes all the two-way interactions of the IV with 6 most important covariates.}
    \label{tab:specifications for v_2}
\end{table}

We implement \texttt{TSCI} with random forests as detailed in Algorithm \ref{algo: TSCI selection} with specifying $\mathcal{V}_0,\mathcal{V}_1$ as in Section \ref{sec: real} and different $\mathcal{V}_2$ in Table \ref{tab:specifications for v_2}. We observe that the point estimators and confidence intervals are relatively stable even with different specifications of $\mathcal{V}_2.$ %In Table \ref{tab: different V2 comp}, we demonstrate the inference and basis selections of \texttt{TSCI} when we use different specifications of $\mathcal{V}_2$. When Prob($Q_{\text{max}}>\widehat{q}_{c}$) is large, \texttt{TSCI}'s selection is not limited by the small IV strength after adjusting the violation forms. For example, if $Q_{\text{max}}=2$ and $\widehat{q}_c=1$, \texttt{TSCI} compares $\widehat{\beta}_q$ among $q=0,1,2$ and chooses $q=1$; while with $Q_{\text{max}}=1$ and $\widehat{q}_c=1$, \texttt{TSCI} only compares among $q=0,1$ and it would give up considering $q=2$ as a potential selection due to the weak IV strengths at $q=2$. Based on this, the selection results of \texttt{TSCI} with $\mathcal{V}_2^{(1)}$ and $\mathcal{V}_2^{(3)}$ might be affected by the weak IV strength issue, while the selection results of \texttt{TSCI} with $\mathcal{V}_2$ and $\mathcal{V}_2^{(2)}$ are unaffected. In all of four identifications, \texttt{TSCI} chooses $q=0$ or $q=1$ in most of the cases and seldom chooses $q=2$ even though it includes $q=2$ into comparison in general when we apply $\mathcal{V}_2$ and $\mathcal{V}_2^{(2)}$. Therefore, the basis set being identified as $\mathcal{V}_1$ is robust. 

\begin{table}[ht!]
\resizebox{\linewidth}{!}{
\begin{tabular}{|c|cc|ccc|c|c|}
        \hline
        \multirow{2}{*}{Identifications} & \multicolumn{2}{c|}{\texttt{TSCI}} & \multicolumn{3}{c|}{Proportions of selection} & \multirow{2}{*}{\makecell{Prob($Q_{\text{max}}>\widehat{q}_c$)}} & \multirow{2}{*}{IV str.} \\
	\cline{2-6}
		& Estimate & CI & $\widehat{q}_c=0$ & $\widehat{q}_c=1$ & $\widehat{q}_c=2$ &  &  \\
	\hline
	\texttt{TSCI}-$\mathcal{V}_2$ & 0.0604 & (0.0294, 0.0914) & 59.2\% & 38.2\% & 2.6\% & 93.6\% & 112.7950 \\
	\texttt{TSCI}-$\mathcal{V}_2^{(1)}$ & 0.0575 & (0.0263, 0.0886) & 40.0\% & 58.6\% & 1.4\% & 2.3\% & 111.8835 \\
	\texttt{TSCI}-$\mathcal{V}_2^{(2)}$ & 0.0614 & (0.0303, 0.0916) & 55.0\% & 40.0\% & 5.0\% & 72.9\% & 111.6233 \\
	\texttt{TSCI}-$\mathcal{V}_2^{(3)}$ & 0.0575 & (0.0268, 0.0891) & 39.2\% & 60.6\% & 0.2\% & 0\% & 113.0134 \\
	\hline
\end{tabular}}
\caption{\small Inference and basis selection of \texttt{TSCI} with different identifications of $\mathcal{V}_2$. The columns indexed with ``Inference'' correspond to the point estimators and the CI reported by \texttt{TSCI} in Algorithm \ref{algo: TSCI selection} with $\mathcal{V}_{\widehat{q}_c}$. The columns indexed with ``Proportions of selection'' correspond to proportions of different selections over 500 rounds of simulations.  The column indexed with ``Prob($Q_{\text{max}}>\widehat{q}_{c}$)'' indicates the proportion of $Q_{\text{max}}>\widehat{q}_c$. This case means that \texttt{TSCI} is able to selects $\widehat{q}_c$ without any weak IV constraints.  The last column shows the average IV strength for \texttt{TSCI} estimators. }
\label{tab: different V2 comp}
\end{table}

\subsection{Falsification Argument of Condition \ref{cond: identification}}
\label{sec: fal}
We propose in the following a falsification argument regarding Condition \ref{cond: identification} for the data analysis. Particularly, we demonstrate that the regression model using covariates belonging to $\mathcal{V}_1$ provides a good approximation of $g(z,x)$ but not $f(z,x)$. With the TSCI estimator $\widehat{\beta}_{\mathcal{V}_{\widehat{q}_c}}$, we construct the pseudo outcome $\widetilde{Y}=Y-D\widehat{\beta}_{\mathcal{V}_{\widehat{q}_c}}$ and if $\widehat{\beta}_{\mathcal{V}_{\widehat{q}_c}}$ is a reasonably good estimator of $\beta$, then the pseudo outcome $\{\widetilde{Y}_i\}_{1\leq i\leq n}$ can be used as proxies of $\{g(Z_i,X_i)\}_{1\leq i\leq n}.$ Then we implement two OLS regressions of the pseudo outcome on the covariates in $\mathcal{V}_1$ and $\mathcal{V}_2$ as defined in Table \ref{tab:basis real data}. In addition, we build a random forests prediction model for the pseudo outcome with the predictors $Z_i$ and $X_i$. To evaluate the performance, we split the data into a training set $\mathcal{A}_2$ and a test set $\mathcal{A}_1$ as in Section \ref{eq: split RF}, where test set $\mathcal{A}_1$ is used to estimate the out-of-sample MSE of the model constructed with the training set $\mathcal{A}_2$. We randomly split the data 500 times and report the violin plot of 500 MSE on the left panel of Figure \ref{fig: sigmaSq D pseudo_Y}. Since our specified basis set $\mathcal{V}_1$ leads to a nearly similar prediction performance as the random forests prediction model, this suggests that $\mathcal{V}_1$ provides a good approximation of the function $g(\cdot)$ if the \texttt{TSCI} estimator is accurate. In comparison, we use the treatment variable to replace the pseudo outcome and compare the MSE of the three prediction models. On the right panel of Figure \ref{fig: sigmaSq D pseudo_Y}, the random forests prediction model performs much better than the OLS with covariates in $\mathcal{V}_1$ and $\mathcal{V}_2$, indicating that $\mathcal{V}_1$ does not provide a good approximation for $f(\cdot)$ in the treatment model.

\begin{figure}[ht!]
\includegraphics[width=0.7\linewidth]{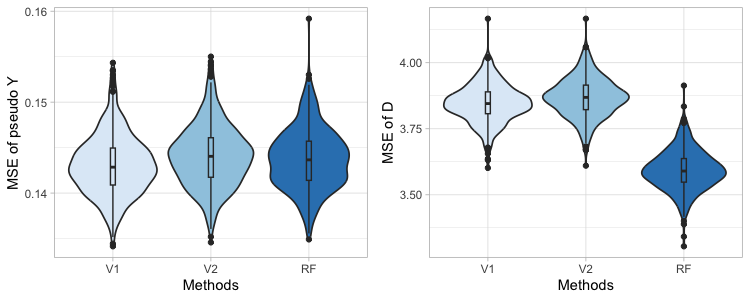}
\caption{\small Violin plot of MSE for different prediction models, where the left and right panels correspond to using $\widetilde{Y}_i=Y_i-D_i\widehat{\beta}_{\mathcal{V}_{\widehat{q}_c}}$ and $D_i$ as the regression outcome variables, respectively. On both panels, ``V1'', ``V2'', and ``RF" respectively stand for OLS regression with $\mathcal{V}_1$, $\mathcal{V}_2$, and random forests with $Z_i$ and $X_i$, where $\mathcal{V}_1$ and $\mathcal{V}_2$ are specified in Table \ref{tab:basis real data}.}
\label{fig: sigmaSq D pseudo_Y}
\end{figure}
\bibliographystylesupp{chicago.bst}
\bibliographysupp{IVRef}

\end{document}